\documentclass{aa}
\usepackage[utf8]{inputenc}
\usepackage{graphicx}
\usepackage{ulem,color}
\usepackage[dvipsnames]{xcolor}
\usepackage{txfonts}
\usepackage{textcomp}
\usepackage{comment}

\usepackage{float}
\usepackage[caption = false]{subfig}
\usepackage[switch]{lineno}
\usepackage{tablefootnote}

\newcommand{\CII}{C\,{\sc ii}}
\newcommand{\HI}{H\,{\sc i}}
\newcommand{\CIII}{C\,{\sc iii}}
\newcommand{\OIII}{O\,{\sc iii}}
\newcommand{\NII}{N\,{\sc ii}}

\newcommand{\kms}{km~s$^{-1}$}
\newcommand{\Lya}{Ly$\alpha$}

\newcommand{\udp}{Instituto de Estudios Astrofísicos, Facultad de Ingeniería y Ciencias, Universidad Diego Portales, Ej$\rm \acute{e}$rcito Libertador 441, 8370191, Santiago, Chile \label{udp}}

\newcommand{\puc}{Instituto de Astrofísica and Centro de Astroingeniería, Facultad de Física, Pontificia Universidad Católica de Chile, Casilla 306, Santiago 22, Chile \label{puc}}
\newcommand{\uc}{Departamento de Astronom\'ia, Universidad de Concepci\'on, Barrio Universitario, Concepci\'on, Chile \label{uc}}
\newcommand{\gu}{Sterrenkundig Observatorium, Ghent University, Krijgslaan 281 - S9, B-9000 Gent, Belgium \label{gu}}
\newcommand{\ugrcos}{Dept. Física Teórica y del Cosmos, Universidad de Granada, 18017, Granada, Spain \label{ugr1}}
\newcommand{\ugrcomp}{Instituto Universitario Carlos I de Física Teórica y Computacional, Universidad de Granada, 18071 Granada, Spain \label{ugr2}}
\newcommand{\texas}{Department of Physics and Astronomy and George P. and Cynthia Woods Mitchell Institute for Fundamental Physics and Astronomy, Texas A\&M University, College Station, TX, USA \label{texas}}
\newcommand{\cambridgecos}{Kavli Institute for Cosmology, University of Cambridge, Madingley Road, Cambridge, CB3 0HA, UK \label{cambridge1}}
\newcommand{\cambridgecomp}{Cavendish Laboratory - Astrophysics Group, University of Cambridge, 19 JJ Thomson Avenue, Cambridge, CB3 0HE, UK \label{cambridge2}}
\newcommand{\sns}{Scuola Normale Superiore, Piazza dei Cavalieri 7, 56126 Pisa, Italy \label{sns}}
\newcommand{\hgu}{Faculty of Engineering, Hokkai-Gakuen University, Toyohira-ku, Sapporo 062-8605, Japan \label{hgu}}
\newcommand{\oxford}{Department of Physics, University of Oxford, Denys Wilkinson Building, Keble Road, Oxford OX1 3RH, UK \label{oxford}}
\newcommand{\cab}{Centro de Astrobiolog\'{i}a (CAB), CSIC–INTA, Cra. de Ajalvir Km.~4, 28850- Torrej\'{o}n de Ardoz, Madrid, Spain \label{cab}}
\newcommand{\forth}{Institute of Astrophysics, Foundation for Research and Technology-Hellas (FORTH), Heraklion, 70013, Greece \label{forth}}
\newcommand{\cyprus}{School of Sciences, European University Cyprus, Diogenes street, Engomi, 1516 Nicosia, Cyprus \label{cyprus}}
\newcommand{\firenze}{Dipartimento di Fisica e Astronomia, Università di Firenze, Via G. Sansone 1, 50019, Sesto F.no (Firenze), Italy \label{firenze}}
\newcommand{\inaf}{INAF - Osservatorio Astrofisico di Arcetri, largo E. Fermi 5, 50127 Firenze, Italy \label{inaf}}
\newcommand{\sokendai}{Department of Astronomy, School of Science, SOKENDAI (The Graduate University for Advanced Studies), 2-21-1 Osawa, Mitaka, Tokyo 181-8588, Japan \label{sokendai}}
\newcommand{\utoc}{Department of Astronomy, The University of Tokyo, 7-3-1 Hongo, Bunkyo, Tokyo, 113-0033, Japan \label{utoc}}
\newcommand{\maxplank}{Max-Planck-Institut für extraterrestrische Physik, Gießenbachstraße 1, 85748 Garching \label{maxplank}}
\newcommand{\nao}{National Astronomical Observatory of Japan, 2-21-1 Osawa, Mitaka, Tokyo 181-8588, Japan \label{nao}}

\title{The ALMA-CRISTAL Survey:}\subtitle{Complex kinematics of the galaxies at the end of the Reionization Era}

\titlerunning{Complex kinematics of the galaxies at the end of the Reionization Era}

\author{K.~Telikova\inst{\ref{udp}},
J.~González-López\inst{\ref{puc}},
M.~Aravena\inst{\ref{udp}},
A.~Posses\inst{\ref{texas}},
V.~Villanueva\inst{\ref{uc}},
M.~Baeza-Garay\inst{\ref{uc}},
G.~C.~Jones\inst{\ref{oxford},\ref{cambridge1},\ref{cambridge2}},
M.~Solimano\inst{\ref{udp}},
L.~Lee\inst{\ref{maxplank}},
R.~J.~Assef\inst{\ref{udp}},
I.~De~Looze\inst{\ref{gu}},
T.~Diaz~Santos\inst{\ref{forth},\ref{cyprus}},
A.~Ferrara\inst{\ref{sns}},
R.~Ikeda\inst{\ref{sokendai},\ref{nao}},
R.~Herrera-Camus\inst{\ref{uc}},
H.~Übler\inst{\ref{cambridge1},\ref{cambridge2}},
I.~Lamperti\inst{\ref{firenze},\ref{inaf}},
I.~Mitsuhashi\inst{\ref{utoc},\ref{nao}},
M.~Relano\inst{\ref{ugr1},\ref{ugr2}},
M.~Perna\inst{\ref{cab}},
\and
K.~Tadaki\inst{\ref{hgu}}
}

\authorrunning{K. Telikova et al.}

\institute{\udp \and \puc \and \texas \and \uc \and \oxford \and \cambridgecos \and \cambridgecomp \and \maxplank \and \gu \and \forth \and \cyprus \and \sns \and \sokendai \and \nao \and \firenze \and \inaf \and \utoc \and \ugrcos \and \ugrcomp \and \cab \and \hgu}
\date{\today}

  \abstract
    {The history of gas assembly in early galaxies is reflected in their complex kinematics. While a considerable fraction of galaxies at $z\sim5$ are consistent with rotating disks, current studies indicate that the dominant galaxy assembly mechanism corresponds to minor and major mergers. Despite the important progress, the dynamical classification of galaxies at these epochs is still severely limited by observations’ angular and spectral resolution.
    We present a detailed morphological and kinematic analysis of the far-infrared bright main sequence galaxy HZ10 (CRISTAL-22) at $z = 5.65$, making use of new sensitive high-resolution ($\lesssim0.3\arcsec$) [\CII]~158$\mu$m ALMA and rest-frame optical JWST/NIRSpec integral field spectroscopy observations.
    These observations reveal a previously unresolved complex morphology and kinematics of the HZ10 system.
    Using position-velocity diagrams, we confirm that HZ10 is not a single massive galaxy but consists of at least three components in close projected separation along the east-to-west direction. We find a [\CII] bright central component (C), separated by 1.5~kpc and 4~kpc from the east (E) and west (W) components, respectively. Our [\CII] observations resolve the HZ10-C component resulting in a velocity gradient, which could be produced by either rotation or a close-in merger. We test the rotating disk possibility using \textsf{DysmalPy} kinematical modeling.
    Based on this, we propose three different dynamical scenarios for the HZ10 system: (i) a double merger, in which the companion galaxy HZ10-W merges with the disturbed clumpy rotation disk complex formed by the HZ10-C and E components; (ii) a triple merger, where the companion satellite galaxies, HZ10-W and HZ10-E, merge with the rotation disk HZ10-C; and (iii) a quadruple merger, in which the companion galaxies HZ10-W and HZ10-E merge with the close double merger HZ10-C. Additionally, from the comparison between ALMA [\CII]~158$\mu$m and JWST/NIRSpec data, we find that [\CII]~158$\mu$m emission closely resembles the broad [\OIII]~5007\AA\ emission both spatially and kinematically. The latter reflects the interacting nature of the system and suggests that ionized and neutral gas phases in HZ10 are well mixed.
    }

\begin{document}
\maketitle
\section{Introduction}\label{sec:intro}

Studying early galaxies and early-stage galaxy assemblies is essential for understanding the fundamental processes of galaxy formation and evolution.
While present-day galaxies are mostly quiescent, this picture dramatically differs from the early Universe, when galaxies were undergoing active interactions that shaped their assembly history. 
 
According to observations and simulations, the relative contribution of smooth cold gas accretion from the intergalactic medium and major mergers to the galaxy mass growth evolves across time \citep{Overzier2016}.
While at low redshifts the mass assembly is dominated by the smooth gas accretion, at higher redshifts, $z>3$, merger processes become increasingly important \citep[e.g.][]{Duncan2019,Romano2021}.

The morpho-kinematic analysis of emission lines in the spectra of galaxies is a powerful tool for determining the history of the gas assembly of the galaxies. This method, however, in addition to high spectral resolution, requires also high spatial resolution, which is challenging to achieve at high redshift.
While in the local Universe one of the most accessible tracers used to study the galaxy kinematics is the \HI~21cm emission line~\citep[e.g.][]{Walter2008,Begum2008,Hunter2012,Patra2016,deBlok2024arXiv}, at high redshift such observations are currently unavailable. So far, the most distant  
\HI~21cm emission line detections are at $z\approx0.4$~\citep{Xi2024} from a few non-lensed galaxies and at $z\approx1.3$ from one strongly lensed galaxy~\citep{Chakraborty2023}. While the Square Kilometre Array shows promise in exploring \HI~21cm signal from the early Universe, studying galaxy kinematics at redshifts $z>2$ will continue to be challenging~\citep{StaveleySmith2015}.

However, the [\CII] fine structure $^2P^{\circ}_{3/2} \to$ $^2P^{\circ}_{1/2}$ transition at rest-frame wavelength of 158$\mu$m provides a reasonable alternative. Since this line is easily excited by collisions of the carbon ion with electrons, as well as molecular and atomic hydrogen \citep{Goldsmith2012}, it traces multiple gas phases and is widely regarded as a good probe of galaxy kinematics in the early Universe \citep{deBlok2016}.
Being one of the key coolants of interstellar gas in galaxies and the strongest far-infrared emission lines, [\CII] is detectable up to very high redshifts \citep[e.g., up to $z\sim8.5$,][]{Bakx2020,Bouwens2022,Fujimoto2024}.  

A major step forward in the systematic study of high-redshift galaxies has been possible thanks to the Atacama Large Millimeter/submillimeter Array (ALMA) Large Program to INvestigate [\CII] at Early times (ALPINE)~\citep{LeFevre2020} and Reionization Era Bright Emission Line Survey (REBELS)~\citep{Bouwens2022}. 
Among many other results such as the presence of extended [\CII] halos around early star-forming galaxies~\citep[e.g.][]{Fujimoto2019,Fujimoto2020,Ginolfi2020}, these surveys revealed a significant fraction of merging galaxies at $z\sim5$ \citep{Romano2021}, yet with a considerable fraction of possible rotating disks, even at $z\sim7$~\citep{Jones2021,Smit2018Natur}. The ALPINE and most of the REBELS observations, however, were limited by their coarse angular resolution ($\sim1\arcsec$), yielding the need for targeted higher-resolution observations to investigate the morphology and kinematics of galaxies at $z>4$.
Indeed, these previously barely resolved observations were not suited for dynamical analysis and could lead to a considerable misclassification between rotation disks and mergers~\citep[e.g.][]{Rizzo2022}.

Thanks to the unprecedented spatial resolution reachable with ALMA and James Webb Space Telescope (JWST), we are entering the era of kpc-scale studies of galaxies within the first billion years of the Universe \citep[e.g.][]{HerreraCamus2022,Posses2023,Mitsuhashi2023,Solimano2024AGN,Solimano2024,Posses2024,Villanueva2024arXiv,Jones2024,Ikeda2024}. High-resolution ALMA observations confirmed the presence of disk-like galaxies at very high redshifts $z\sim6-7$~\citep[e.g.][]{Neeleman2023,Rowland2024}, however, frequently unveiling a complex morphology and interactions instead \citep[e.g.][]{Solimano2024,Posses2024,Ikeda2024}. Moreover, these observations
confirmed the existence of extended [\CII] emission and its possible connection to mergers and/or outflows~\citep[e.g.][]{Pizzati2020,Akins2022,Ikeda2024}.
However, such high-resolution observations remain scarce, and even with the advent of JWST, observations of [\CII] are essential as 
[\CII] emission in galaxies predominantly originates from photodissociation regions with contribution from atomic neutral and ionized gas \citep[e.g.,][]{Pineda2013,Pallottini2017,Olsen2017} and typically traces galaxy kinematics \citep[e.g.,][]{Kohandel2019}. In contrast, rest-frame optical line kinematics such as H$\alpha$ and [\OIII] might be dominated by outflows.

Using new high angular resolution ALMA observations ($\sim0.3\arcsec$) presented in this work, complemented by JWST data, we aim to 
perform detailed morphological and kinematic analyses of the cool and ionized gas in HZ10, a far-infrared bright and gas-rich galaxy in the COSMOS field at $z=5.65$. 
This source is part of a protocluster with recent or ongoing major merging processes \citep{Pavesi2018}, and thus provides an exceptional opportunity to look at the complex picture of starburst-driven phenomena and interactions occurring in overdensity environments at high redshifts~\citep{Oteo2018,Guaita2022,Lewis2018}.

The paper is organized as follows. In section~\ref{sec:targets_prior_knowledge} we describe previous studies of the HZ10 and its neighborhood. In section~\ref{sec:observations} we describe the sensitive high-resolution ALMA [\CII] observations used for the analysis and briefly describe Hubble Space Telescope (HST) and JWST data used for the comparison. In section~\ref{sec:overview} we present moment maps of HZ10 derived from new high-resolution ALMA [\CII] observations, along with a 2D parametric modeling of the moment-0 map and a kinematic analysis based on position-velocity (PV) diagrams and rotation curves. In section~\ref{sec:modeling} we perform kinematic modeling of the brightest [\CII] component of HZ10, assuming this component represents a rotating disk. We also examine the validity of this assumption, considering alternative dynamic scenarios for the system, such as a triple merger or a merger involving a disturbed disk and a satellite galaxy.  In section~\ref{sec:JWST_comparison} we compare ALMA observations with those from JWST/NIRSpec.
Lastly, we discuss the results and present
our conclusions in sections~\ref{sec:discussion} and ~\ref{sec:conclusions}, respectively.

Throughout the paper, we adopt a standard flat $\Lambda$CDM cosmology with the matter and dark energy density parameters $\Omega_m = 0.3$, and $\Omega_\Lambda = 0.7$, respectively, and with the reduced Hubble constant $h=0.7$, which gives a conversion scale of 5.899~kpc/\arcsec\ at redshift $z=5.65$.

\section{Early studies}\label{sec:targets_prior_knowledge}
The HZ10 galaxy was reported for the first time as a \Lya\ emitter candidate found in the Subaru narrowband imaging survey of the COSMOS field with Suprime-Cam~\citep{Murayama2007}. It was later confirmed at a spectroscopic redshift of $5.65$ with Keck/DEIMOS observations~\citep{Mallery2012}. This galaxy was found to be bright in the rest-frame ultraviolet (UV) emission and thus was selected for the 
ALMA survey of [\CII]~158$\mu$m and dust continuum emission conducted by ~\cite{Capak2015}.

These observations revealed in close proximity (at a projected distance of $\approx70$~kpc) and at the same redshift another massive gas-rich galaxy -- the dusty starburst galaxy CRLE, hidden in the previous Subaru and HST images by a foreground galaxy.
Photometric and spectroscopic studies of this field showed that HZ10 and CRLE reside in an overdensity of Ly$\alpha$ emitters, resembling an early protocluster at  $z\sim5.7$~\citep{Pavesi2018}, hence providing a unique opportunity to shed light on the link between star formation and the environment at the end of the Reionization Epoch.

HZ10 was found to be a ``normal'' massive main-sequence galaxy at  $z=5.6548$ with a stellar mass of $\log M_{\star}/M_{\odot}=10.39\pm0.17$ \citep{Capak2015}.
In addition to [\CII] emission, further detection of \CIII]~1909$\AA$ line emission was reported by \cite{Markov2022} thanks to VLT/X-Shooter spectroscopy. These observations showed the metal-rich nature of HZ10 with the gas-phase metallicity with respect to the solar value of $Z/Z_\odot=0.6^{+0.3}_{-0.5}$.
Using ALMA \cite{Pavesi2016} revealed a weak [\NII]~205$\mu$m emission. \cite{Pavesi2016} concluded that
while the smooth [\NII] velocity gradient of HZ10 favored a disk nature of the galaxy, a tentative indication of the spatial and kinematic offset between [\NII] and [\CII] emission was reminiscent rather of a clumpy disk or a recent/ongoing merger.
In addition to atomic gas, the molecular phase associated with HZ10 was detected via CO(2-1) emission using the Karl G. Jansky Very Large Array (VLA)~\citep{Pavesi2019}\footnote{Seeking higher CO(5-4) and CO(6-5) transitions with the Northern Extended Millimeter Array (NOEMA) resulted in non-detection for HZ10~\citep{Vieira2022}.}. These observations showed that HZ10 is remarkably rich in molecular gas with the estimated\footnote{\cite{Pavesi2019} estimated gas mass from CO luminosity assuming a Galactic conversion factor $\alpha_{\rm CO}\sim4.5$.} $\log M_{\rm gas}/M_\odot=11.1$. Moreover, the star-formation efficiency with a gas depletion timescale of $\sim1$~Gyr of HZ10 seems comparable to lower-redshift (from nearby up to $z\sim3$) disk main-sequence galaxies~\citep{Pavesi2019}.

Despite the significant progress made in characterizing HZ10 with the previous barely resolved observations ($\gtrsim 0.6$\arcsec), the kinematic properties of this galaxy could only be addressed with higher angular resolution data.
Recent high-angular resolution JWST/NIRSpec ($\sim0.15\arcsec$) integral field spectroscopy (IFS) observations of rest-frame optical emission~\citep{Jones2024} and ALMA ($\sim0.3\arcsec$) observations of dust emission~\citep{Villanueva2024arXiv} unveiled the complex morphology of HZ10. In this work, we address the kinematic properties of HZ10 based on the high angular and spectral resolution [\CII] 158$\mu$m emission. 

\section{Observations and data reduction}\label{sec:observations}

\subsection{ALMA}
High-resolution ALMA observations of the [\CII]~158$\mu$m line and dust continuum emission toward the HZ10 system, including the dusty star-forming galaxy CRLE, were obtained in Cycle 7 under program 2019.1.1075.S (PI: M. Aravena). These ALMA band 7 observations were obtained in C43-4/5 configuration during March 2021, under standard observatory weather conditions. Data calibration was performed using CASA pipeline version 6.4.1.12. No additional flagging was necessary. The brightness of the neighboring CRLE galaxy within the field of view of HZ10 yielded high signal-to-noise continuum emission. We thus conducted standard continuum self-calibration in two stages, masking out the CRLE and HZ10 in the process. The background rms decreased by $21\%$ from 19 to 15 $\mu{\rm Jy}\,{\rm beam}^{-1}$ for the continuum image. 
 
The signal-to-noise level achieved was not enough to perform line self-calibration. Imaging was performed using the \textsf{tclean} task in CASA following the procedures described in Herrera-Camus et al. (in prep.). 

Due to the high spatial resolution (synthesized beam size of $0.34\arcsec \times0.27\arcsec$, positional angle of 54.9 degrees, natural weighting) and sensitivity these observations were included as part of the extended sample of the [\CII] Resolved ISM in Star-forming ALMA Large program (CRISTAL). The HZ10 system is also denoted as CRISTAL-22 as part of the extended CRISTAL sample. CRISTAL is an ALMA Cycle-8 large program (2021.1.00280.L; PI: R. Herrera-Camus) aimed at getting spatially resolved [\CII] and dust continuum observations of main-sequence star-forming galaxies at $z \sim 4 - 6$ (Herrera-Camus et al. (in prep.), \cite{Ikeda2024,Mitsuhashi2024}).

We examined whether our conclusions are sensitive to the JvM effect correction \citep{JvM1995} by cross-checking all the results based on the data cubes with and without JvM correction applied. We conclude that in both cases the flux density estimations are consistent. Therefore, all the figures throughout the paper are based on the ALMA cubes without JvM correction.

\subsection{HST}
To compare the ALMA data of dust continuum and [\CII] emission with the rest-frame UV  HST images we retrieved the available data for the HZ10 field from the Barbara A. Mikulski Archive for Space Telescopes (MAST\footnote{https://mast.stsci.edu/portal/Mashup/Clients/Mast/Portal.html}). 
We used archival observations with the WFC3/F105W filter which covers the wavelength range of $1300-1800$\AA\ in the rest frame at $z = 5.65$. Images were processed using the
standard pipeline, co-added, and aligned to Gaia DR2 \citep{GaiaDR2}. The image was drizzled with a square kernel and a pixel fraction of 0.5 using the \textsf{AstroDrizzle} routine~\citep{AstroDrizzle} of \textsf{DrizzlePac}~\citep{DrizzlePac} to a pixel size of 0.06\arcsec.

\subsection{JWST}

HZ10 is also a part of the JWST/NIRSpec Guaranteed Time Observation GA-NIFS project 1217 (PI: N. L\"{u}tzgendorf). Detailed analysis of this data is presented in \citet{Jones2024}. \citet{Jones2024} compared JWST data astrometry with the Gaia DR3 \citep{gaia16,gaia21} reference frame and found no significant offset.
Due to similar spatial resolution (fiducial point spread function is $0.15\arcsec$) and high spectral resolution (G395H/F290LP; $R\approx 2700$) of JWST/NIRSpec observation we compare them with our ALMA data.

\section{High-resolution ALMA view on HZ10}\label{sec:overview}

\begin{figure}[h]
{\includegraphics[trim={0 0.7cm 0 0.3cm},clip,width = 1.0\columnwidth]{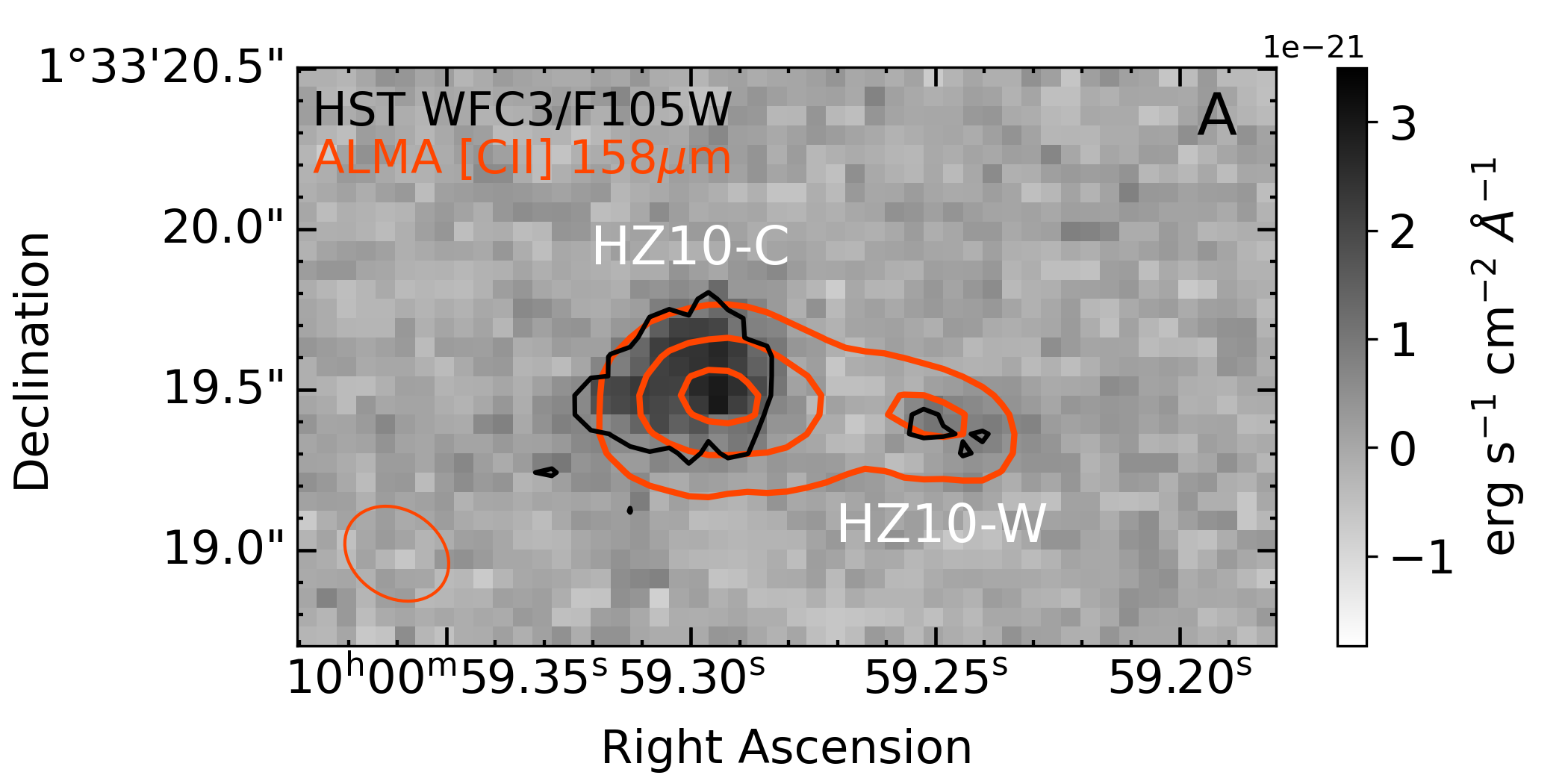}} 
{\includegraphics[trim={0 0.7cm 0 0.3cm},clip,width = 1.0\columnwidth]
{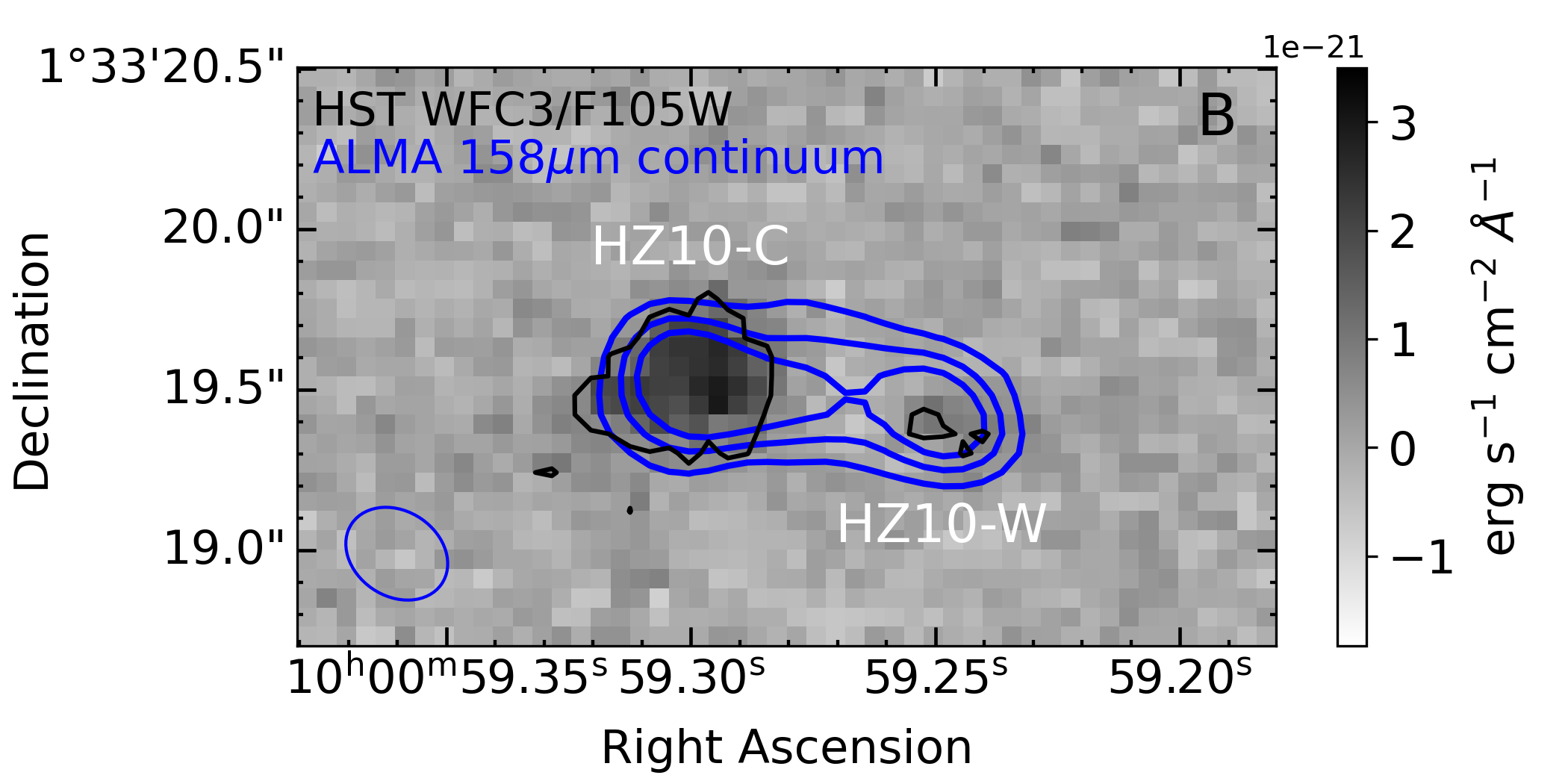}}

\caption{HST WFC3/F105W image of HZ10 is shown on panels A and B in greyscale with the black 3-$\sigma$ contours.
A: [\CII] 158 $\mu$m integrated intensity of HZ10 is shown by 5-, 10-, and 15-$\sigma$ red contours. 
B: 158 $\mu$m continuum of HZ10 is shown by 3-, 5-, and 7-$\sigma$ blue contours.
The beam size of the ALMA observations is shown by an ellipse in the bottom left corner of each panel.}\label{fig:hst-alma-HZ10}
\end{figure}

The comparison between the HST WFC3/F105W image with [\CII] 158$\mu$m line and 158$\mu$m dust continuum emission of HZ10 is shown in Fig.~\ref{fig:hst-alma-HZ10}. 

The new ALMA observations of HZ10 reveal two spatially separated components seen in the [\CII] 158$\mu$m integrated intensity map, HZ10-C (center) and HZ10-W (west). The HZ10-W component in the HST image is barely detected, likely due to the significant dust attenuation. 
In addition to the HZ10-C and HZ10-W components, \cite{Villanueva2024arXiv} also reported a tentative third component seen as a 158~$\mu$m dust continuum emission excess, a dusty bridge connecting the central and west components of HZ10. We will address the possible presence of this bridge in [\CII] line emission data in the following subsections.

\subsection{Moment maps}

\begin{figure*}[h]
{\includegraphics[trim={0.3cm 0 0 0},clip,width = 0.33\textwidth]{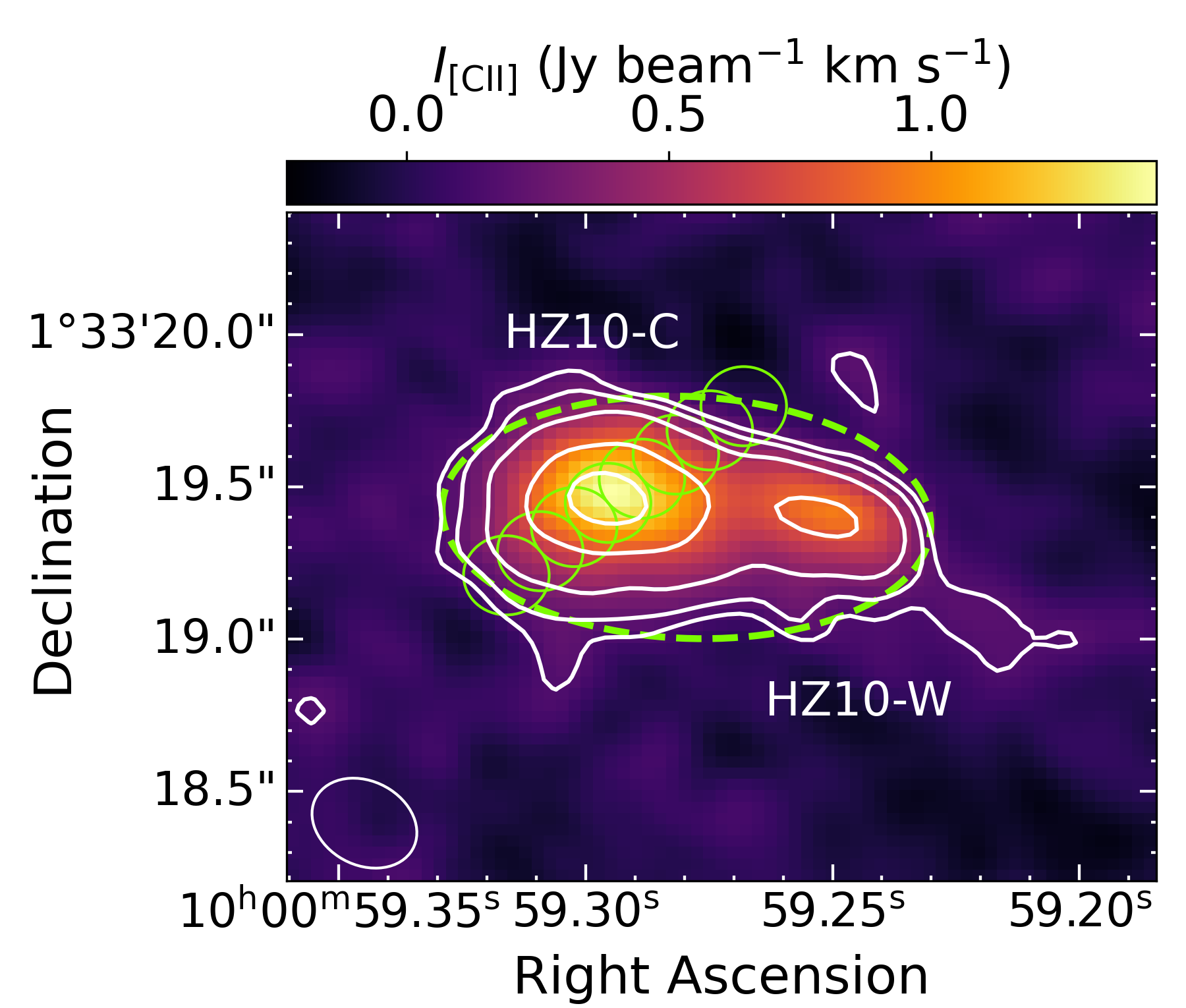}} 
{\includegraphics[trim={0.3cm 0 0.0cm 0},clip,width = 0.33\textwidth]{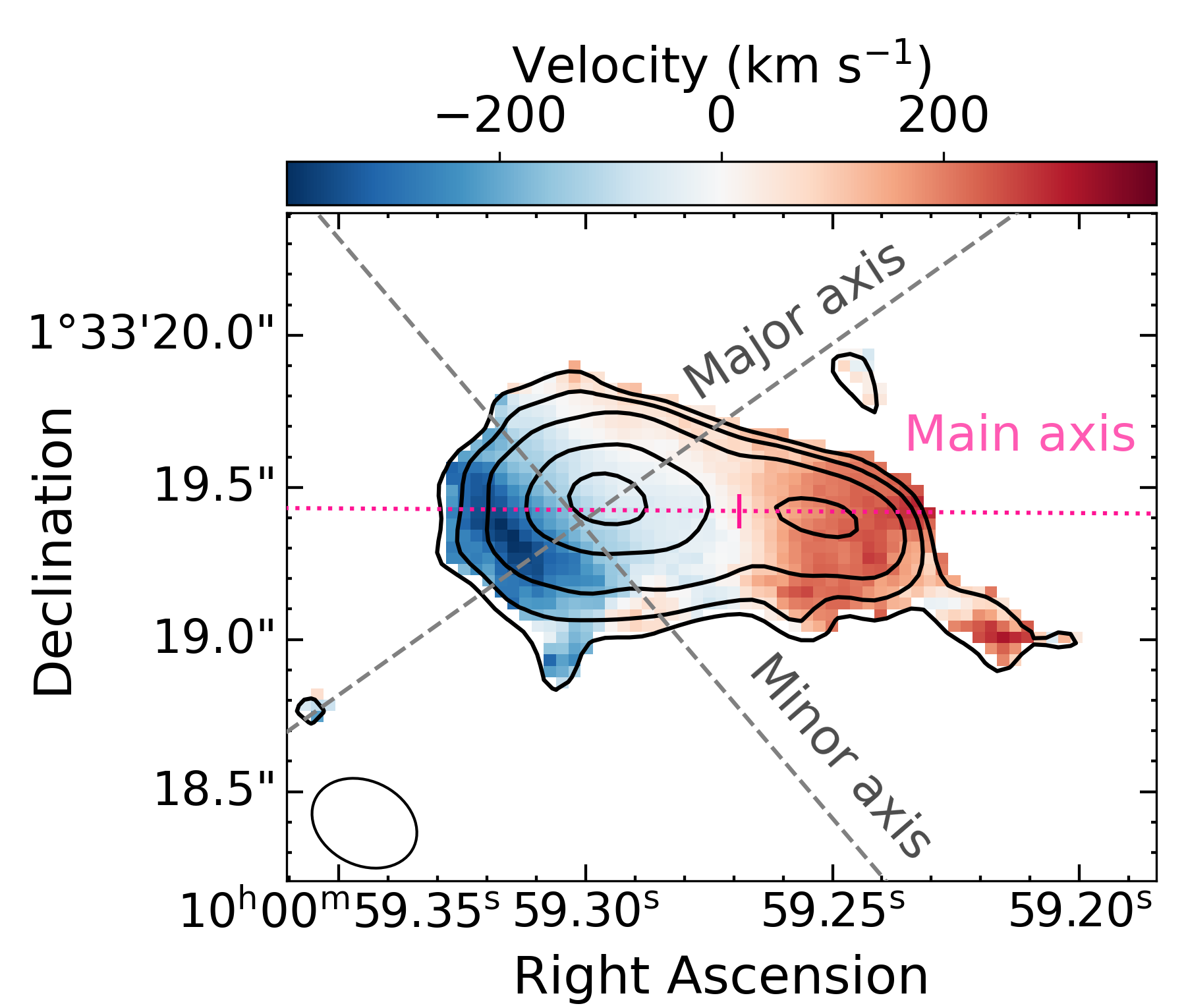}}
{\includegraphics[trim={0.3cm 0 0 0},clip,width = 0.33\textwidth]{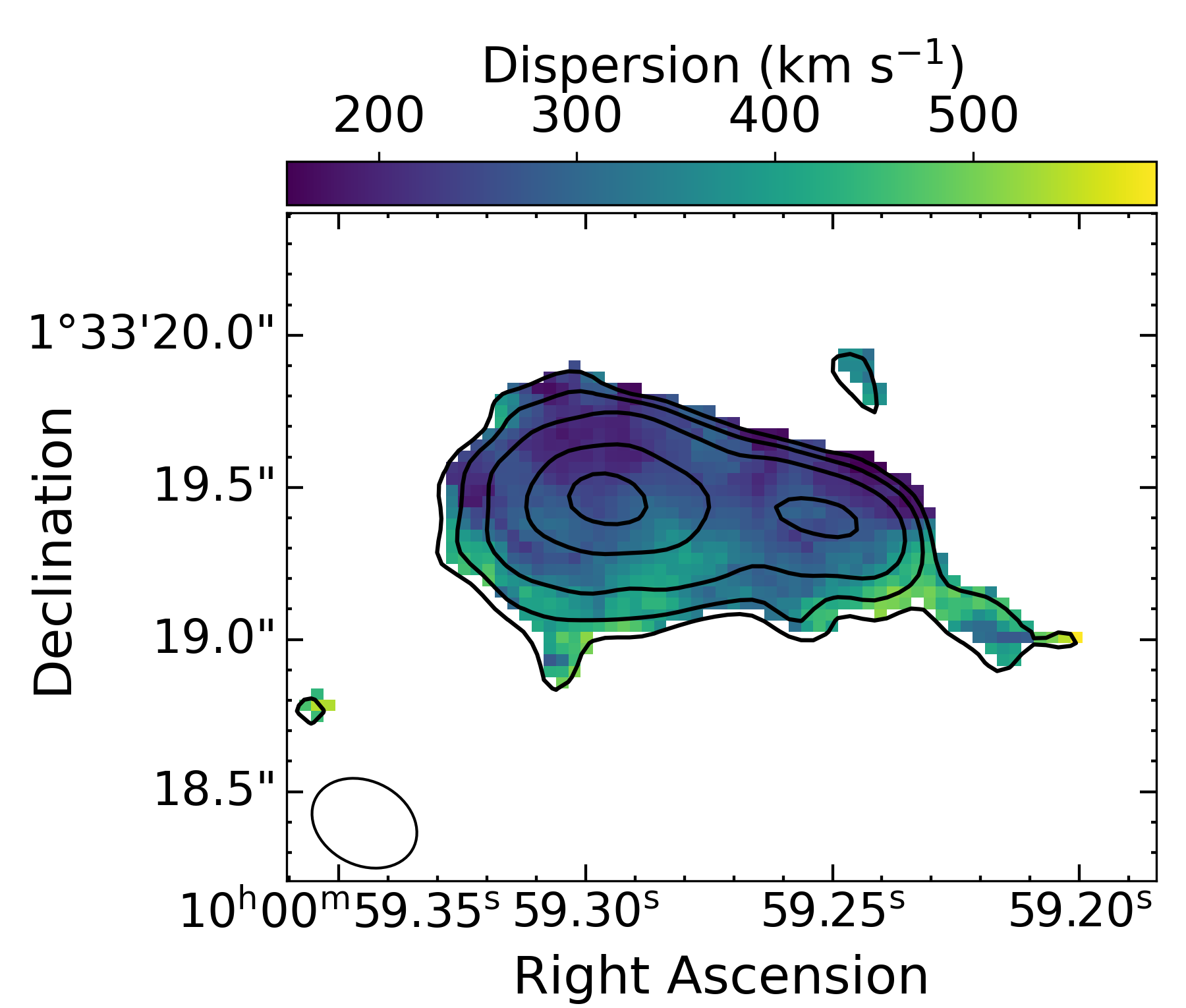}}
\caption{[\CII] integrated intensity (moment-0, left panel), line-of-sight velocity (moment-1, middle panel), and velocity dispersion (moment-2, right panel) maps of HZ10.
Contours correspond to 2-, 3-, 5-, 10-, and 15-$\sigma$ noise levels estimated from the corresponding moment-0 map. 
Zero velocity corresponds to the rest-frame [\CII] frequency at $z=5.6548$. The dashed and solid green regions on the moment-0 map correspond to the apertures for the spectra extraction; see the text for details.
The dashed grey and dotted pink lines on the moment-1 map correspond to the axes for the position-velocity diagrams calculations, see text for the details. The center of the main axis of HZ10 is indicated with the pink vertical line.
An ellipse shows the beam size of the ALMA observations in the bottom left corner of each panel.}\label{fig:moments-HZ10}
\end{figure*}

The [\CII] integrated intensity (or moment-0 map) is shown on the left panel of Fig.~\ref{fig:moments-HZ10}. 
We perform 2D parametric modeling of [\CII] integrated intensity map assuming a two-component S\'ersic profile and using \textsf{PyAutoGalaxy}~\citep{pyautogalaxy}, which is based on the probabilistic programming language \textsf{PyAutoFit}~\citep{pyautofit}. The results along with the other main properties of HZ10 are summarized in Table~\ref{tab:hz10_properties}. Details of the 2D parametric modeling are presented in Appendix~\ref{a:2D_morphological_modeling}.

\begin{table}[t]
\begin{minipage}{\columnwidth}  
\centering
\caption{General properties of HZ10 system. 
    }
    
    \begin{tabular}{c|c}    
    \hline
         &HZ10 \\
          &HZ10$-$C | HZ10$-$W \\
         \hline \hline
redshift\footnote{Systemic redshift.}         & 5.6548  \\
$I_{\text{[\CII]}}$ (Jy \kms)&$4.9\pm0.2$\\  
\hline
Center RA (10$^{\rm h}$00$^{\rm m}$59$^{\rm s}$)    
&0$^{\rm s}$.298 $\pm$ 0$^{\rm s}$.038 | 0$^{\rm s}$.254 $\pm$ 0$^{\rm s}$.039\\
Center Dec ($+$01$^{\rm d}$33$^{\rm m}$19$^{\rm s}$)         &0$^{\rm s}$.419 $\pm$ 0$^{\rm s}$.037 | 0$^{\rm s}$.349 $\pm$ 0$^{\rm s}$.038 \\

S\'ersic index\footnote{Since this study utilizes ALMA data products with natural weighting, the results of the 2D parametric analysis differ from those in \cite{Villanueva2024arXiv}, where the authors applied Briggs (\textsf{robust~$=$~0.5}) weighting, which leads to a higher angular resolution, but lower sensitivity in comparison with the Natural weighting.} & $0.53 \pm 0.05$ | $0.59 \pm 0.05$\\
Effective radius (kpc) & $1.5 \pm 0.2$ | $1.0 \pm 0.2$\\
Axis ratio (min/maj)& $0.77 \pm 0.01$ | $0.55 \pm 0.02$\\
$I_{\text{[\CII]}}^{(r=0)}$ (Jy beam$^{-1}$ \kms)& $0.75 \pm 0.02$ | $0.58 \pm 0.04$\\
\hline
    \end{tabular}

    \label{tab:hz10_properties}
\end{minipage}    
\end{table}

To reduce the impact of noisy pixels on the resulting velocity and velocity dispersion maps, we masked those pixels out before calculating moment-1 and moment-2. To do that we exclude pixels with intensity lower than the 2-$\sigma$ noise level of the respective channel. 
To ensure that channel masking does not result in underestimating the velocity dispersion, we compared the dispersion map calculated from the masked data with that derived from the unmasked data cube and found the values consistent. The resulting velocity and dispersion maps are shown on the middle and right panels of Fig.~\ref{fig:moments-HZ10}. 

The combination of the two spatially separated components, HZ10-C (central, the brightest component on the [\CII] moment-0 map) and HZ10-W (to the west of HZ10-C with dimmer, and more compact [\CII] emission) results in the smooth velocity gradient on the moment-1 map.

\begin{figure}[h]
{\includegraphics[trim={0 0cm 0 0},clip, width = 1.0\columnwidth]{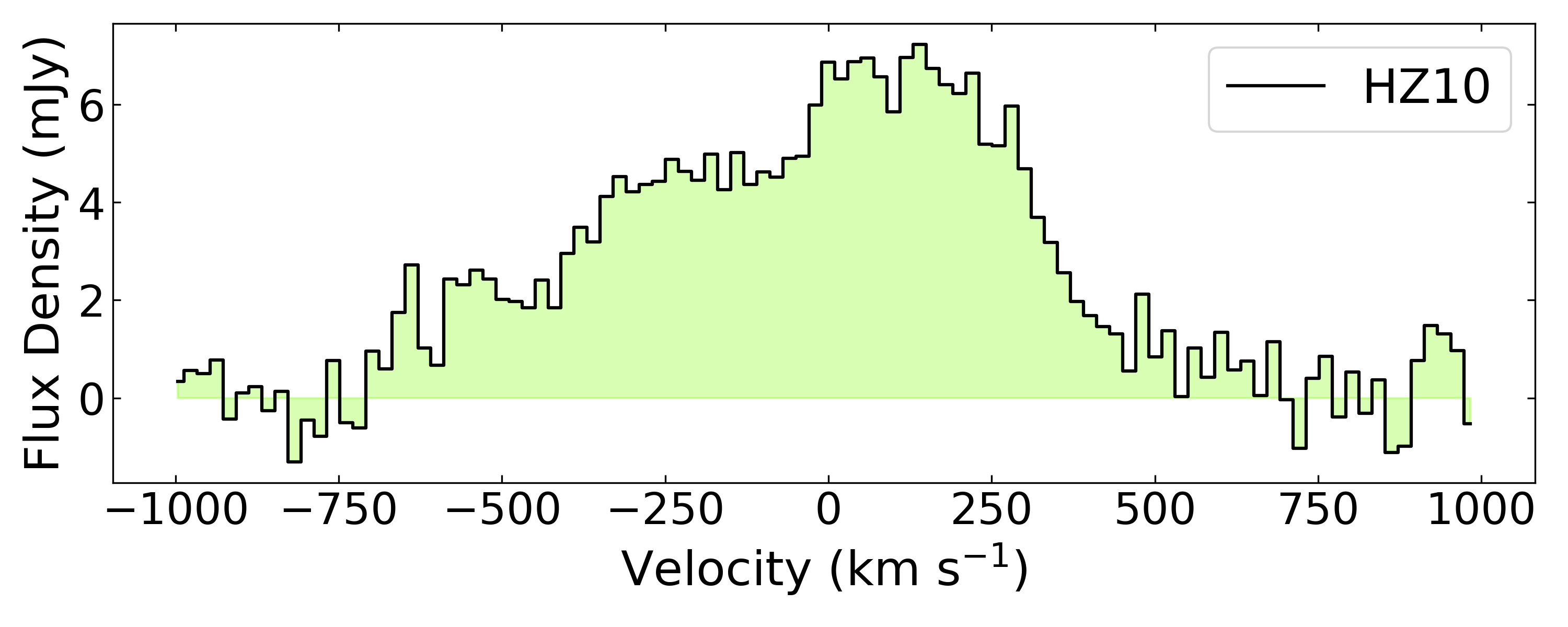}}
\caption{[\CII] spectrum of HZ10 complex, extracted from the elliptical aperture, shown on the left panel of Fig.~\ref{fig:moments-HZ10}. Zero velocity corresponds to the rest-frame [\CII] frequency at $z=5.6548$}\label{fig:hz10_spectra}
\end{figure}

For reference, in Fig.~\ref{fig:hz10_spectra}, we show the [\CII] spectra of the HZ10 complex, extracted from the elliptical aperture shown on the left panel of Fig.~\ref{fig:moments-HZ10}. Zero velocity corresponds to the rest-frame [\CII] frequency at $z=5.6548$.
The integrated [\CII] flux within this aperture is $I_{\text{[\CII]}}=4.8\pm0.2$~Jy~\kms. This flux is in agreement with the previously reported value estimated from the barely resolved [\CII] observations of \cite{Pavesi2016}.

\subsection{Kinematic analysis}\label{sec:analysis}

In contrast with the relatively compact HZ10-W component,  HZ10-C is resolved along its major kinematic axis in $\approx5$ independent beams. This allowed us to perform a more detailed kinematic analysis of the HZ10-C component using position-velocity diagrams along the different directions, and by measuring rotation velocity and dispersion curves.

\begin{figure}[h]
{\includegraphics[width = 1.0\columnwidth]{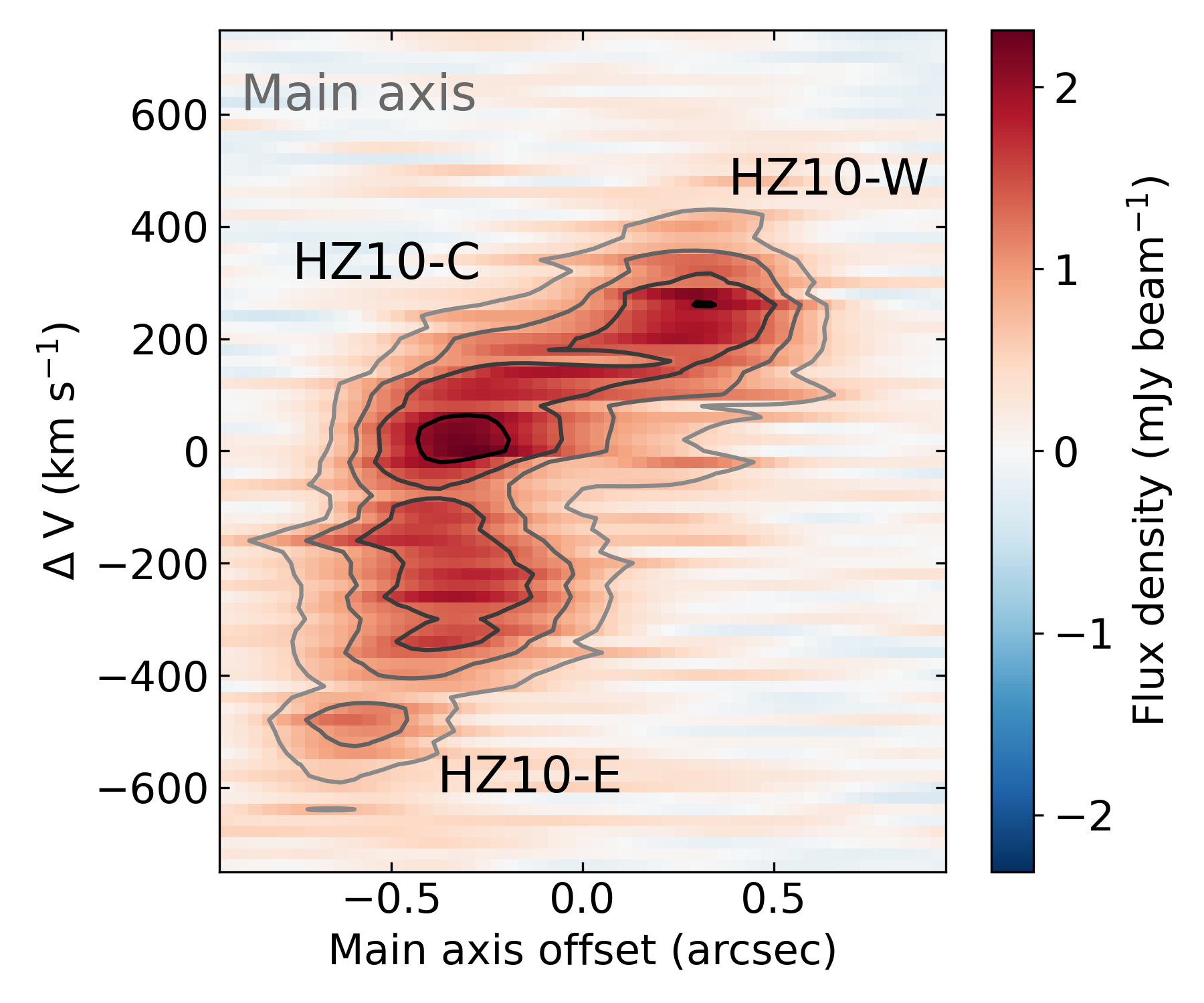}}
\caption{Position-velocity diagram calculated along the pseudo slit of 10-pixel width placed along the main axis in Fig.~\ref{fig:moments-HZ10}) characterized by the largest velocity gradient of the whole HZ10 complex. Smoothed grey contours correspond to the 3-, 5-, 7-, and 10-$\sigma$ noise levels. Zero velocity is calculated for the rest-frame [\CII] frequency at $z=5.6548$. }\label{fig:PV-HZ10-whole}
\end{figure}

Firstly, we look at the different kinematic components of HZ10 as a whole complex, including the HZ10-W, and calculate the PV diagram along the ``main'' axis connecting HZ10-C and HZ10-W components (the axis of the highest velocity gradient of the whole system, see the pink dotted line on the middle panel of Fig.~\ref{fig:moments-HZ10}).
We placed a pseudo slit of 10-pixel width, which is comparable with the beam size of the observations (pixel size of the data cube is 0.0372\arcsec), along the main axis of HZ10. The result is shown in Fig.~\ref{fig:PV-HZ10-whole}. 
 
Recent high-resolution studies of HZ10 dust continuum emission demonstrated the presence of a dusty bridge seen in 158$\mu$m continuum emission, connecting HZ10-C and HZ10-W components \citep{Villanueva2024arXiv}. The presence of such a bridge could result from the interacting nature of the HZ10 complex. Interestingly, at the spatial location of the ``bridge'' component seen in the 158 $\mu$m dust continuum, we do not see such a distinctive component in the residuals of [\CII] line emission 2D parametric fit with two S\'ersic profiles (see Fig~\ref{fig:pyautofit}). However, on the PV diagram in Fig.~\ref{fig:PV-HZ10-whole} we see [\CII] emission connecting HZ10-C and HZ10-W, which could be associated with the dusty ``bridge'' component or potentially attributed to beam smearing.

\begin{figure}[h]
{\includegraphics[trim={0 2.9cm 0 0},clip, width = 1.0\columnwidth]{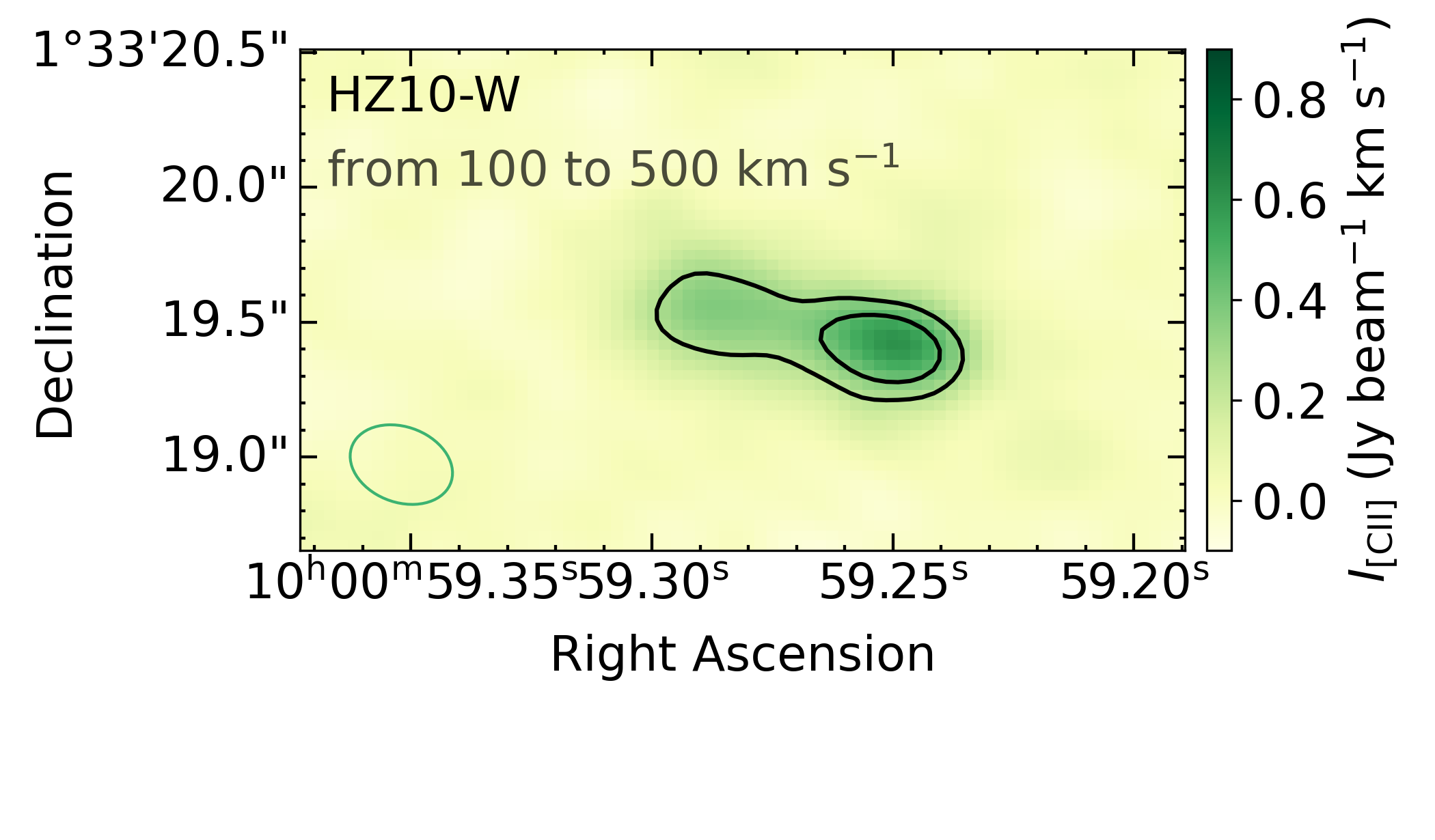}}
{\includegraphics[trim={0 2.9cm 0 0},clip,width = 1.0\columnwidth]{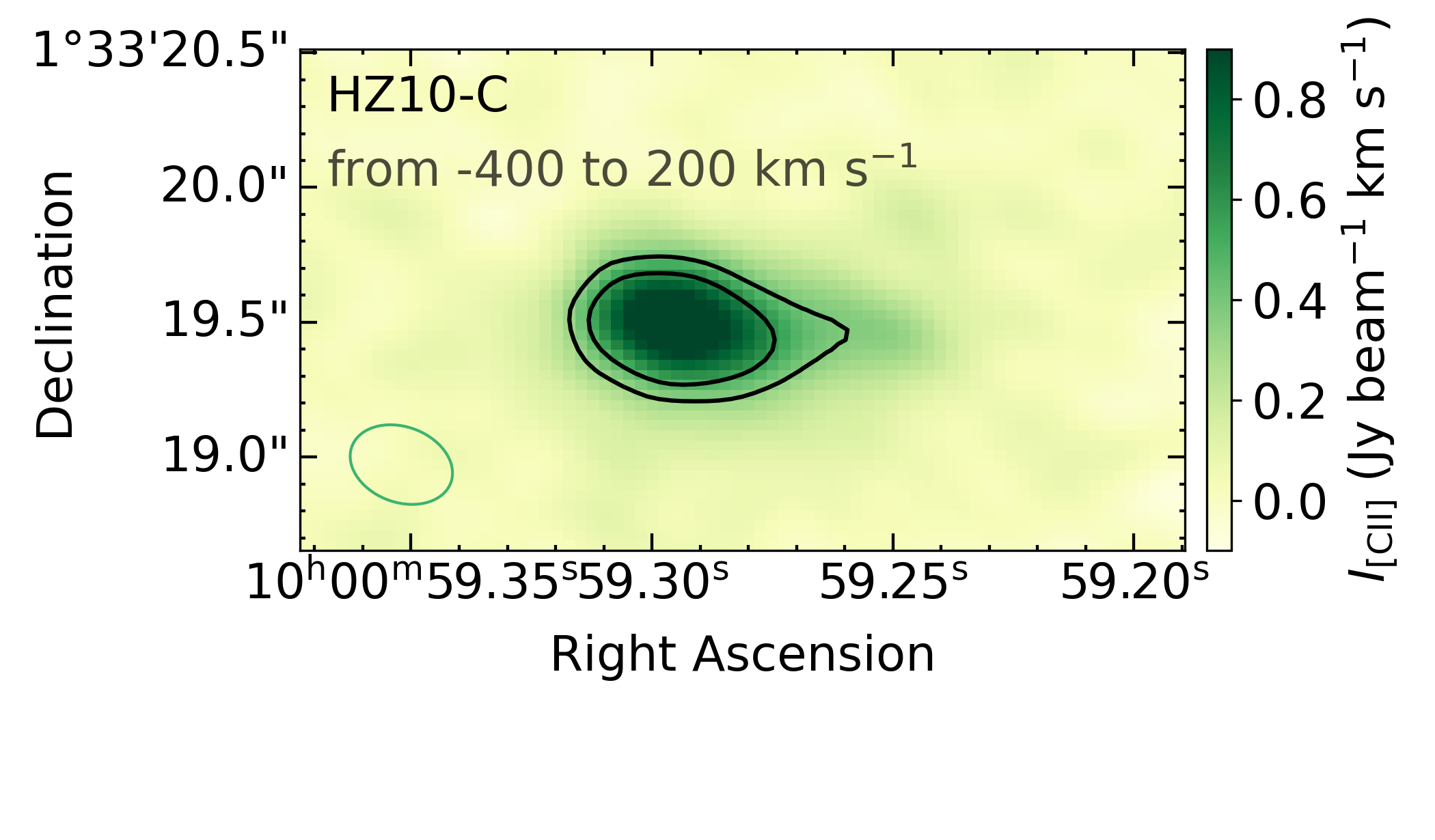}}
{\includegraphics[trim={0 1.5cm 0 0}, clip, width = 1.0\columnwidth]{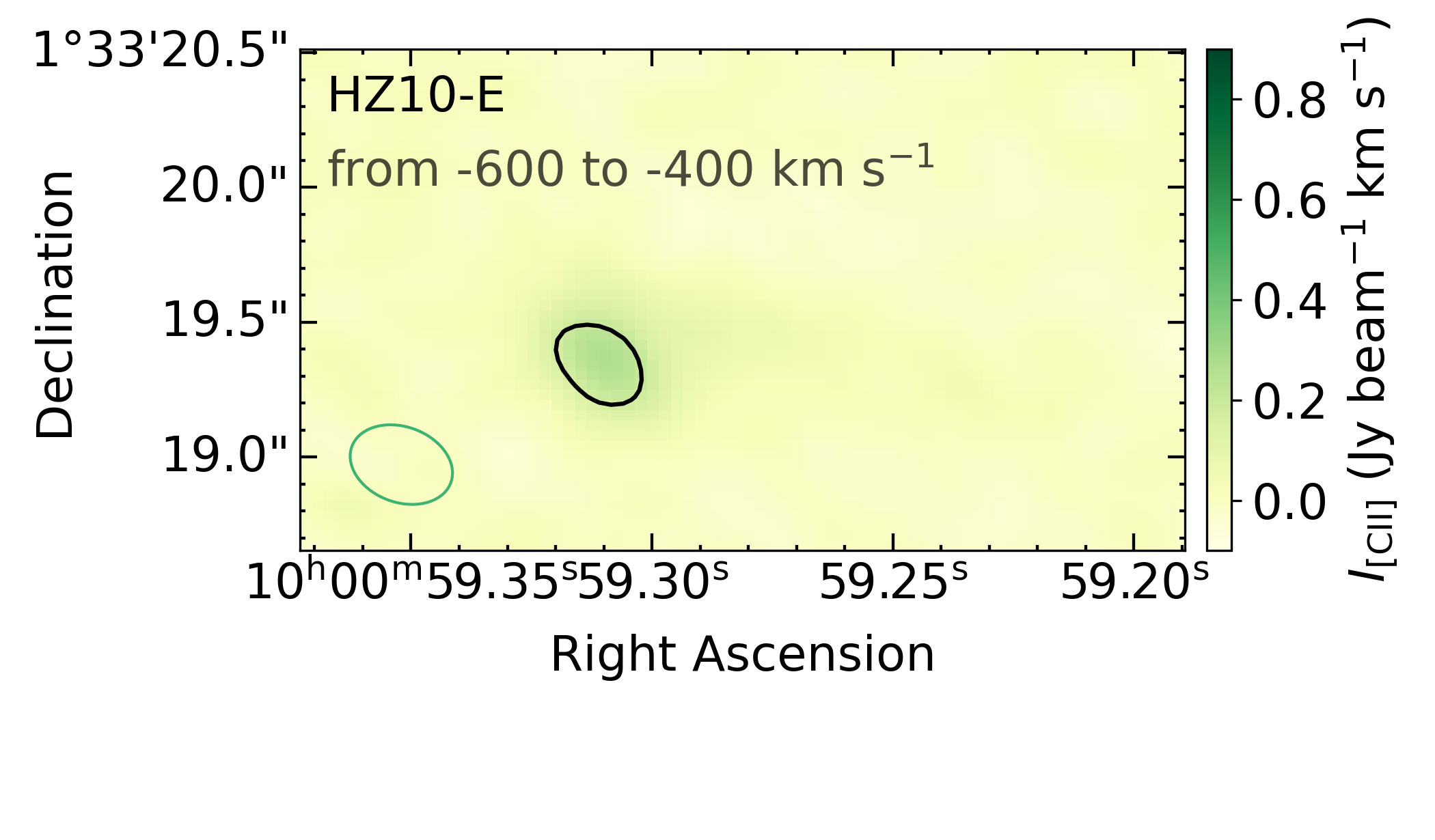}}
\caption{Integrated [\CII] intensity calculated within specific velocity ranges, defined from the PV diagram along the main axis of the HZ10 complex. Contours correspond to the 10-, and 15-$\sigma$ noise levels of the corresponding moment-0 maps. The beam size of the ALMA observations is shown in the green ellipse at the bottom left corner of each panel.}\label{fig:hz10_mom0_comps}
\end{figure}

The HZ10-W component appears in the PV diagram within the velocity range of $\Delta V$ from 100 to 500~\kms, while the HZ10-C component appears within the velocity range of $\Delta V$ from $-600$ to 200~\kms\ showing a hint of two different sub-components, a bright one with $\Delta V$ from $-400$ to 200~\kms\ consisting of two bright blobs typical for rotation disks or mergers, and a dimmer one with $\Delta V$ from $-600$ to $-400$~\kms. To illustrate the spatial positions of these three components, we show moment-0 maps, calculated within the specified velocity ranges in Fig.~\ref{fig:hz10_mom0_comps}. 

Considering the velocity and spatial separation between the HZ10-C and HZ10-W components, along with the JWST/NIRSpec integral-field spectroscopic study of the HZ10 system by \cite{Jones2024}, which shows that HZ10-W has physical properties such as UV slope and color excess distinct from those of HZ10-C, we further regard HZ10-W as a separate galaxy merging with HZ10-C.

The presence of a dim emission blob to the east of the main HZ10-C component suggests that the bright central complex on the moment-0 map of the whole system could be either a merger or a clumpy disk consisting of two sub-components -- HZ10-C and HZ10-E (east). 

\begin{figure}[h]
{\includegraphics[width = 1.0\columnwidth]{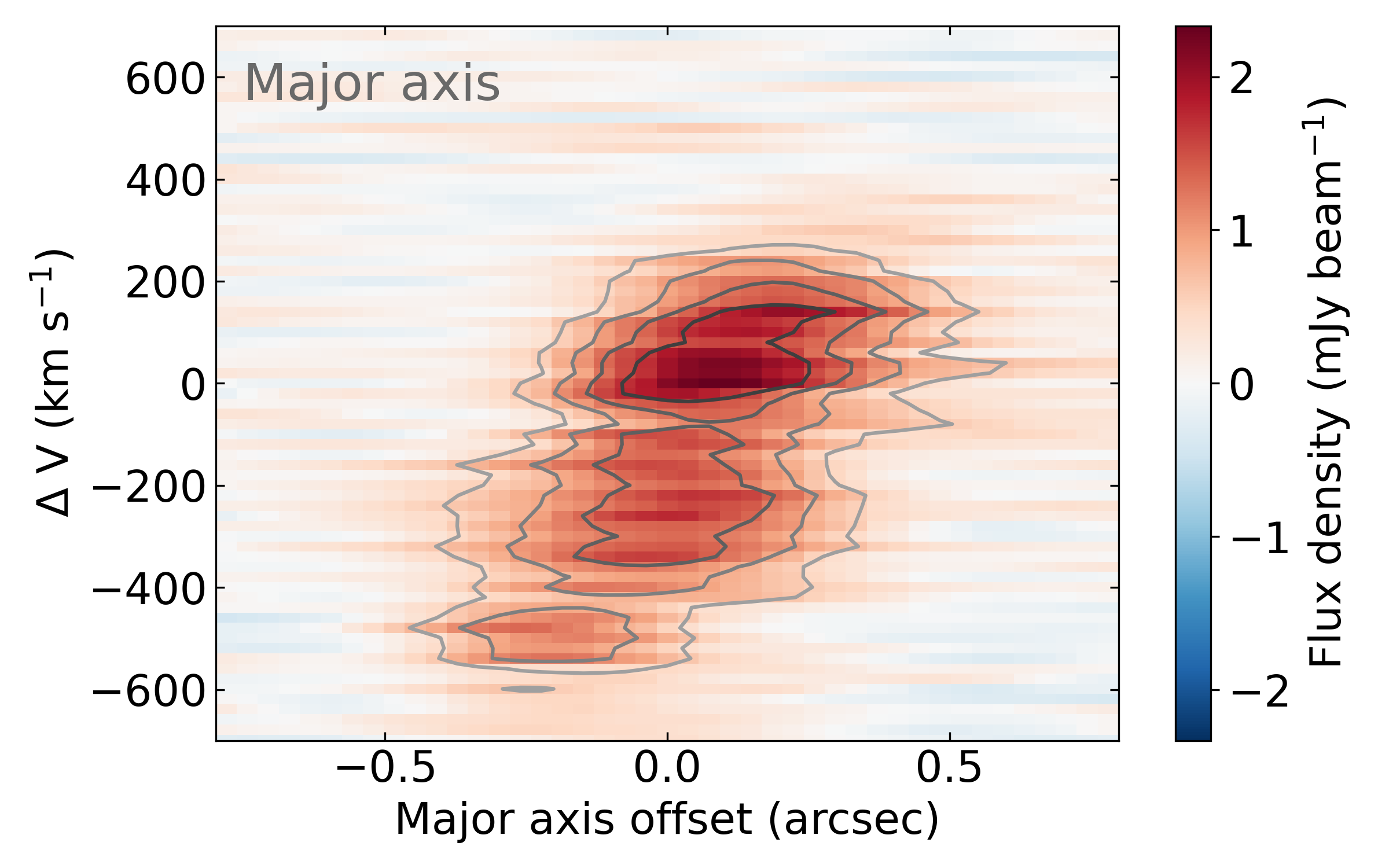}} 
{\includegraphics[width = 1.0\columnwidth]{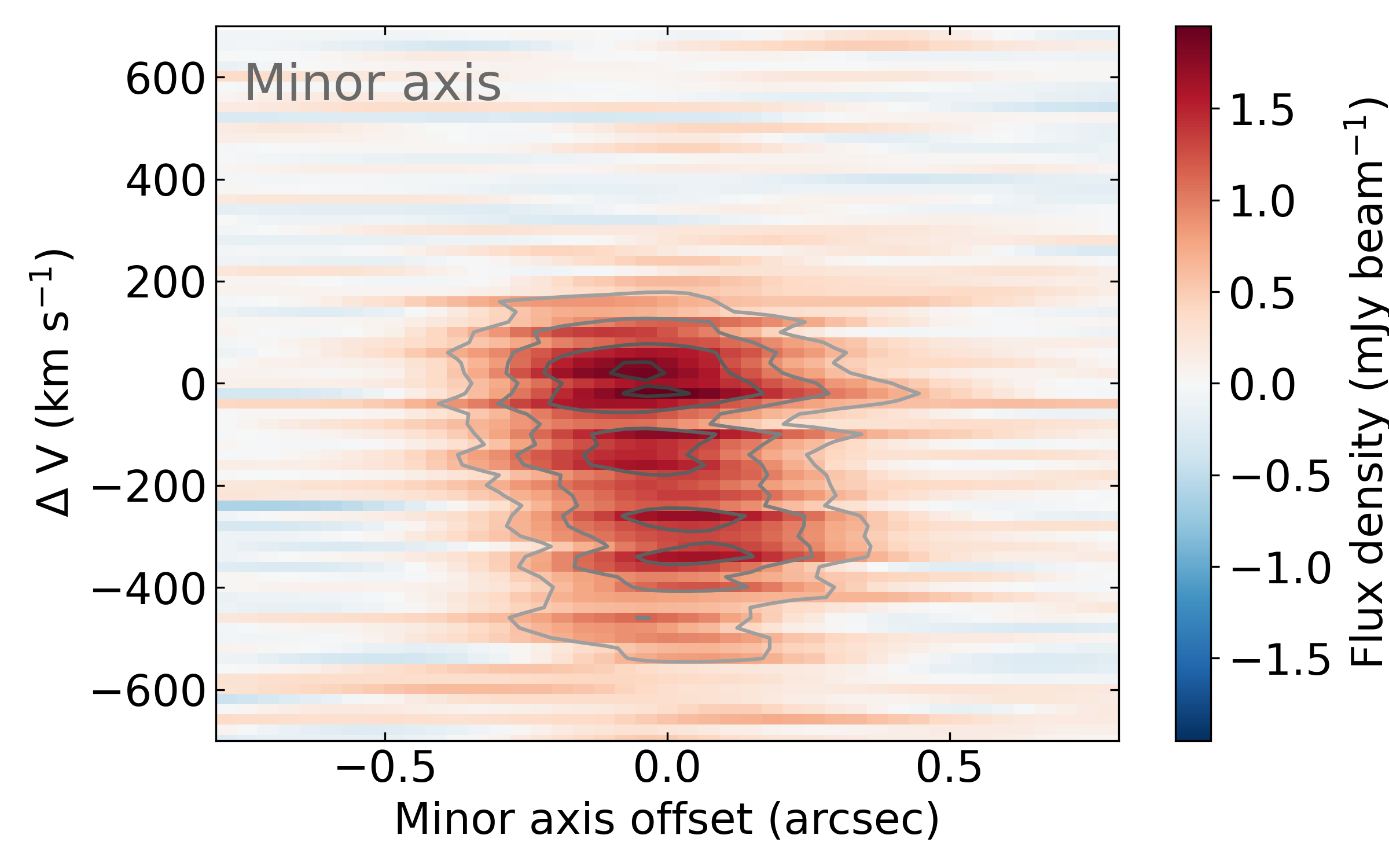}}
\caption{Position-velocity diagrams calculated using a pseudo slit of 10-pixel width placed along the major (top panel) and minor (bottom panel) axes of HZ10-C$+$HZ10-E complex. Zero velocity is calculated for the rest-frame [\CII] frequency at $z=5.6548$. Smoothed grey contours correspond to the 3-, 5-, 7-, and 9-$\sigma$ noise levels of the PV-diagrams. 
}\label{fig:PV-HZ10}
\end{figure}

To calculate PV diagrams for the HZ10-C$+$HZ10-E complex, we placed a pseudo slit of 10-pixel width along its major and minor axes. The major axis corresponds to the direction of the higher velocity gradient on the moment-1 map for these two components (excluding the HZ10-W component). It is centered on the intensity peak of the moment-0 map. To estimate the position of the minor kinematic axis, we extracted the velocity rotation curve (discussed later in this section) along the major axis and found it to be symmetrical about the systemic velocity of $V_{\rm sys}=-133$~\kms.
We use this systemic velocity to define the position of the minor kinematic axis. These axes are shown by the grey dotted lines on the middle panel of Fig.~\ref{fig:moments-HZ10}). The resulting PV diagrams are shown in Fig.~\ref{fig:PV-HZ10}. 

The PV diagram extracted along the major axis of the HZ10-C$+$HZ10-E system only tentatively resembles the classical S-shape typical for disks with an ordered rotation. While it was shown that the S-shape of disk-like galaxies becomes less pronounced with decreasing spatial resolution \citep{Rizzo2022}, the distribution of the various blobs on the PV diagram can still be explained by a disturbed clumpy disk or a merger nature of the system. 
Moreover, the asymmetry of the PV-diagram, extracted along the minor axis, suggests non-circular motions which could be attributed to the minor interactions or/and nuclear/stellar feedback~\citep[e.g.][]{Price2021,Rizzo2022}. 

\begin{figure*}[h]
\centering
{\includegraphics[width = 0.45\textwidth]{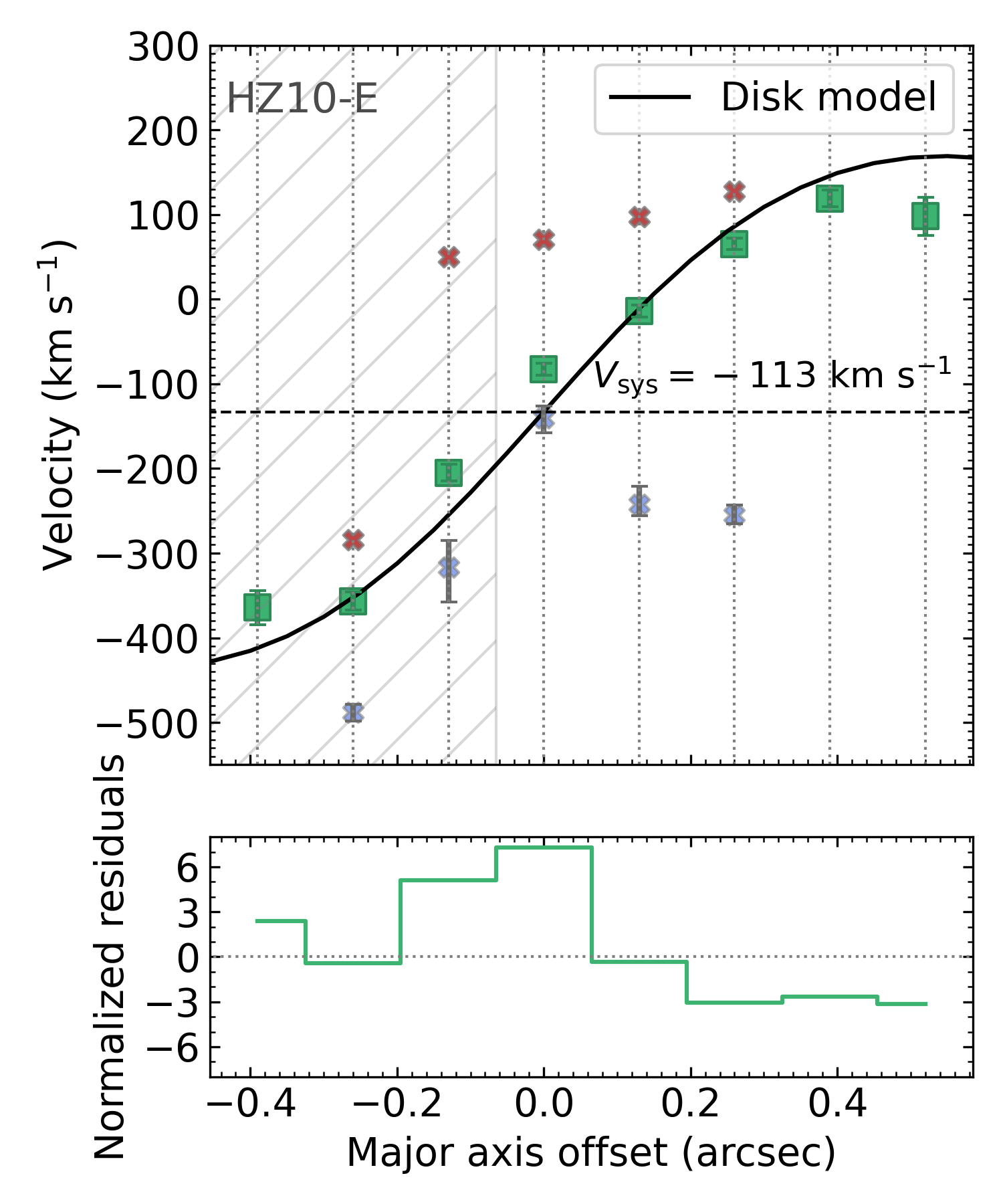}} 
{\includegraphics[width = 0.45\textwidth]{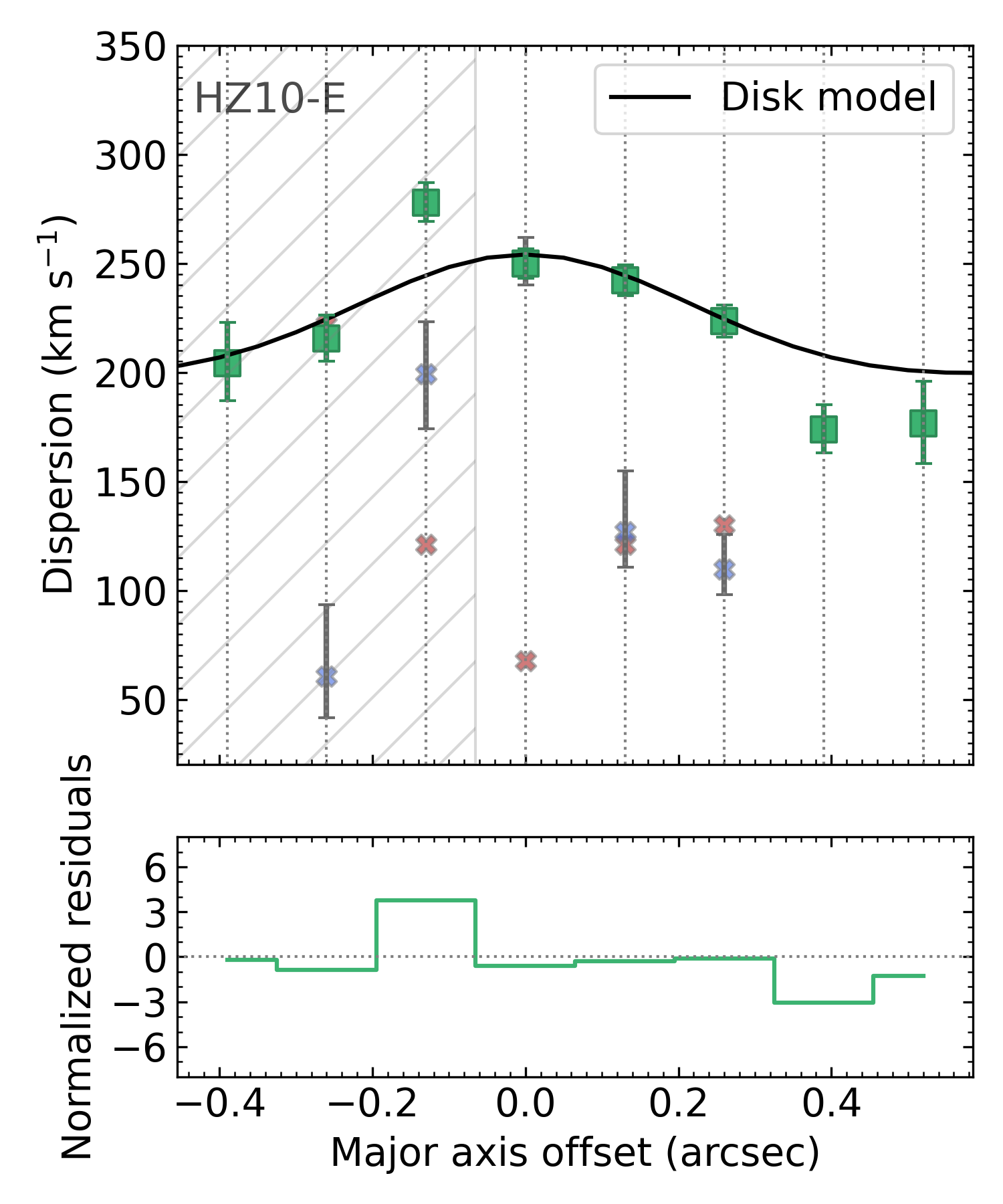}}
\caption{Rotation curves calculated along the major axis of HZ10-E$+$HZ10-C complex. Data points correspond to the velocity centroids (left panel) and velocity dispersion (right panel) of a single (green) and double (blue and red) Gaussian fit of the spectra within the circular apertures of a radius of 0.13\arcsec\ as a function of the relative aperture position. Apertures were placed along the major axis with the step of 0.13\arcsec. Green squares result from the single Gaussian spectral fit within the apertures. Zero velocity is calculated for the rest-frame [\CII] frequency at $z=5.6548$. Zero offset corresponds to the aperture position at the same zero point as for the respective PV diagram. Aperture positions where the HZ10-E component impacts the spectra are shown by the shaded area. Black curves represent the best-fit disk model done with \textsf{DysmalPy}, see text for the details. Normalized residuals represent the difference between data points and the model, divided by the data uncertainties.}\label{fig:rotation-curves-hz10}
\end{figure*}

To further investigate the nature of the HZ10-C$+$HZ10-E system, we extracted rotation curves (velocity and dispersion as a function of the spatial offset) along the major axis, which are shown in Fig.~\ref{fig:rotation-curves-hz10}. 

We extract spectra along the major axis using circular apertures with a diameter of 0.26\arcsec\ (solid green regions on the moment-0 map in Fig.~\ref{fig:moments-HZ10}), which roughly matches the beam size. The corresponding aperture spectra are shown in  
Appendix~\ref{a:aperture_spectra_fit}.
The aperture spectra show distinct non-symmetric profiles, partly due to blending velocity components within the apertures from the beam-smearing effect. 
Therefore, we fit the extracted spectra with single and double Gaussian profiles to calculate velocity centroids and dispersion in each aperture. In cases where the double-component profile fitting effectively results in a single-component solution (i.e. one of the components is poorly constrained), we use only a single-Gaussian model fit (which is common for apertures with a low signal-to-noise ratio).
The resulting velocity centroids and dispersion as a function of the aperture position along the major axis are shown in Fig.~\ref{fig:rotation-curves-hz10}.

The resulting velocity rotation curve exhibits an S-shape, while the dispersion is characterized by an almost constant value of $200$~\kms, which slightly increases at the center. These findings suggest consistency of the HZ10-C$+$HZ10-E system with a rotating disk.
Note that due to the relatively weak [\CII] emission of the HZ10-E component, it only contributes to the first three apertures, and thus does not affect the suggested conclusion that the HZ10-C component itself could be a rotating disk showing the interaction with the close component HZ10-E.

In the following section, we will fit the rotation curves of HZ10-C$+$HZ10-E, assuming the system to be a rotating disk.  

\section{Kinematic modeling: testing the disk scenario}\label{sec:modeling}

The kinematic analysis presented in the previous section suggests the possibility that the HZ10-C$+$HZ10-E system (or HZ10-C component alone) has a rotating disk nature. We test the feasibility of this scenario by approximating the HZ10-C$+$HZ10-E system with a disk model.

We perform kinematic modeling of HZ10-C$+$HZ10-E complex using \textsf{DysmalPy}\footnote{https://www.mpe.mpg.de/resources/IR/DYSMALPY/} (DYnamical Simulation and Modelling ALgorithm in PYthon) code (\citealt{Davies2004a,Davies2004b,Davies2011,Cresci2009,Wuyts2016,Lang2017,Price2021}, Lee et al. in prep.). \textsf{DysmalPy} is a Python-based forward modeling code designed for analyzing galaxy kinematics, specifically for disk galaxies. The modeling is based on the 
parametrization of the mass distribution of a galaxy, generating a 3D mock cube capturing composite kinematics and accounting for observational effects such as beam smearing and instrumental line broadening. 
This allows us to compare the observed rotation curves with those extracted from the simulated data in the same way. 
For the fitting parameters estimation, we use the Bayesian framework with the affine Markov Chain Monte Carlo (MCMC) sampler \textsf{emcee} \cite{Foreman-Mackey2013} implemented into \textsf{DysmalPy}.

We use a model consisting of a baryon disk parametrized by a S\'ersic profile and a dark-matter halo with a Navarro–Frenk–White profile \citep{Navarro1996}.
During the fitting, we treat total baryon mass, $M_{\rm bar}$, disk effective radius, $r_{\rm eff}$, disk S\'ersic index, $n$, dark matter fraction within the effective radius, $f_{\rm DM}(r_{\rm eff})$, intrinsic velocity dispersion, $\sigma_0$, and galaxy inclination, $i$, as free parameters. 
As prior Gaussian distributions of the HZ10-C$+$HZ10-E disk effective radius and S\'ersic index, we use the results of the 2D parametric modeling of the HZ10\footnote{Due to their close separation and weak emission from HZ10-E, HZ10-E and HZ10-C cannot be distinguished in the integrated [\CII] map. As a result, these components were modeled as a single S\'ersic profile, with the best-fit parameters listed in Table~\ref{tab:hz10_properties} under the HZ10-C column.}  integrated [\CII] map with a S\'ersic profile using \textsf{PyAutoGalaxy} software (see Table.~\ref{tab:hz10_properties}). In addition, we use an estimate from the axis ratio derived during the 2D parametric modeling as a Gaussian prior distribution of the system inclination. 
The resulting parameter values are $\log M_{\rm bar}/M_{\odot} = 11.1^{+0.2}_{-0.3}$, $r_{\rm eff} = 1.3^{+0.2}_{-0.2}$~kpc, $n=0.55^{+0.05}_{-0.05}$, $f_{\rm DM}(r_{\rm eff})=0.3^{+0.3}_{-0.2}$, $\sigma_0=196^{+8}_{-9}$~\kms, and $i=40^{+1}_{-1}$ degrees. The point estimations correspond to the median of the posterior distributions and the uncertainties are 68 percent credible intervals, respectively. The estimated baryon mass of HZ10-C$+$HZ10-E is in agreement with the gas mass of the HZ10 estimated from the CO luminosity in \cite{Pavesi2019}.
In Fig.~\ref{fig:rotation-curves-hz10}, we overplot modeled rotation velocity and dispersion curves based on the best-fit results with the black curves.
Details of the \textsf{DysmalPy} modeling are presented in Appendix~\ref{a:dysmal}.

Although a disk model could well describe the HZ10-C$+$HZ10-E system, we cannot discard the other scenarios. To examine the significance of the rotation support for the HZ10-C$+$HZ10-E system, we estimated the ratio between inclination-corrected rotation velocity and derived from modeling intrinsic dispersion corrected by beam smearing. We calculated $V_{\rm rot}$ as an average between the absolute values of minimum and maximum observed velocities along the major kinematic axis (green squares in Fig.~\ref{fig:rotation-curves-hz10}, left panel) divided by $\sin(i)$, inferred from the modeling.
The ratio for HZ10-C$+$HZ10-E is $V_{\rm rot}/\sigma_0 = 1.9^{+0.1}_{-0.1}$. 
Although the value of $V_{\rm rot}/\sigma_0$ ratio is consistent with that observed in dynamically warm disk galaxies at $z\sim4-8$ \citep[e.g.,][and references threrein]{Kohandel2024}, the intrinsic dispersion of 200~\kms\ is on the higher end of the range typically seen in rotation-dominated systems, even at high redshifts \citep[e.g.,][]{Rizzo2024} and significantly exceeds the values predicted by simulations \citep{Kohandel2024}, which hints at the complex dynamics of HZ10-C.
As an additional test, we tried to separate the HZ10-C component from HZ10-E and fit it with a disk model. To do this, we modeled the HZ10-E contribution to the total spectra in the first three apertures using a Gaussian profile, which allowed us to extract rotation curves attributed to HZ10-C. We then repeated  the \textsf{DysmalPy} analysis for HZ10-C using the same model parameters. The details are provided in Appendix~\ref{a:aperture_spectra_fit} and~\ref{a:dysmal}. We conclude that even when analyzes independently, HZ10-C exhibit a high velocity dispersion, similar to what we obtained for HZ10-C$+$HZ10-E.

Thus, we cannot rule out the dispersion-dominated nature of the HZ10-C+HZ10-E system as well as of the HZ10-C alone. Although the high velocity dispersion could indicate a neutral outflow traced by [\CII] rather than the dispersion-dominated nature of the system, the available data do not allow us to claim the presence of a statistically significant broad [\CII] spectral component.

Below, we summarize the potential scenarios for the HZ10-C+HZ10-E system based on observations and kinematic modeling:

{\noindent (i)} HZ10-C$+$HZ10-E represents a rotationally supported, disturbed galaxy disk. In this scenario, HZ10-E would be a fainter clump within the extended disk, exhibiting brightness asymmetry due to an inhomogeneous dust distribution and/or star formation.
Indeed, numerical simulations have shown that recent merging events can produce a disturbed galaxy disk characterized by rotational support and the presence of multiple gas clumps \citep{Kohandel2019}. 

{\noindent (ii)} HZ10-E is a satellite galaxy undergoing a merger with the rotationally supported, disk-like galaxy HZ10-C.

{\noindent (iii)} HZ10-C is not a rotationally supported system but rather represents a close double merger, and a faint HZ10-E galaxy is merging with it 
suggesting a triple merger scenario of the HZ10-C$+$HZ10-E system, which seems to be common at high redshift \citep[e.g.,][]{Diaz-Santos2018}.

\begin{figure*}[th]
{\includegraphics[trim={0.0cm 0cm 0cm 0cm},clip, width = 0.32\textwidth]{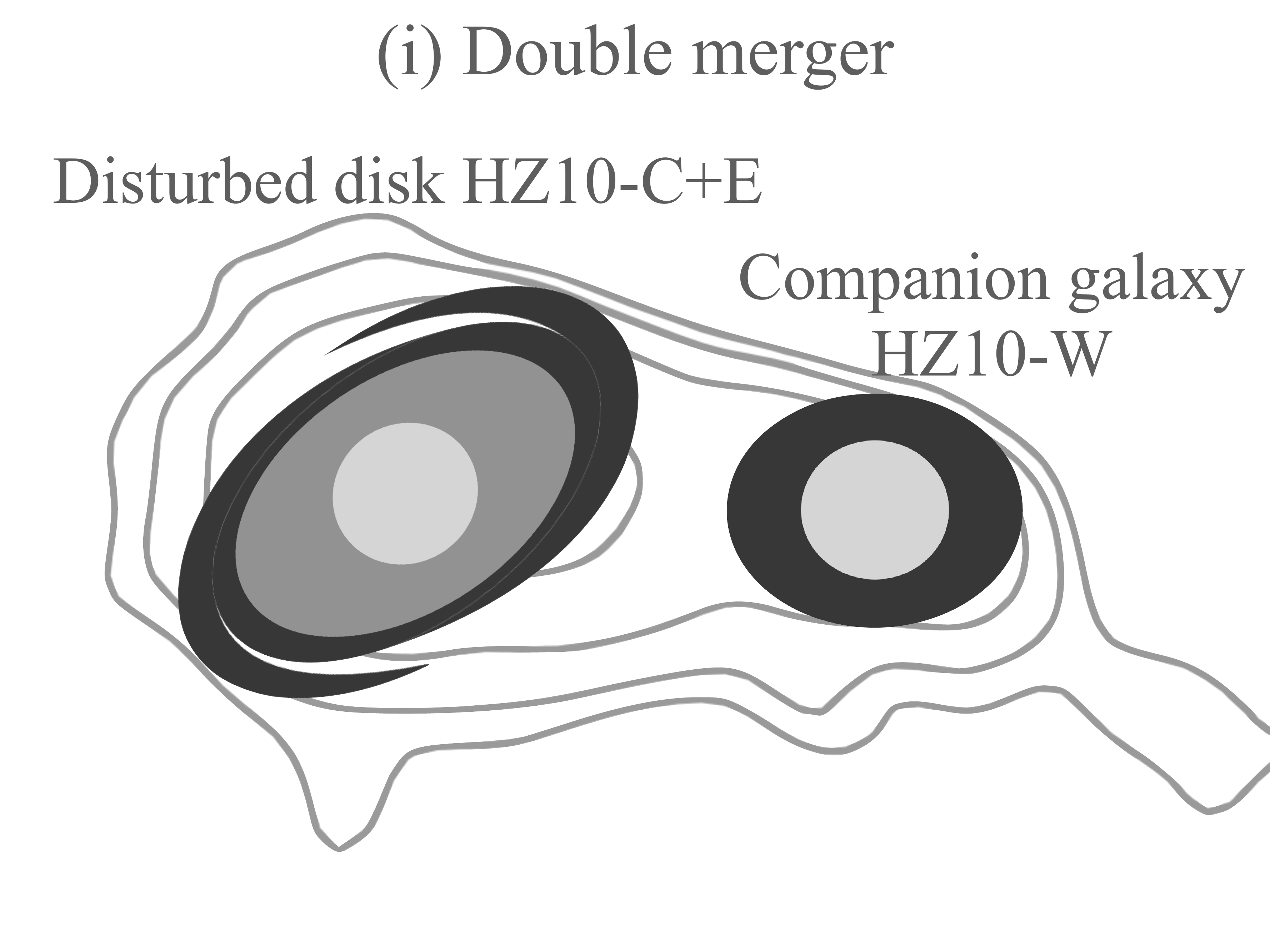}}
{\includegraphics[trim={0.0cm 0cm 0cm 0cm},clip, width = 0.32\textwidth]{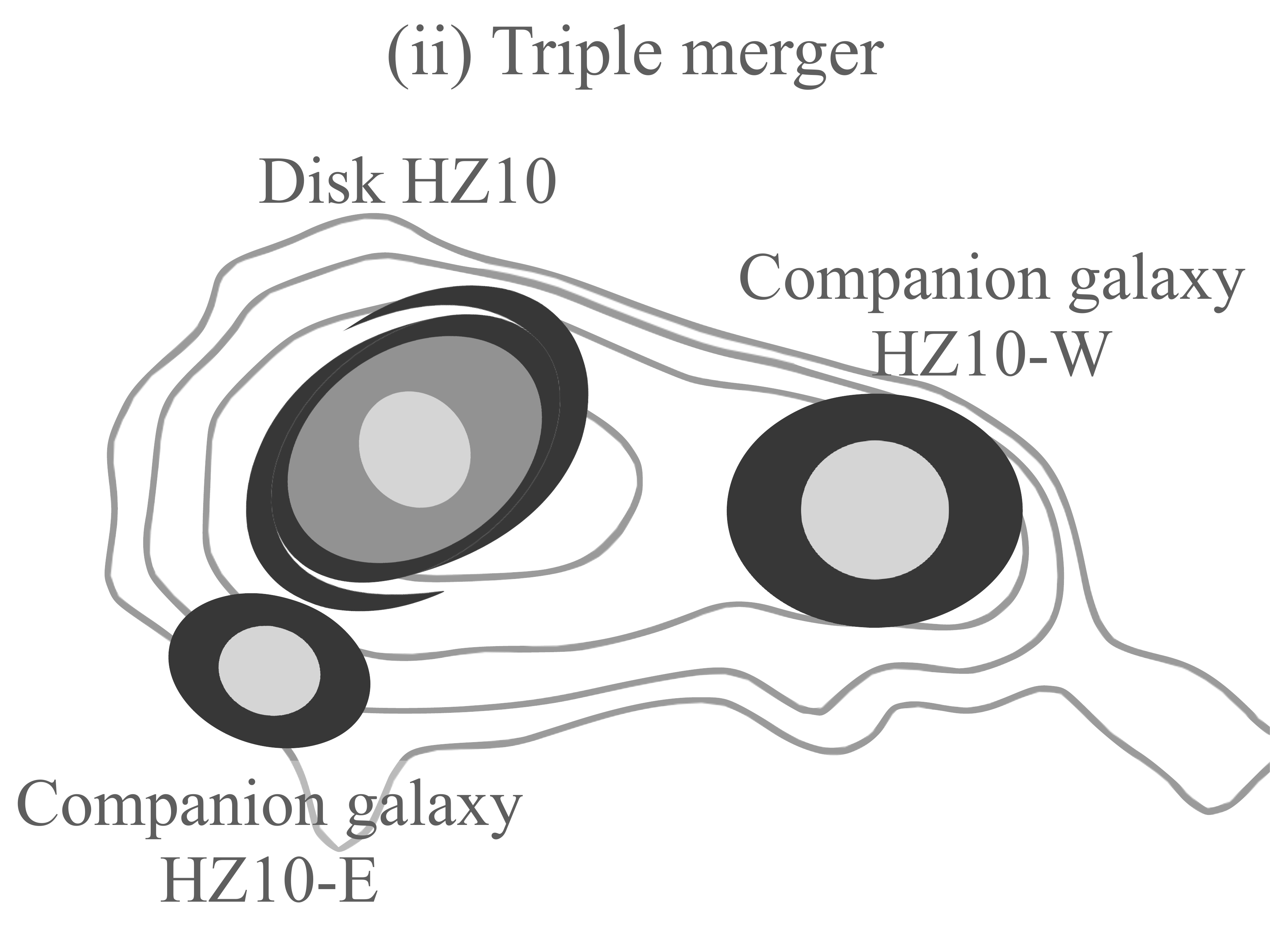}}
{\includegraphics[trim={0.0cm 0cm 0cm 0cm},clip, width = 0.32\textwidth]{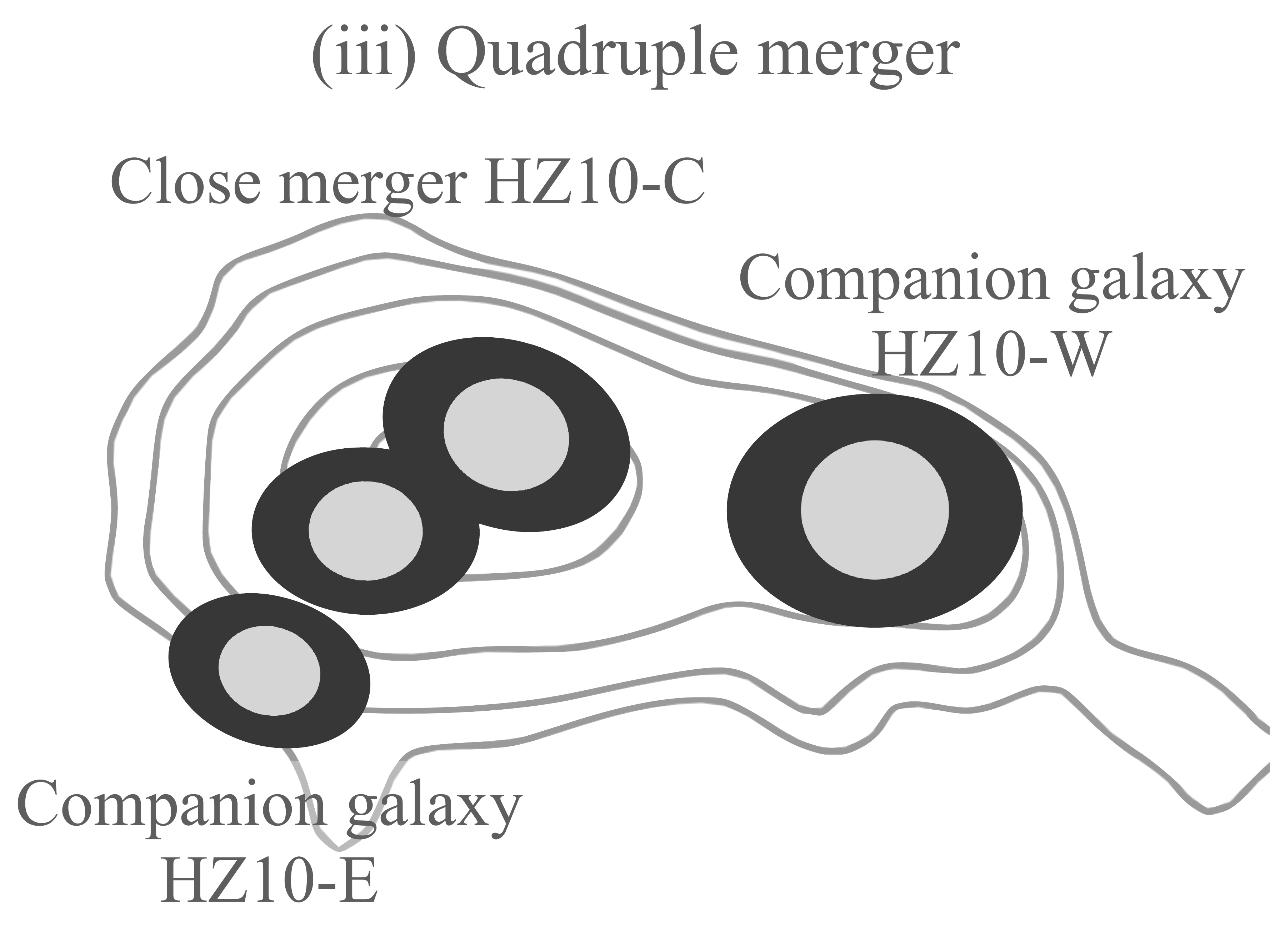}}
\caption {Illustration of three suggested dynamical scenarios for the HZ10 system consistent with the kinematic modeling of [\CII] emission and observational properties derived from rest-frame optical emission.}\label{fig:scenarios} 
\end{figure*}

We illustrate these scenarios in Fig.~\ref{fig:scenarios} including there the companion galaxy HZ10-W.

\section{Comparison with JWST/NIRSpec}\label{sec:JWST_comparison}

\begin{figure}[th]
{\includegraphics[trim={0 2.0cm 0 0},clip,
width = 1\columnwidth]{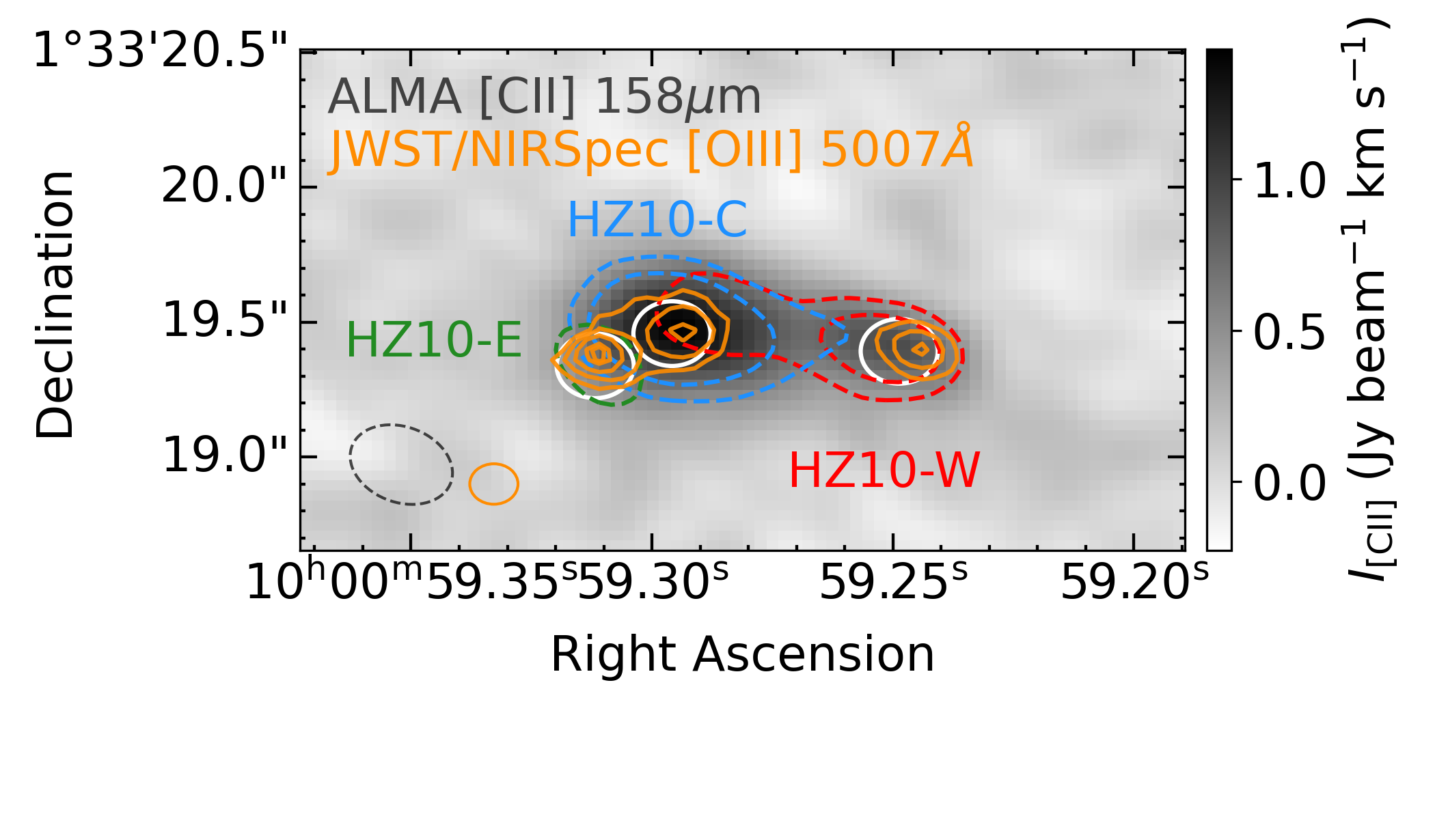}}
{\includegraphics[trim={0 1.0cm 0 0},clip,width = 1\columnwidth]{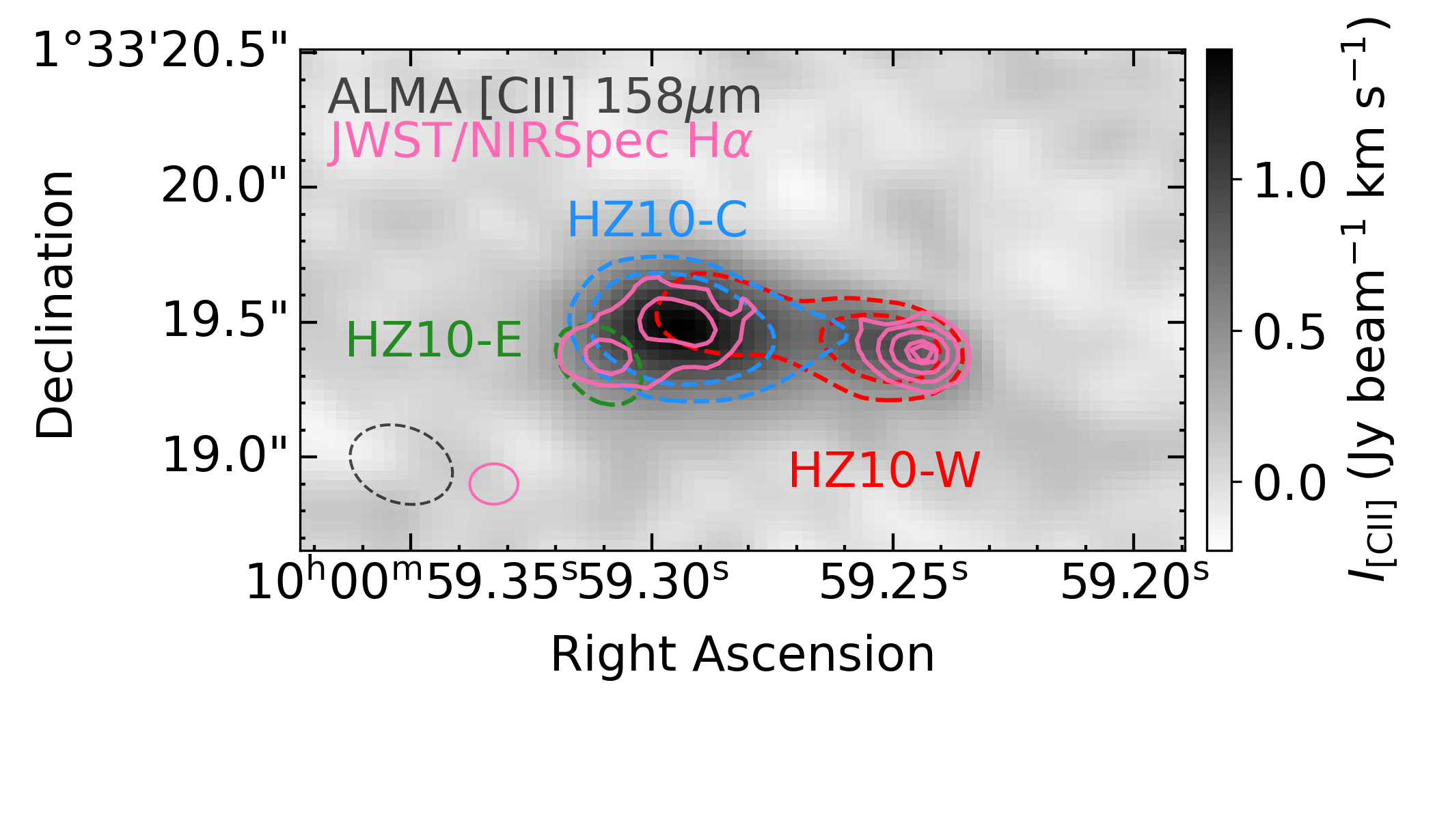}}
\caption {Integrated [\CII] intensity of the whole HZ10 complex is shown in grey scale. To emphasize the spatial positions of the three different components, HZ10-E, HZ10-C, and HZ10-W, seen in [\CII] we overplot contours from Fig.~\ref{fig:hz10_mom0_comps}. The orange (pink) contours on the top (bottom) panel correspond to the [\OIII]~5007$\AA$ (H$\alpha$) relative contours from JWST/NIRSpec observations \citep{Jones2024}. Apertures for the component's spectra (see Fig.~\ref{fig:HZ10-aperture-spectra}) extraction are shown with the white circles on the top panel. The beam size of the
ALMA observations ($0.34\arcsec\times0.27\arcsec$) and JWST/NIRSpec PSF ($0.15\arcsec$) are shown on the bottom left corner of each panel by the dashed grey ellipse and solid orange (pink) circle, respectively.}\label{fig:HZ10-alma-jwst} 
\end{figure}

Recent JWST/NIRSpec observations (with angular resolution of $0.15\arcsec$ and resolving power of $R\approx2700$) support the complex morphology and kinematics of the HZ10 \citep{Jones2024}.
In Fig.~\ref{fig:HZ10-alma-jwst} we show the morphological comparison between the [\CII] components and those detected in [\OIII]~5007\AA\ (top panel) and in H$\alpha$ (bottom panel) with JWST/NIRSpec. 
The spatial positions of the [\CII], [\OIII], and H$\alpha$ components are in good agreement.
As shown in section~\ref{sec:overview} above, the HZ10-E component appears relatively dim in [\CII] emission and can only be detected through its kinematic properties using, for example, PV diagrams (Figs.~\ref{fig:PV-HZ10-whole} and \ref{fig:PV-HZ10}) or adaptive integrated intensity maps (Fig.~\ref{fig:hz10_mom0_comps}). However, the spatial position of this component aligns well with one of the three components clearly revealed in rest-frame optical emission line intensity maps. The latter provides independent confirmation of the reliability of the kinematically detected dim HZ10-E component in [\CII] emission.

Since the spatial distribution of the [\CII] emission resembles the rest-frame optical line emission from JWST/NIRSpec, we also aim to compare them in velocity space. \cite{Jones2024} reported the presence of narrow and broad spectral components for all three resolved systems: HZ10-E, HZ10-C, and HZ10-W. From a spaxel-by-spaxel analysis of [\OIII]~5007$\AA$ and H$\alpha$ spectral profiles, \cite{Jones2024} showed that HZ10-W and HZ10-E mostly exhibit blue and red spectral profile asymmetry, respectively, while HZ10-C shows a gradient of red-to-blue asymmetry.

\begin{figure}[th]
\centering
{\includegraphics[trim={0.3cm 1.5cm 0.cm 0cm},clip, width = 0.83\columnwidth]{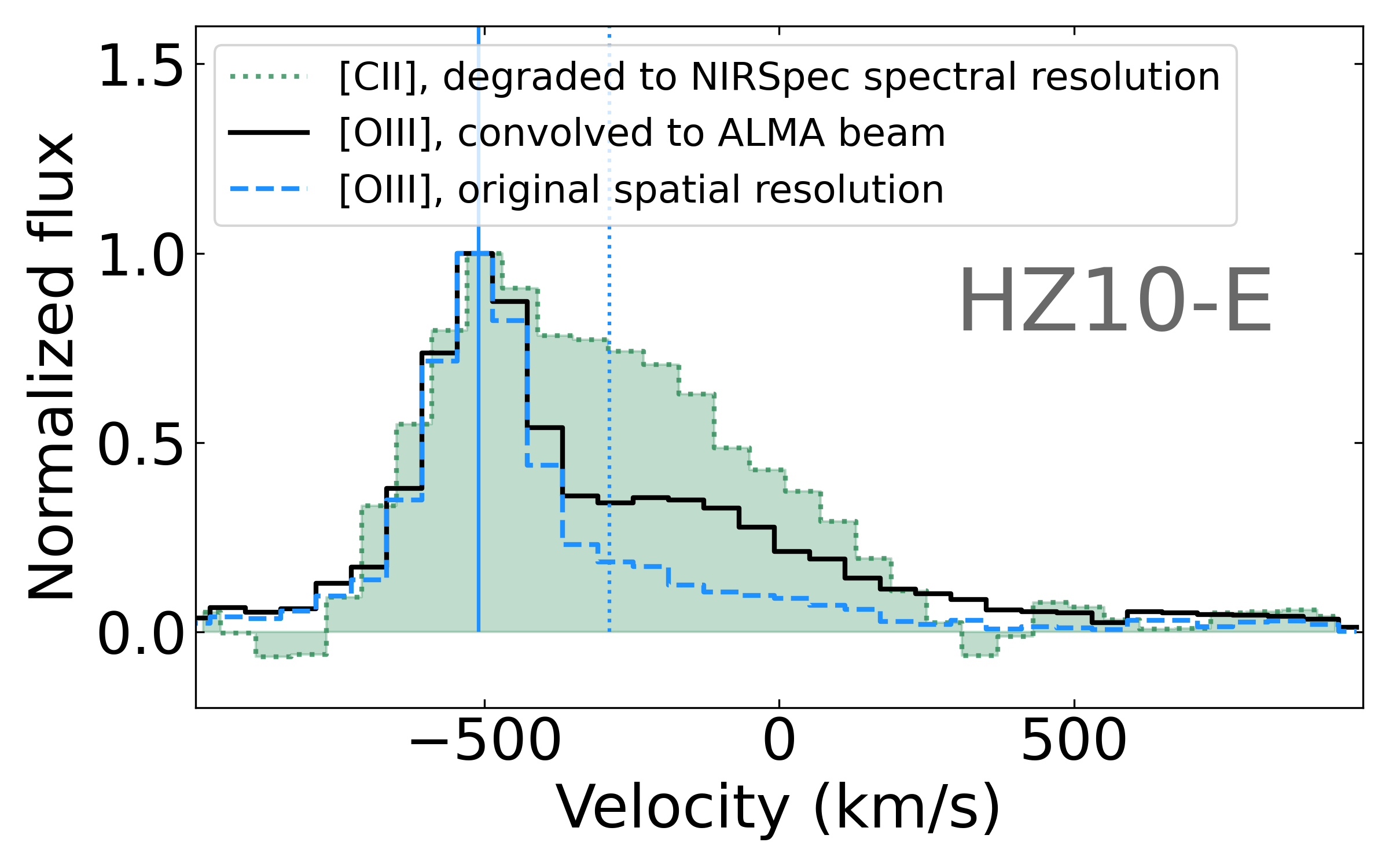}}
{\includegraphics[trim={0.3cm 1.5cm 0cm 0cm},clip, width = 0.83\columnwidth]{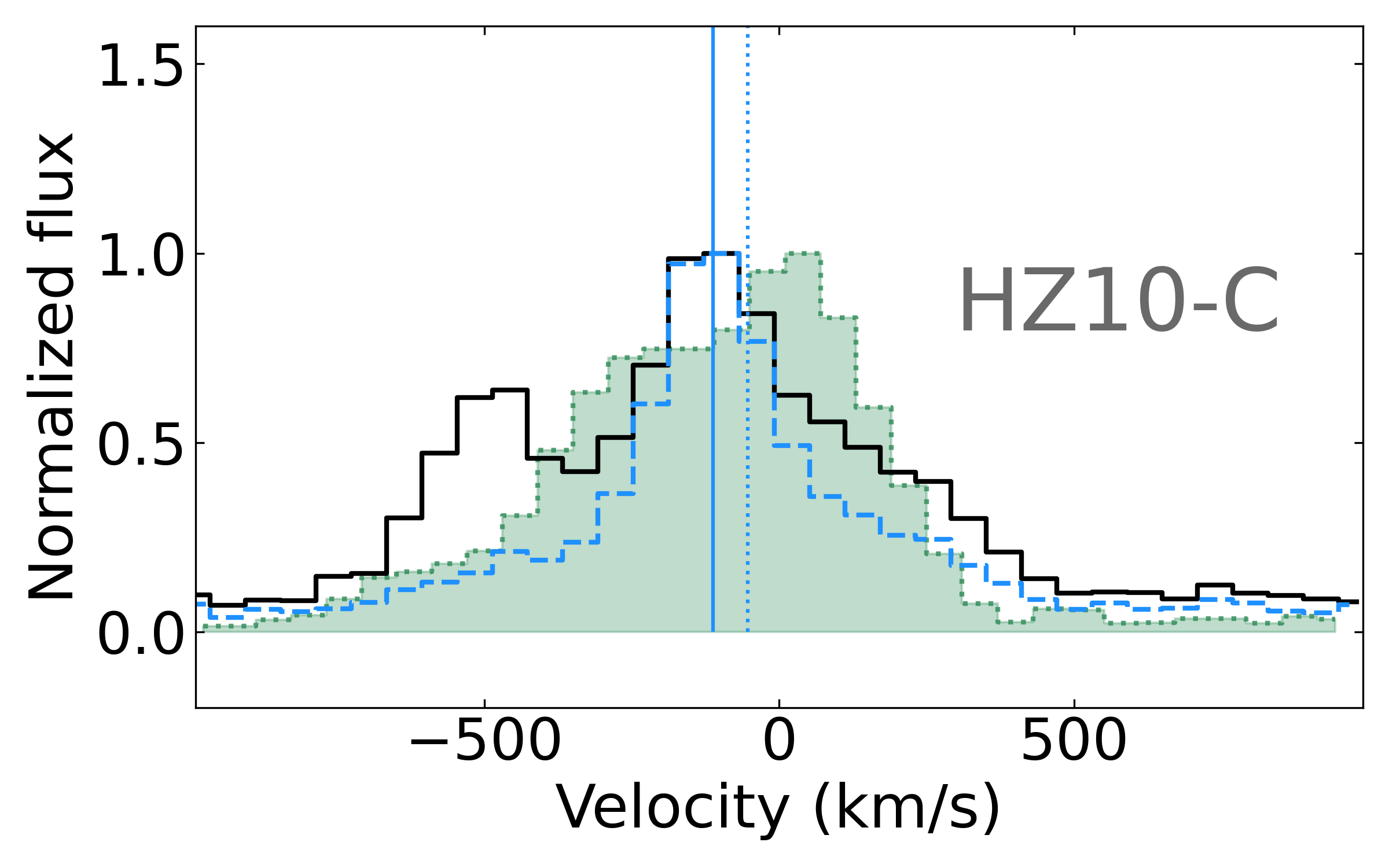}}
{\includegraphics[trim={0.3cm 0.0cm 0cm 0cm},clip, width = 0.83\columnwidth]{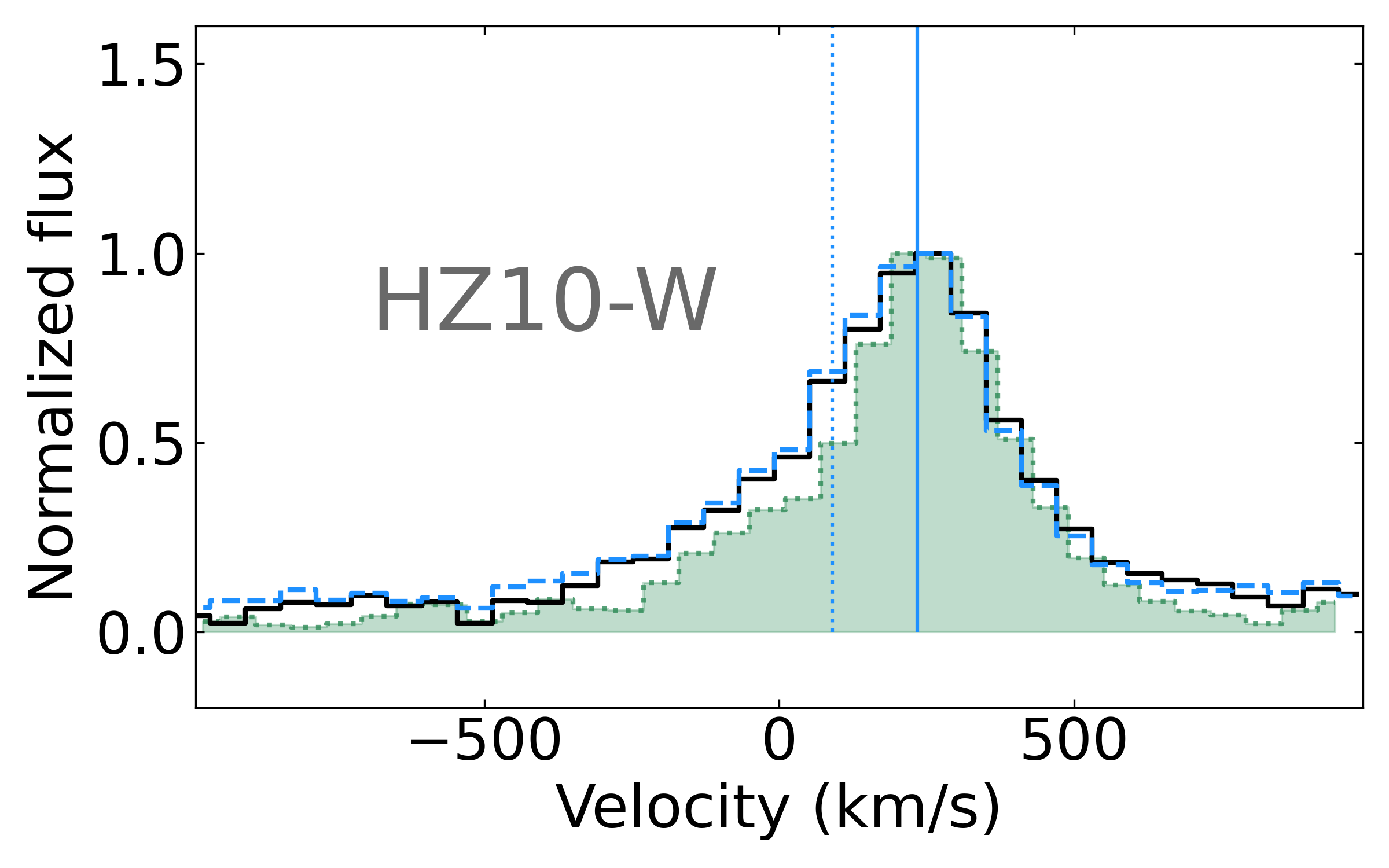}}
\caption {[\OIII]~5007$\AA$ JWST/NIRSpec spectra (solid black, data cube with degraded spatial resolution and dashed blue with the original spatial resolution) and [\CII] ALMA (filled green, spectral resolution is degraded to match that of [\OIII] data) of the HZ10-E (top), HZ10-C (middle), and HZ10-W (bottom) components, extracted from circular apertures, roughly matching the beam size, centered on the spatial positions of the components. 
Zero velocity is calculated at the systemic redshift $z=5.6548$. The solid and dotted blue vertical lines indicate the velocity centroids of the narrow and broad reft-frame optical spectral line components, respectively, taken from \citep{Jones2024}. Note, that we renormalized all spectra to their peak intensity to simplify the kinematic comparison between the profiles, thus the relative fluxes here are not comparable to the observed ones. However, we compare the observed fluxes in the text.}\label{fig:HZ10-aperture-spectra} 
\end{figure}

We extracted [\CII]  and [\OIII]~5007$\AA$   spectra from circular apertures with radii of 0.12\arcsec centered on the HZ10-E, HZ10-C, and HZ10-W components\footnote{Although the aperture sizes are comparable to those used in \cite{Jones2024}, the central position of the HZ10-W aperture differs due to a slight offset (less than the ALMA beam size) between the [\CII] and [\OIII] emission peaks. To accurately capture the [\CII] emission for spectral comparison, we shifted this aperture eastward from the [\OIII] emission peak.} to compare their spectral properties, see Fig.~\ref{fig:HZ10-aperture-spectra}.

Given the lower spatial resolution of ALMA observations compared to those of JWST/NIRSpec, accounting for the blending of different spectral components from HZ10-C and HZ10-E  becomes important during the analysis. To take this into account we also show the [\OIII] spectra extracted from the JWST/NIRSpec data cube with the degraded spatial resolution to match that of ALMA (black solid curves in Fig.~\ref{fig:HZ10-aperture-spectra}). In addition, we averaged the [\CII] spectra to match the lower spectral resolution of [\OIII] data.

We see from the aperture spectra that the velocity centroids and widths of the [\CII]~158$\mu$m emission for all three HZ10 components are similar to those of the [\OIII]~5007$\AA$ emission.

As a next step, we compare the [\CII] velocity map with those of the narrow and broad components of [\OIII].  
For this, we perform a spaxel-by-spaxel [\OIII] spectral fitting with single and double Gaussian profiles taking into account the NIRSpec instrumental broadening. 

The spatial resolution of ALMA [\CII] observations does not allow us to perform the same spaxel-by-spaxel [\CII] spectral fitting with a double Gaussian profile to search for the presence of a broad emission component, since the multi-component Gaussian profile of [\CII] emission is dominated by the beam smearing. 
However, we can test which [\OIII] line component, narrow or broad, more closely resembles the [\CII] emission kinematics by comparing the [\OIII] velocity and dispersion maps with the [\CII] moment-0 and moment-1 maps. For the direct comparison, we also rebinned the [\CII] moment-0 and moment-1 maps to match the pixel size of the [\OIII] maps and applied the same spatial mask. 

\begin{figure*}[h]
\centering
{\includegraphics[trim={1.1cm 1.cm 1cm 1cm},clip,width = 0.05\textwidth]{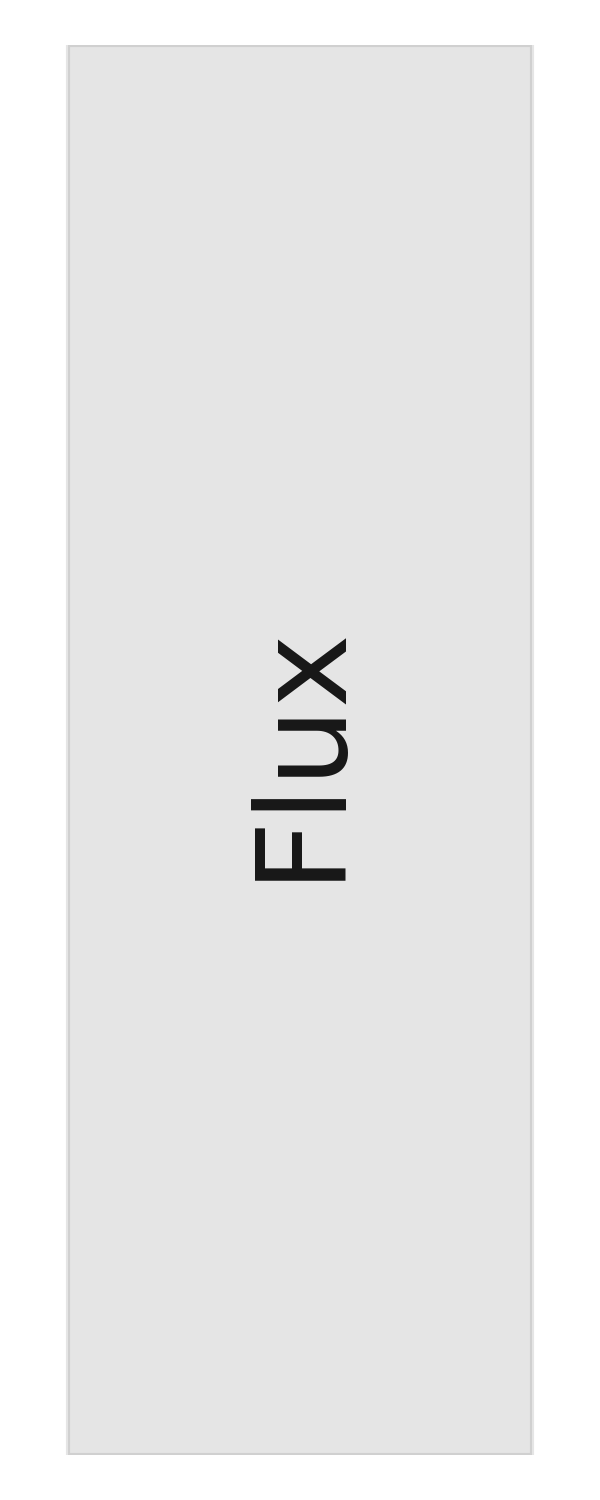}}
{\includegraphics[trim={0.0cm 1.77cm 0 -0.5cm},clip,width = 0.28\textwidth]{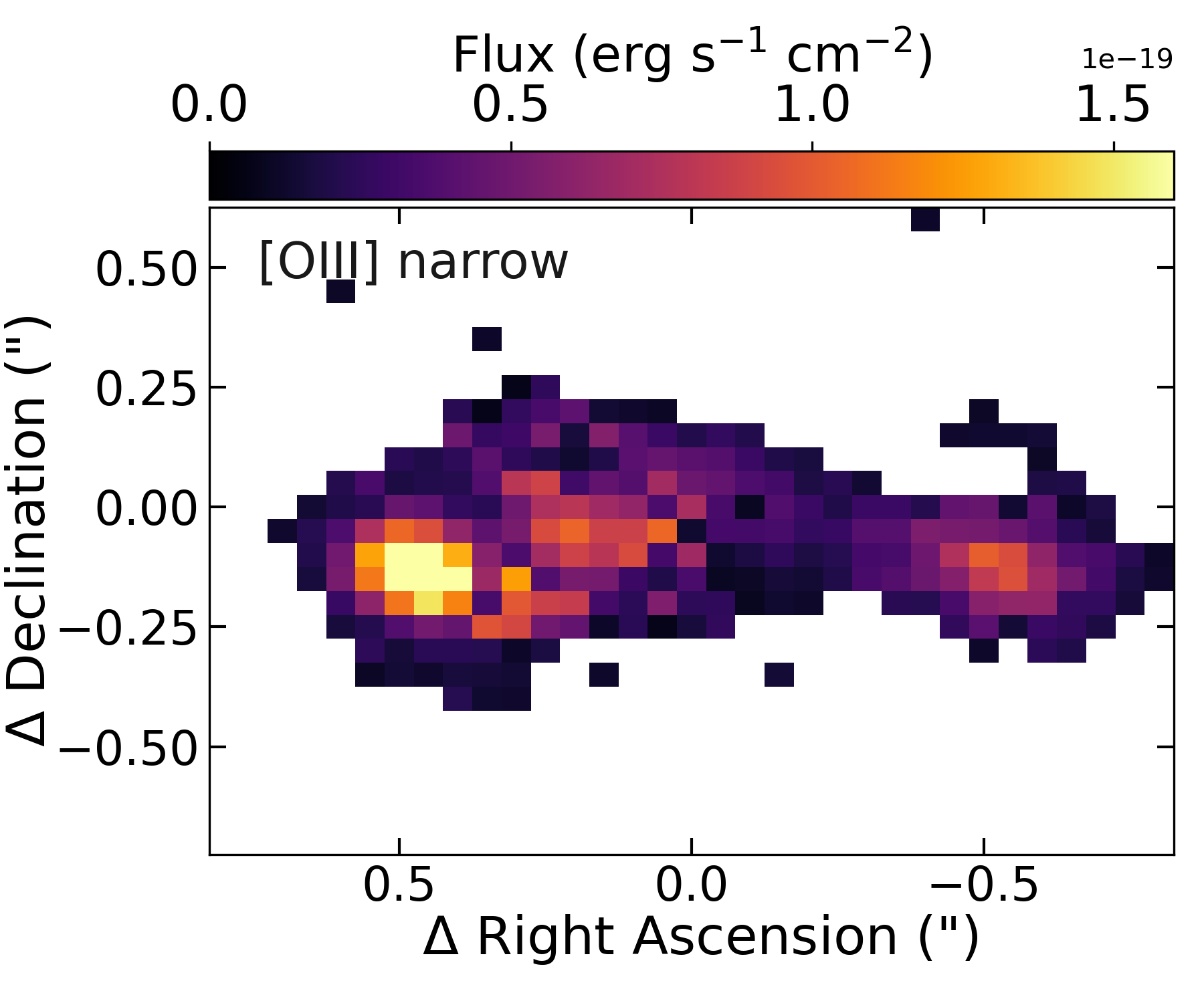}}
{\includegraphics[trim={0.0cm 1.77cm 0 0},clip,width = 0.28\textwidth]{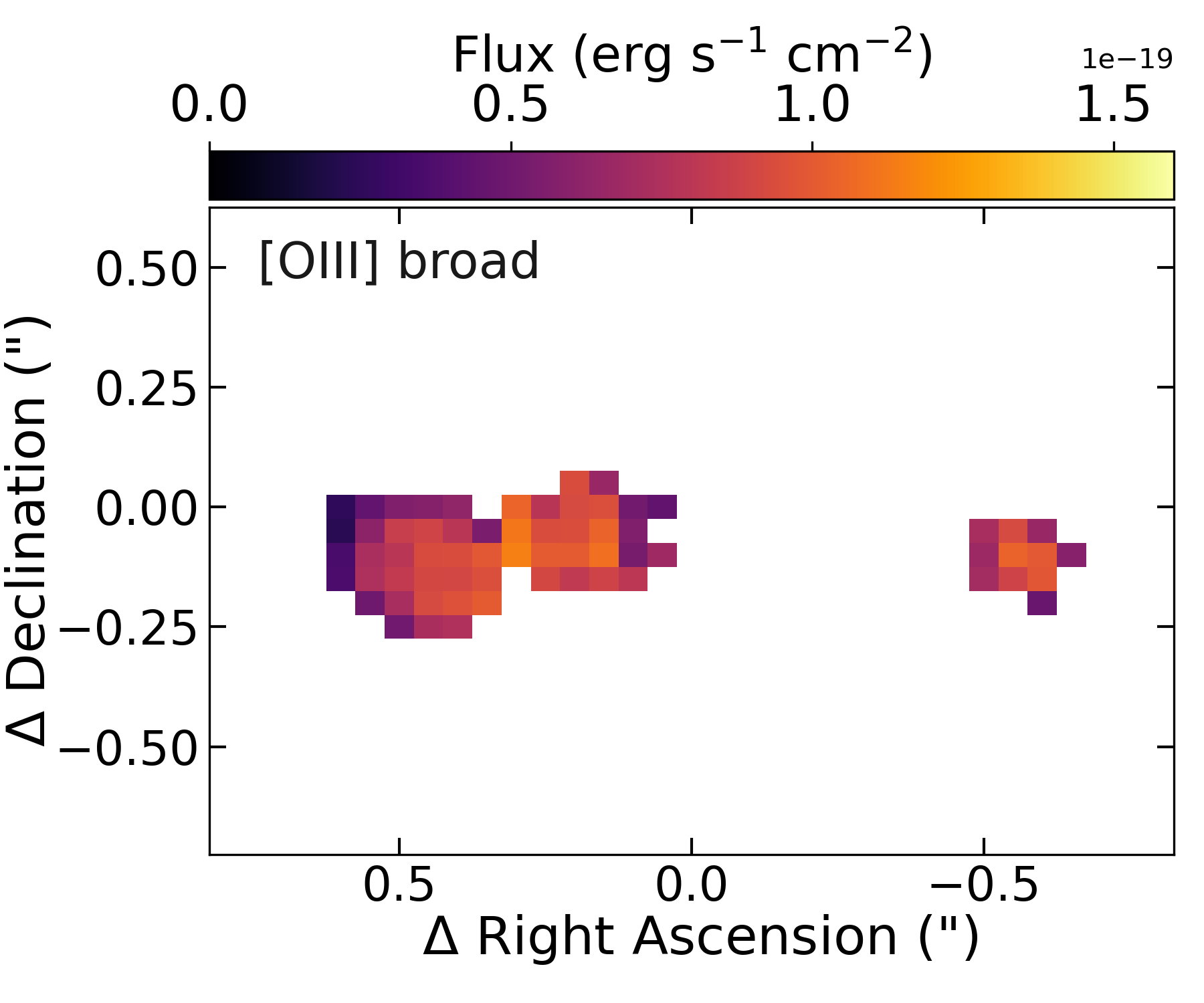}}
{\includegraphics[trim={0.0cm 1.77cm 0.0cm 0cm},clip,width = 0.28\textwidth]{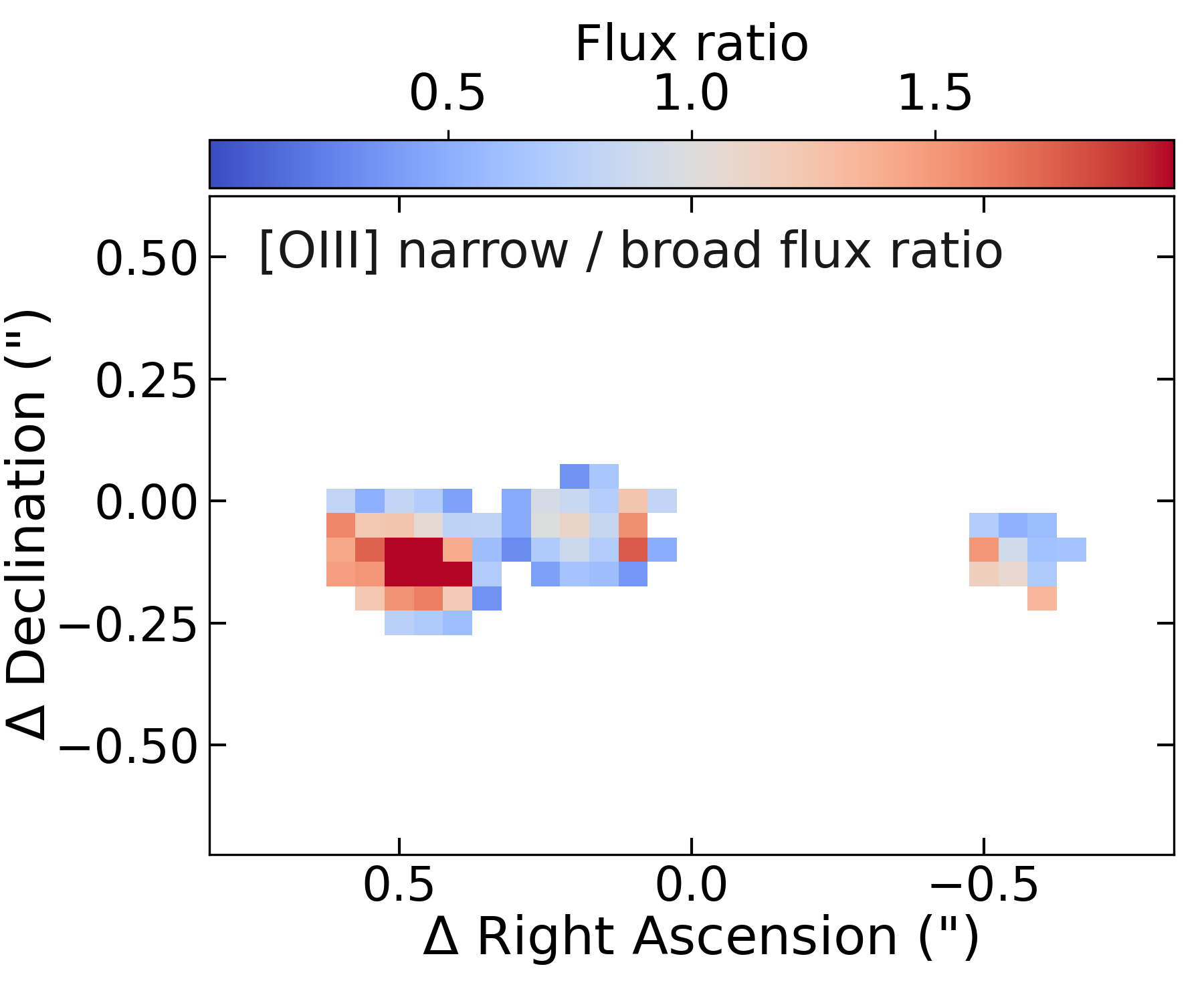}}

{\includegraphics[trim={1.1cm 1.cm 1cm 1cm},clip,width = 0.05\textwidth]{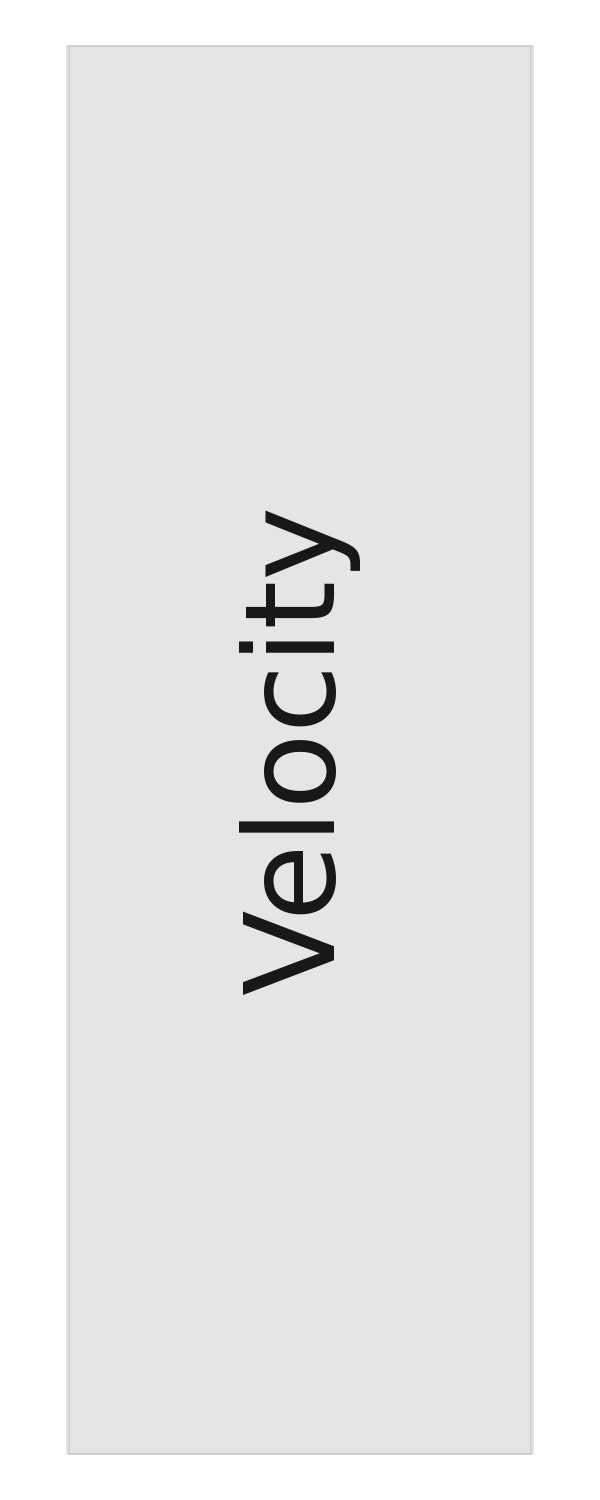}}
{\includegraphics[trim={0 1.75cm 0 0 0},clip,width = 0.28\textwidth]{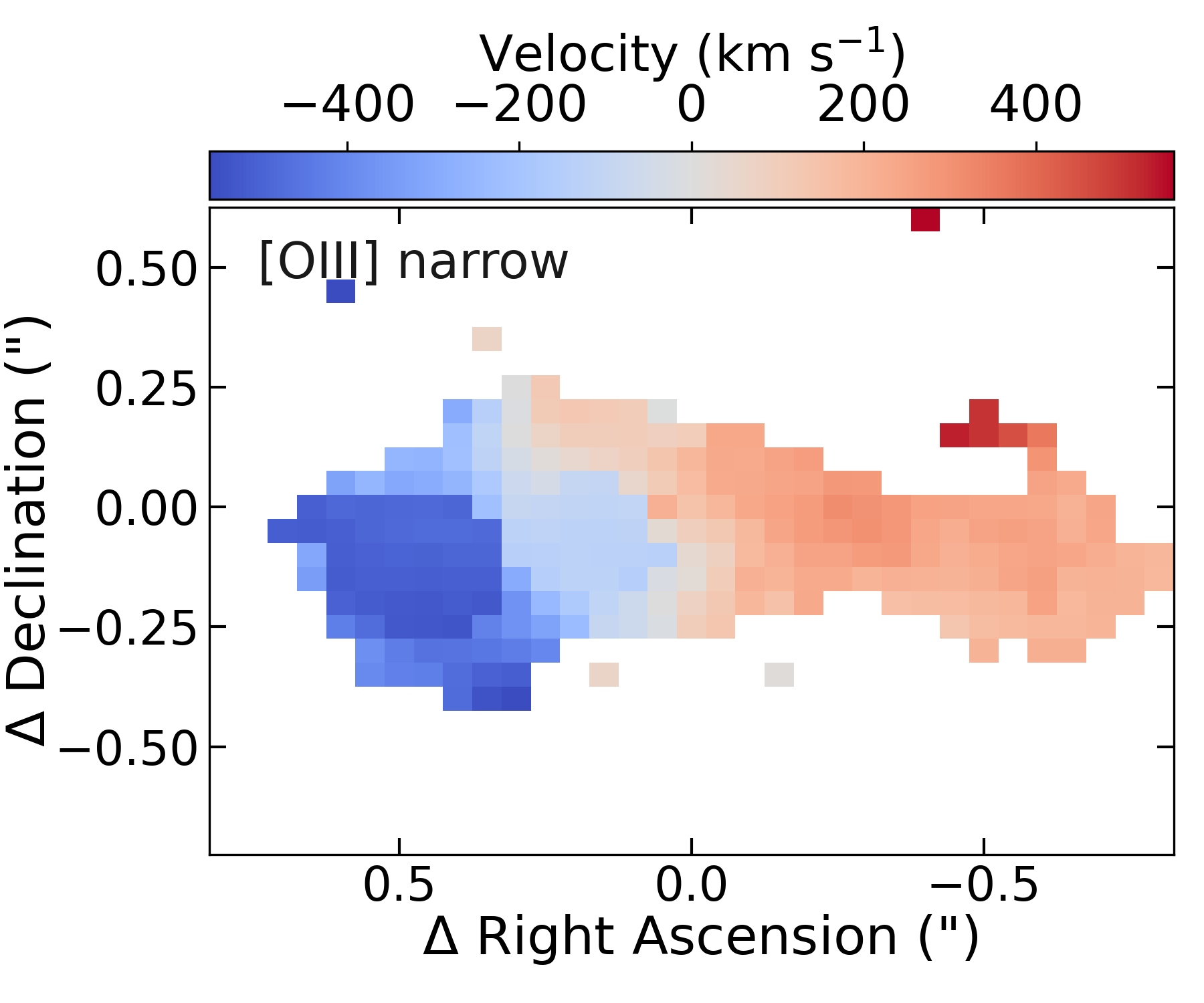}} 
{\includegraphics[trim={0.0cm 1.75cm 0.0cm 0},clip,width = 0.28\textwidth]{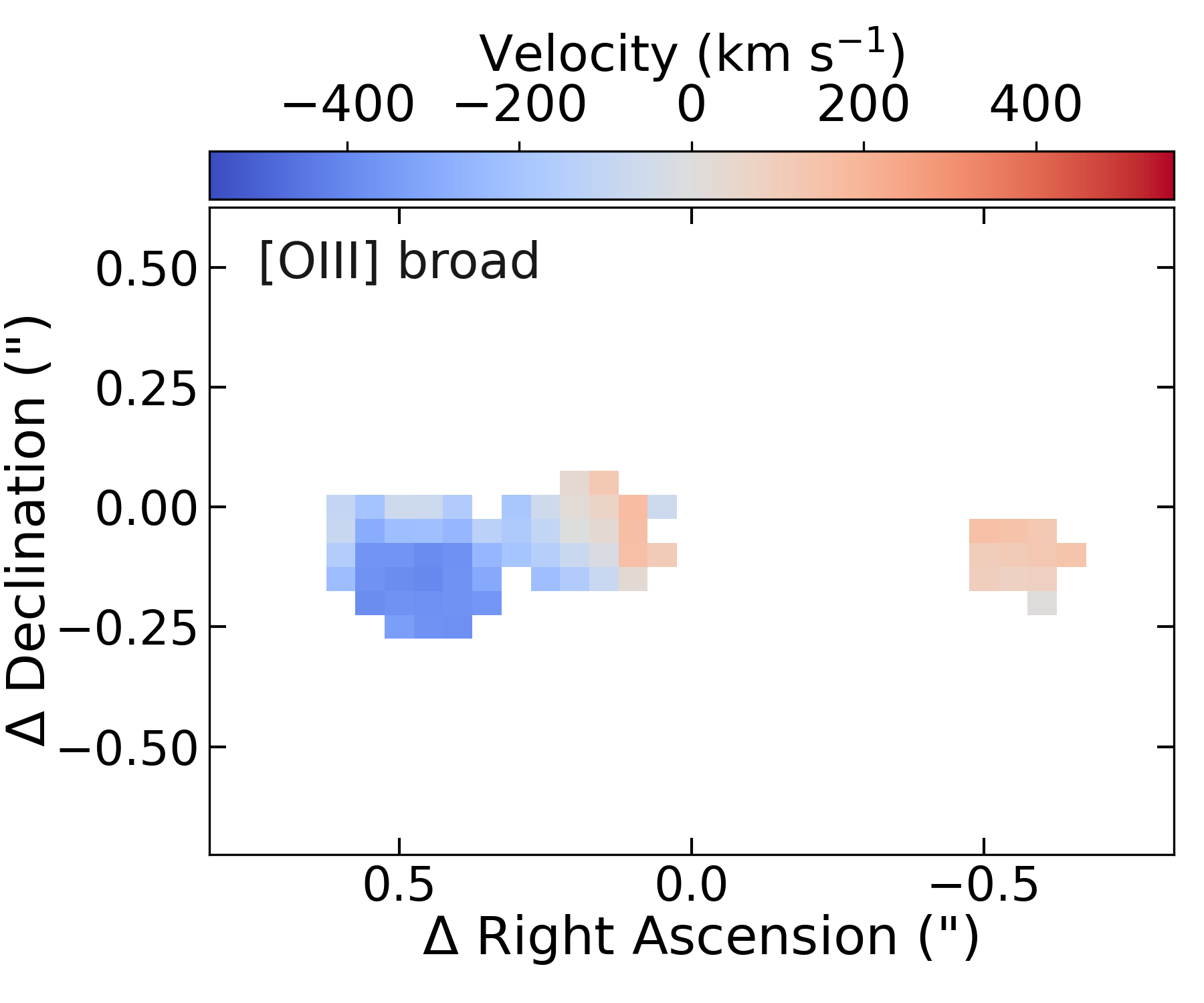}}
{\includegraphics[trim={0.0cm 1.75cm 0 0},clip,width = 0.28\textwidth]{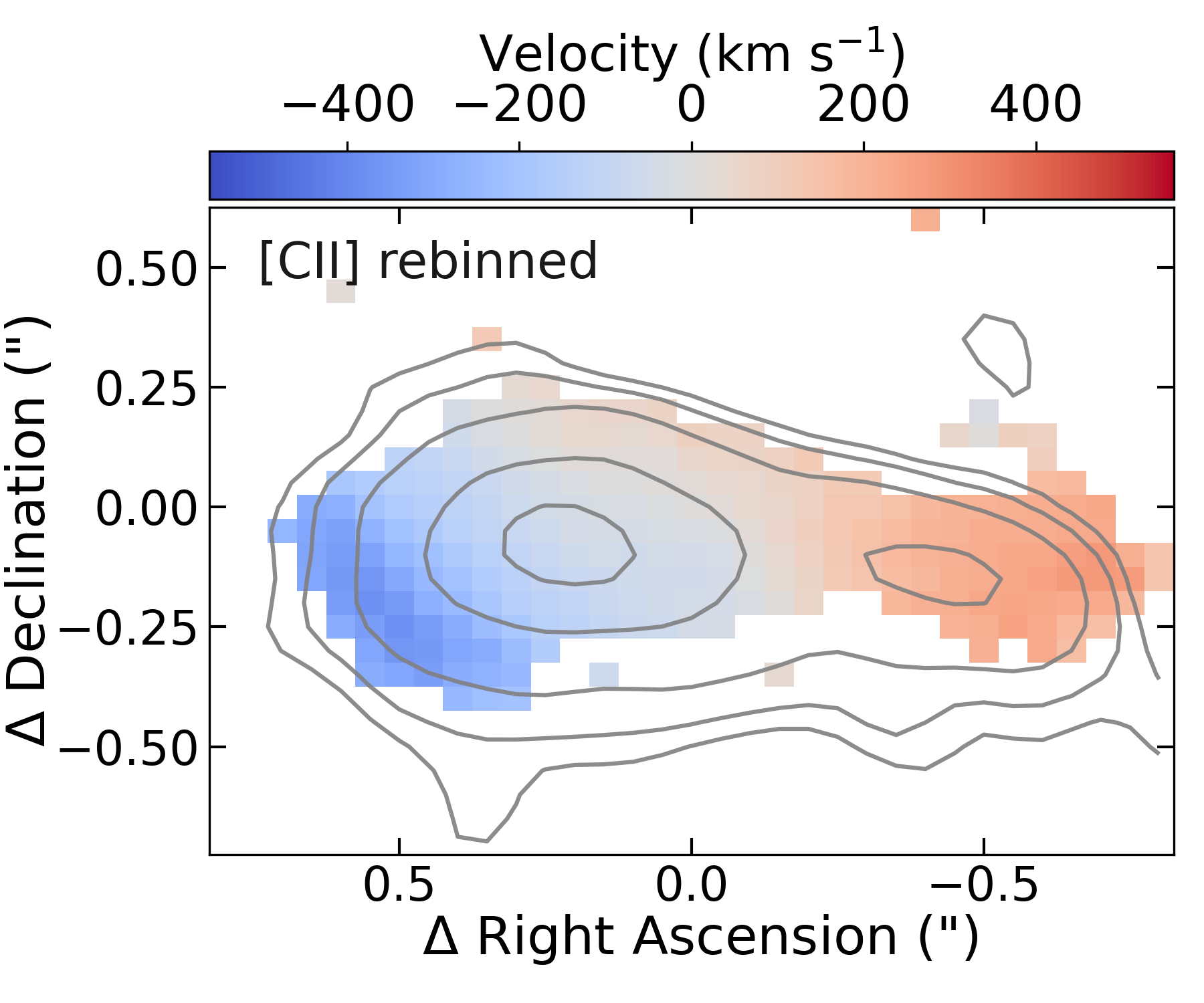}}

{\includegraphics[trim={1.1cm 1.cm 1cm 1cm},clip,width = 0.05\textwidth]{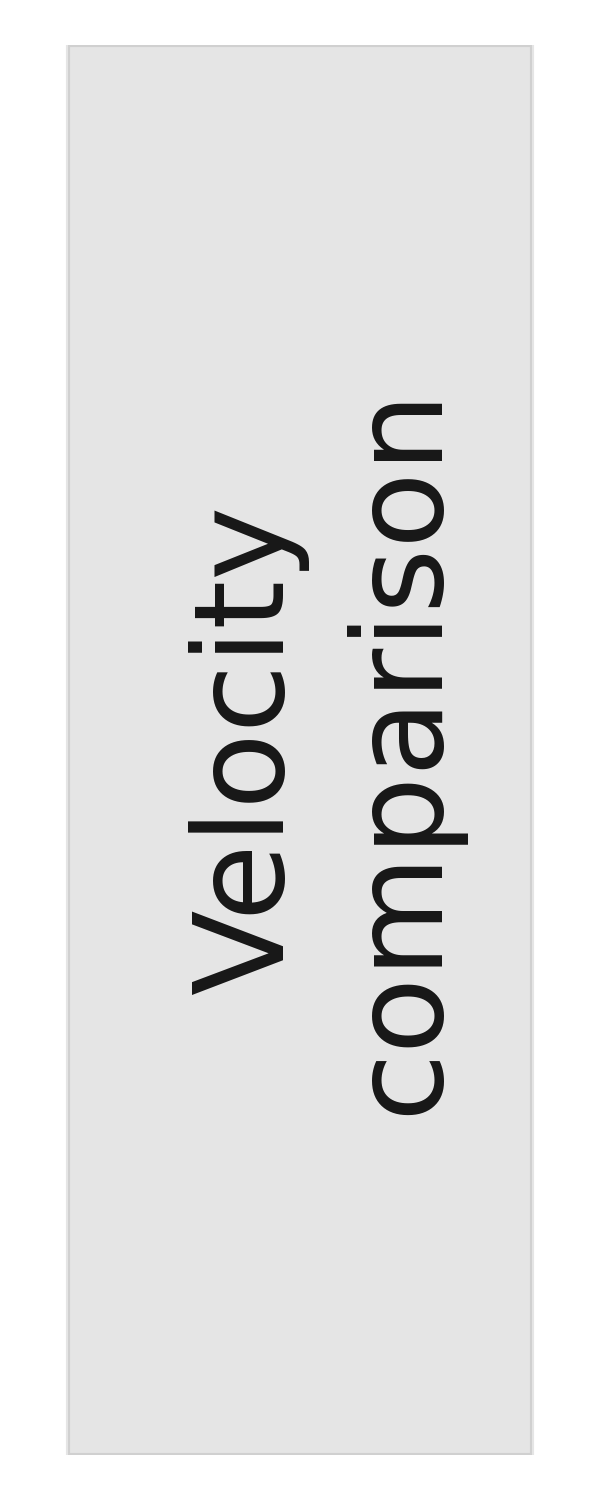}}
{\includegraphics[trim={0.0cm 1.75cm 0 0},clip, width = 0.28\textwidth]{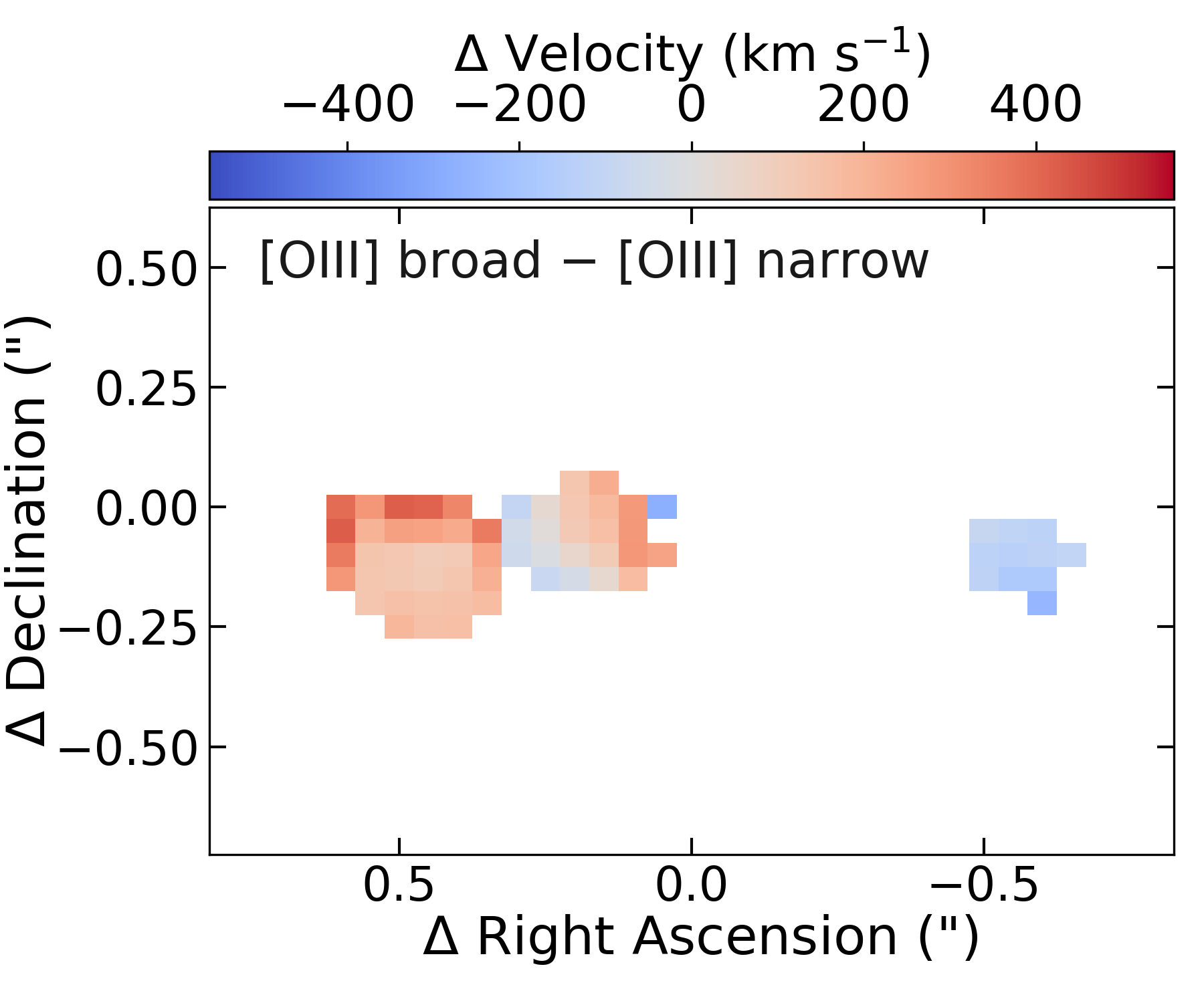}}
{\includegraphics[trim={0.0cm 1.75cm 0.0cm -0.5cm},clip,width = 0.28\textwidth]{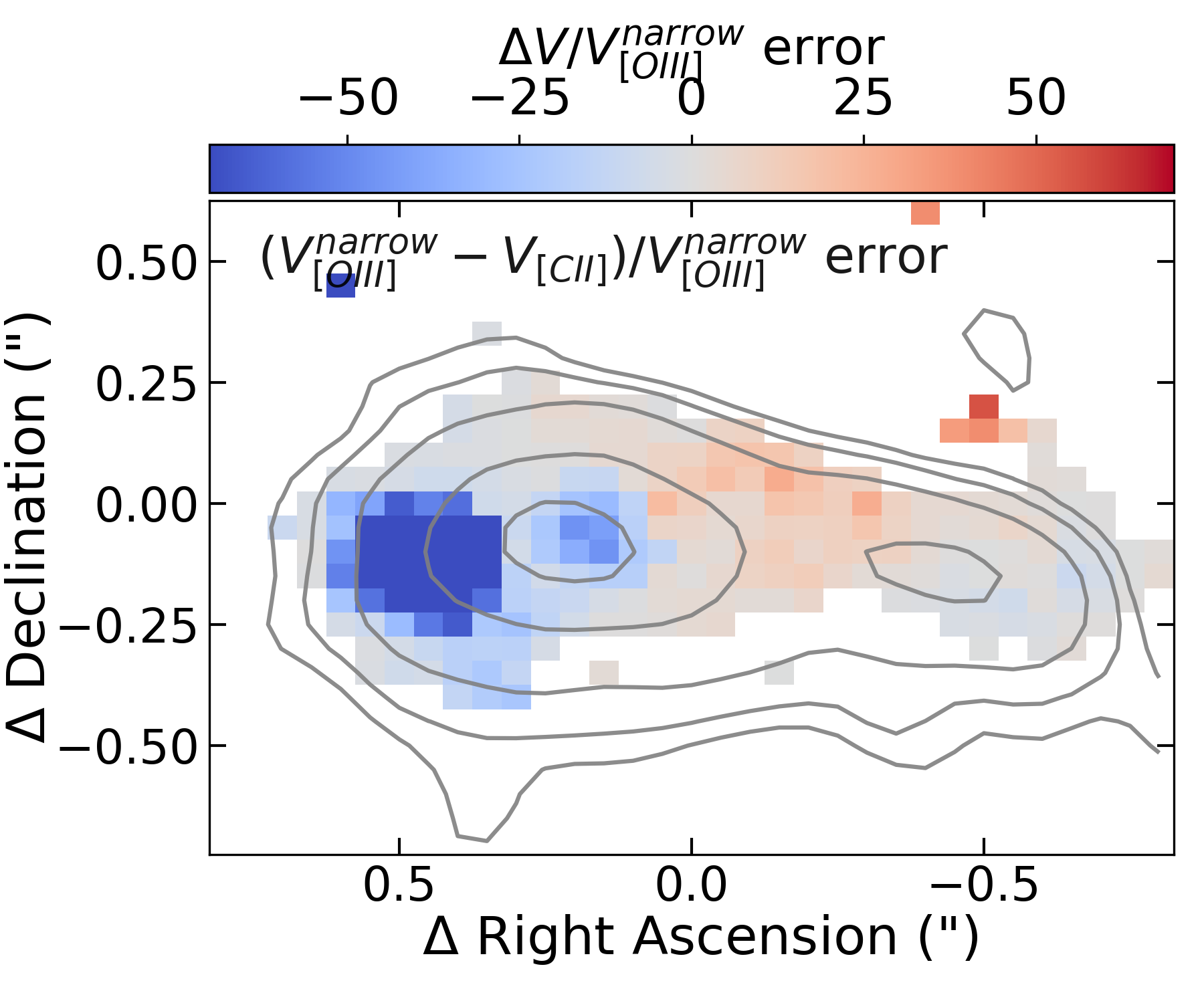}}
{\includegraphics[trim={0.0cm 1.75cm 0 0},clip,width = 0.28\textwidth]{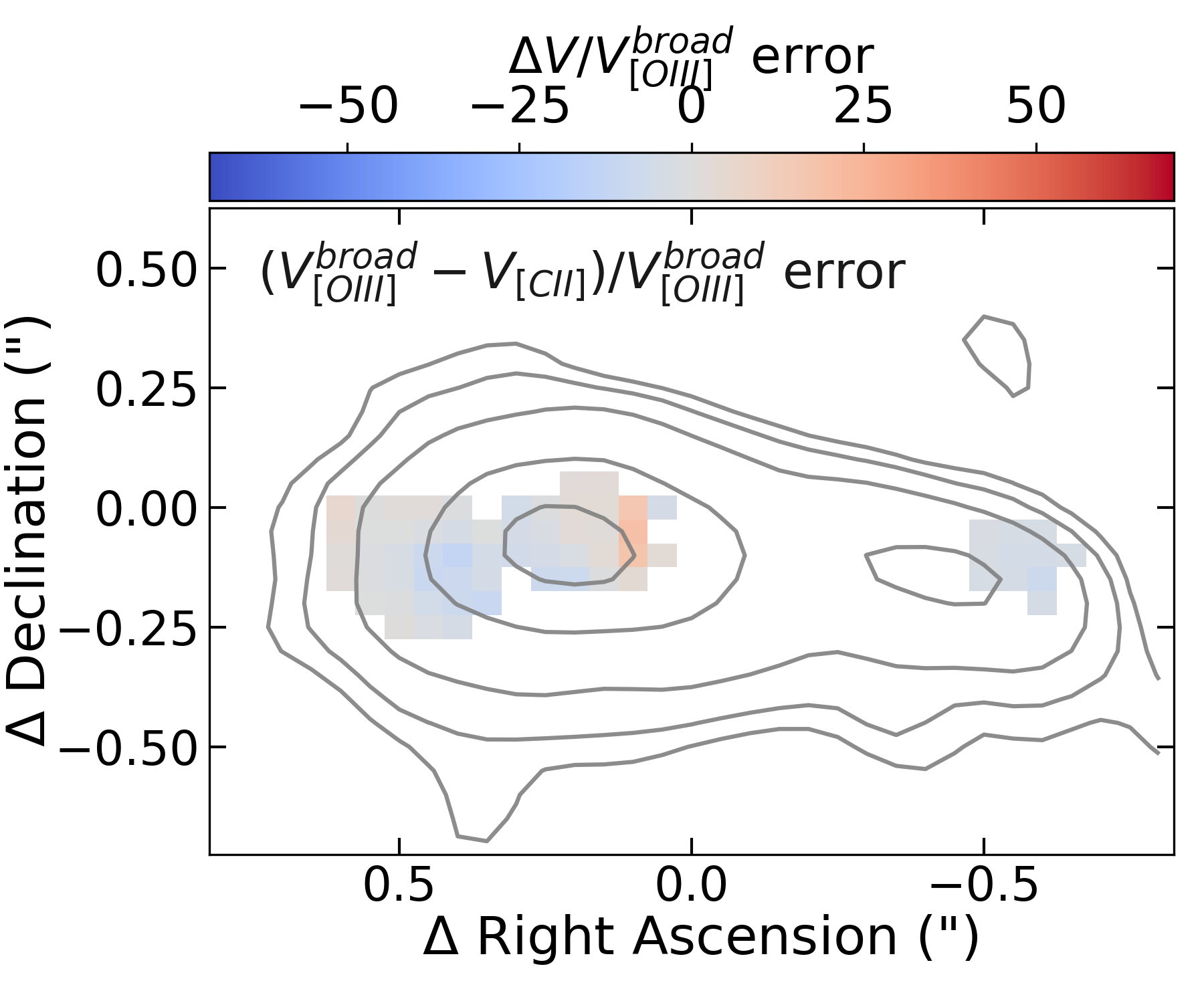}}

{\includegraphics[trim={1.1cm 1.cm 1cm 1cm},clip,width = 0.05\textwidth]{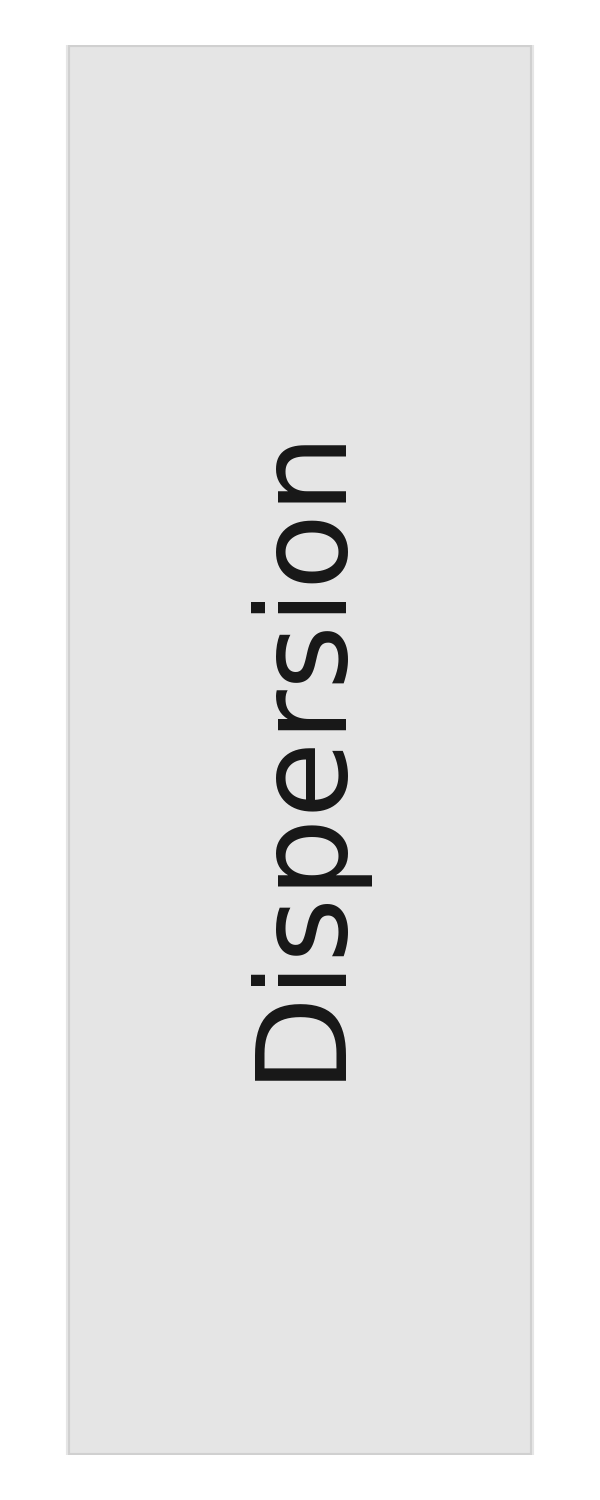}}
{\includegraphics[trim={0.0cm 1.75cm 0 -0.5cm},clip,width = 0.28\textwidth]{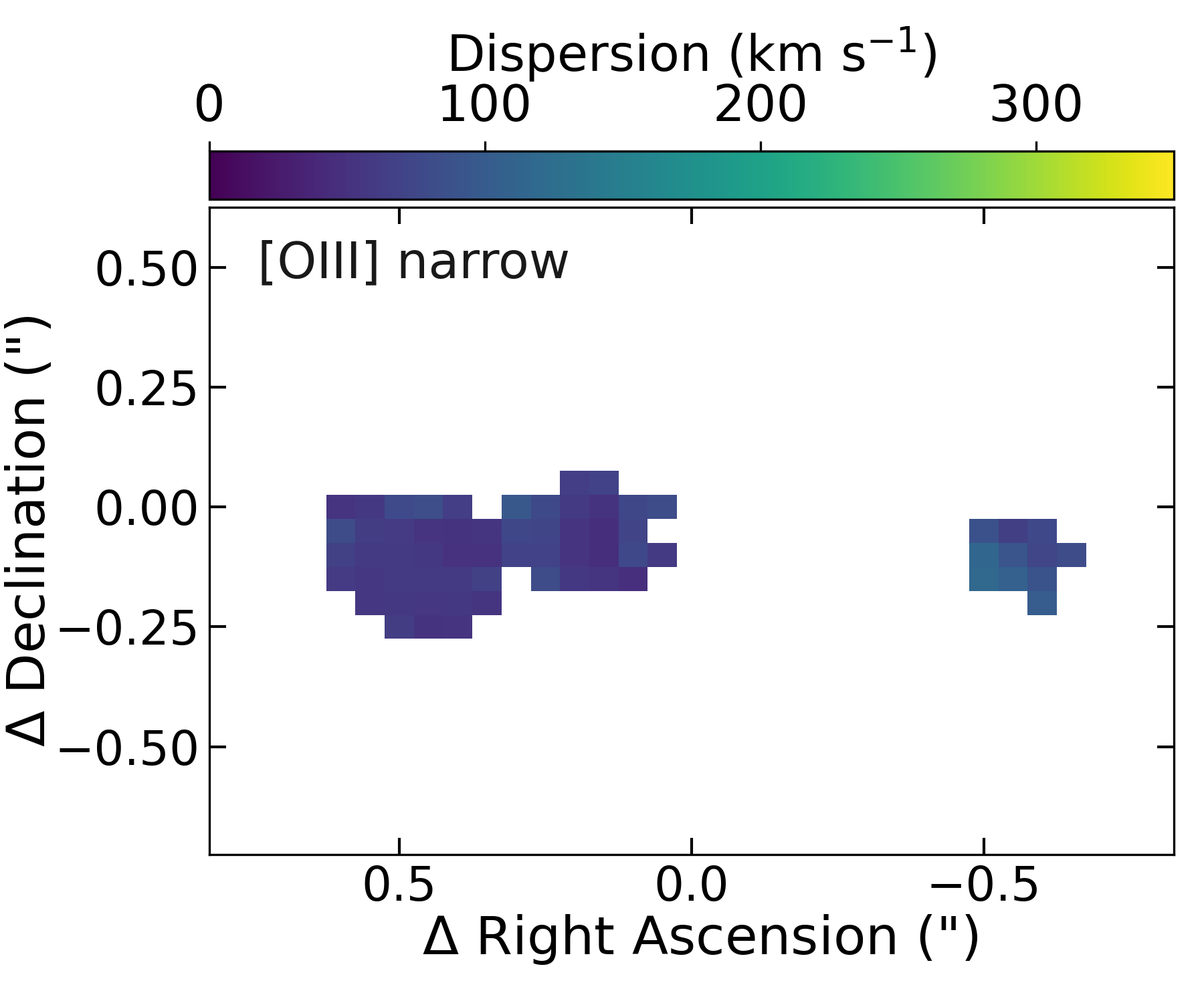}}
{\includegraphics[trim={0.0cm 1.75cm 0 0},clip,width = 0.28\textwidth]{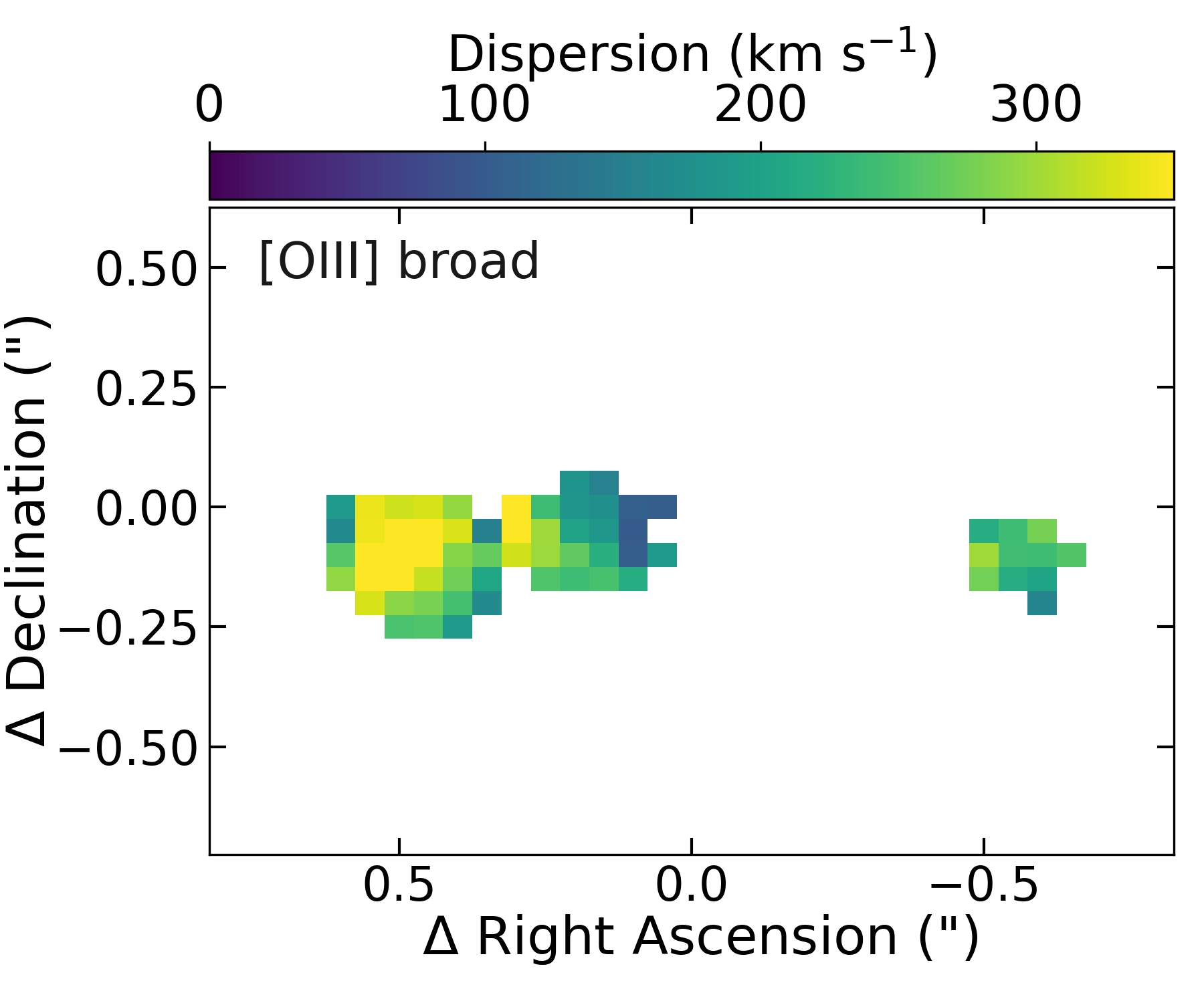}}
{\includegraphics[trim={0.0cm 1.75cm 0 0},clip,width = 0.28\textwidth]{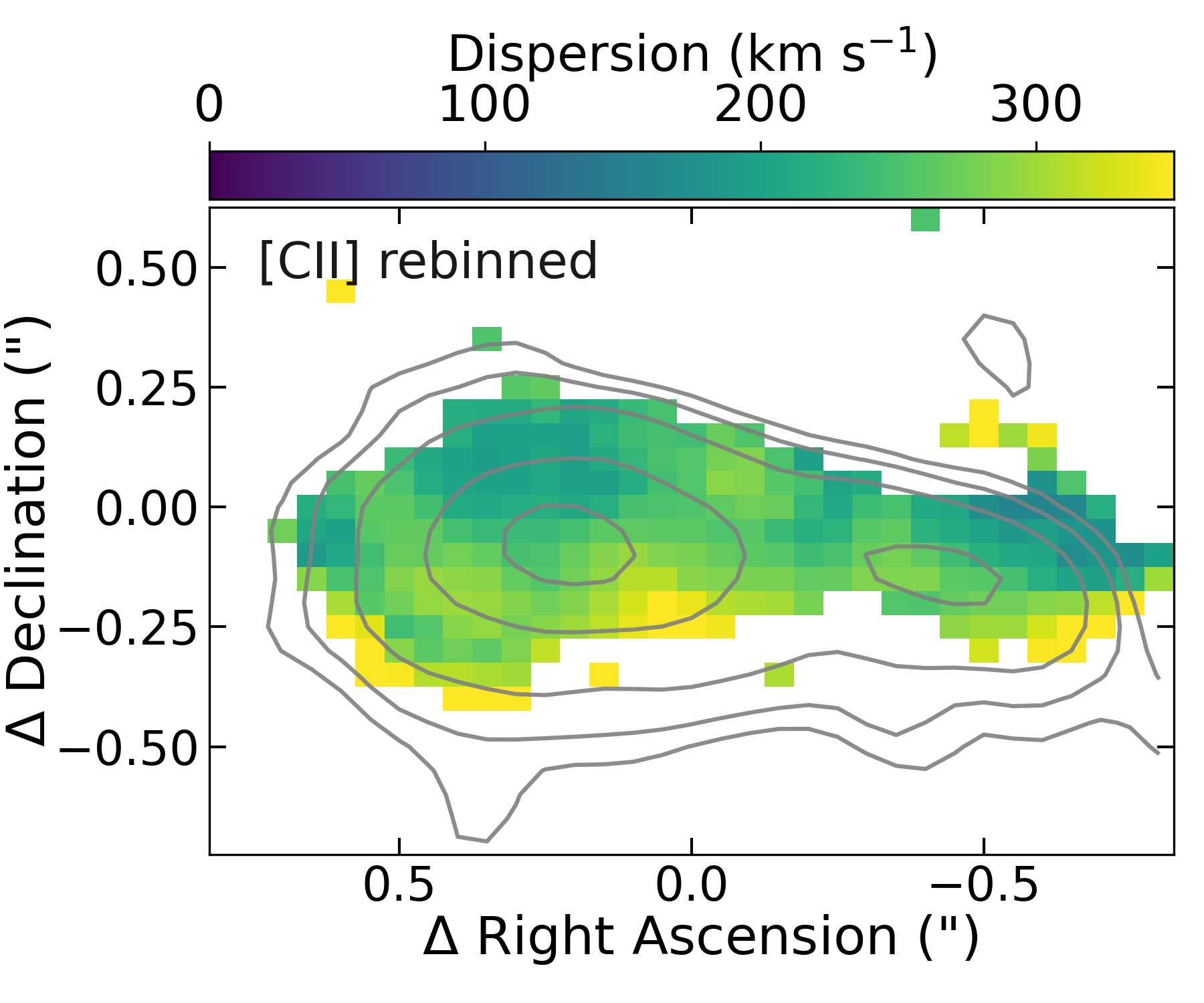}}

{\includegraphics[trim={1.1cm -0.75cm 1cm 1cm},clip,width = 0.05\textwidth]{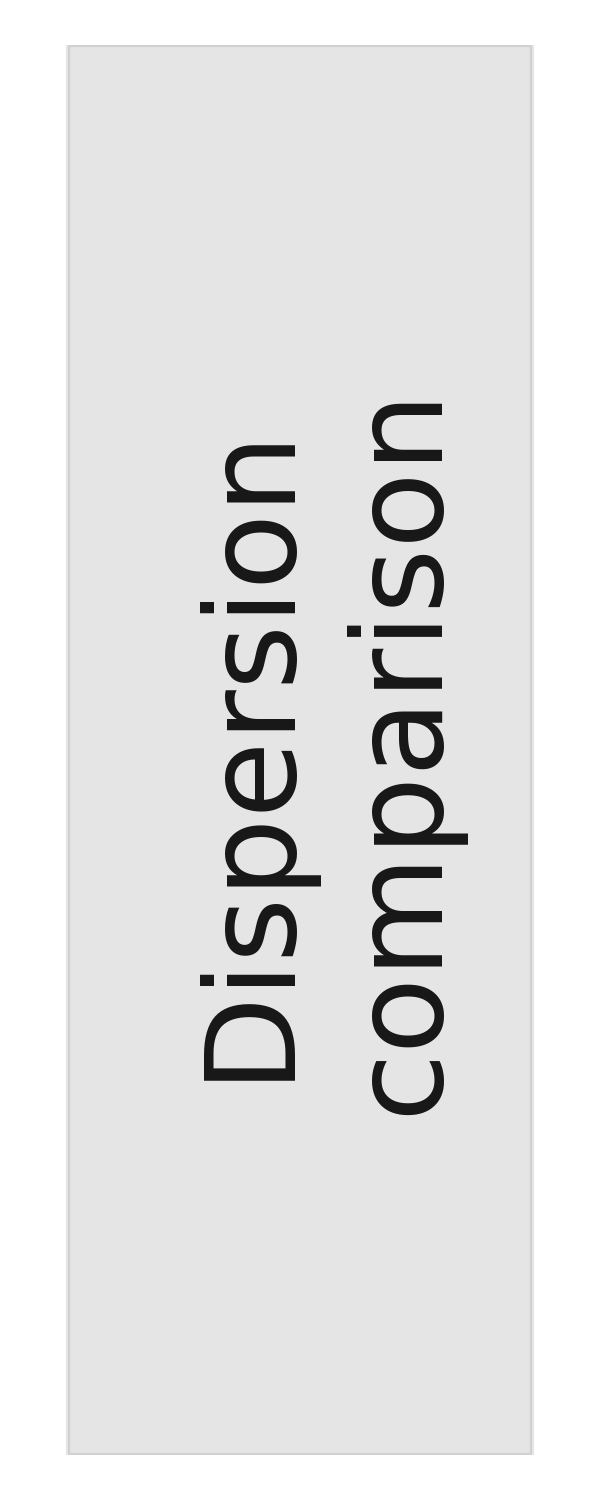}}
{\includegraphics[trim={0.0cm 0cm 0 -0.5cm},clip,width = 0.28\textwidth]{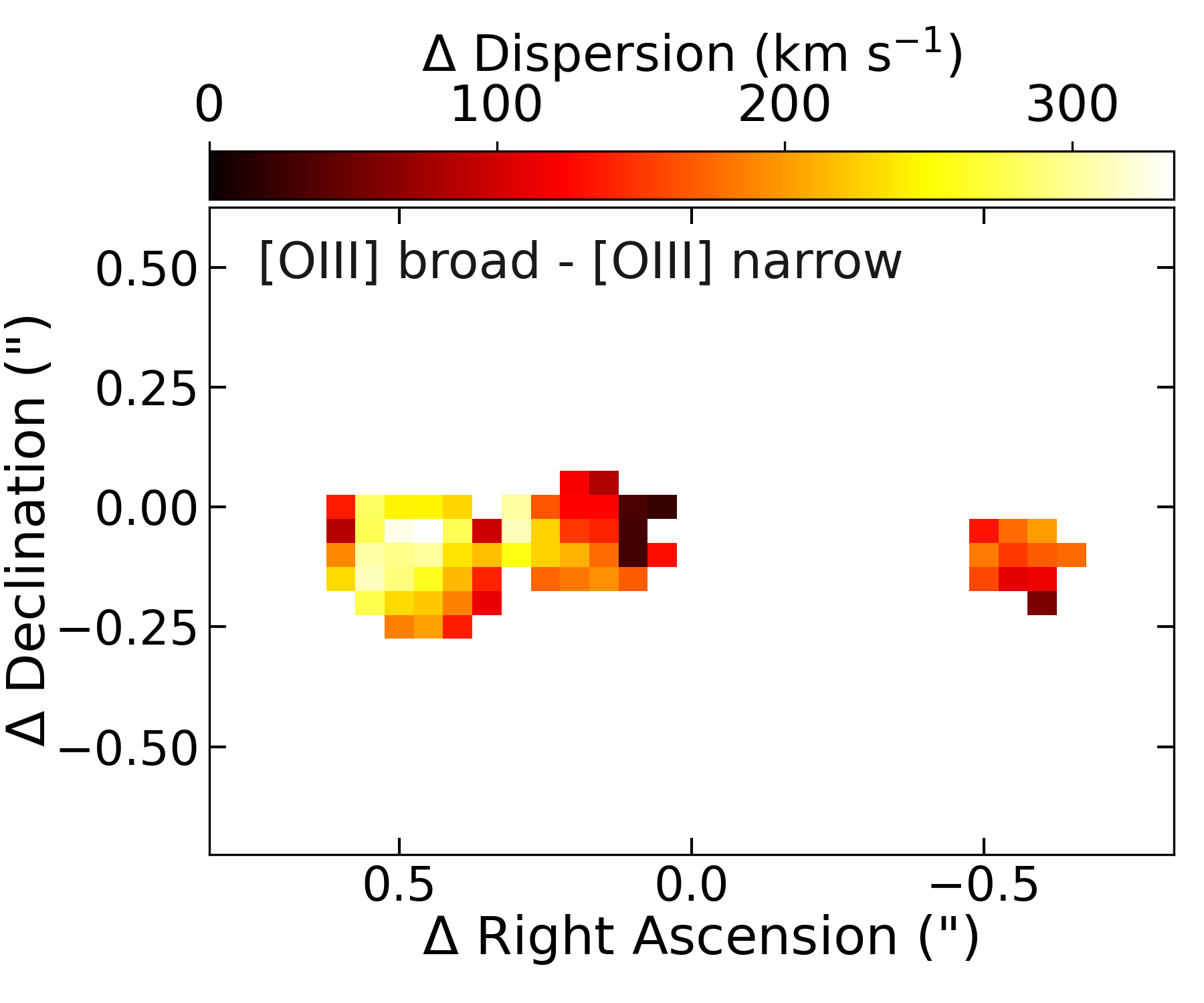}}
{\includegraphics[trim={0.0cm 0cm 0 0},clip,width = 0.28\textwidth]{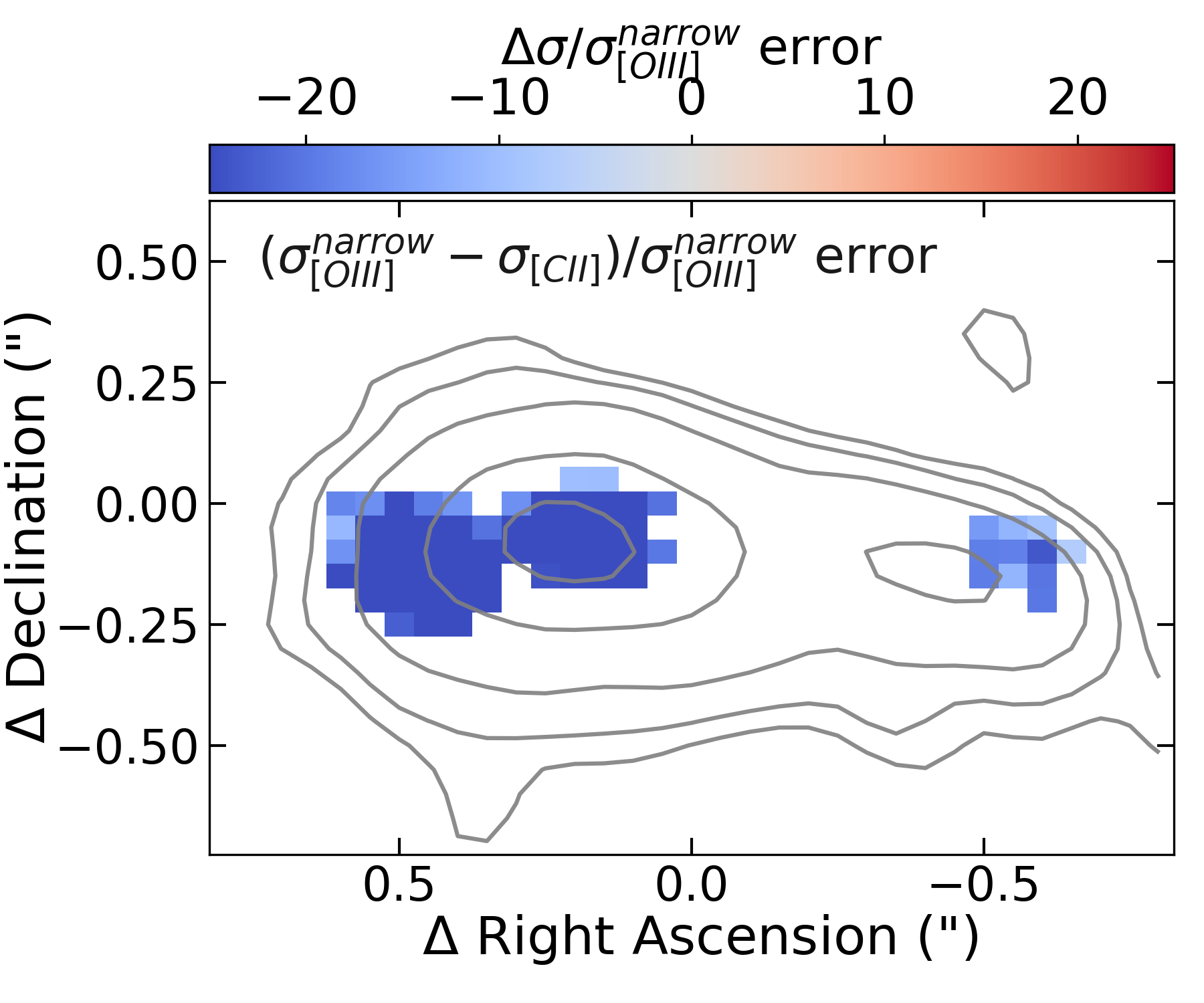}}
{\includegraphics[trim={0.0cm 0cm 0 0},clip,width = 0.28\textwidth]{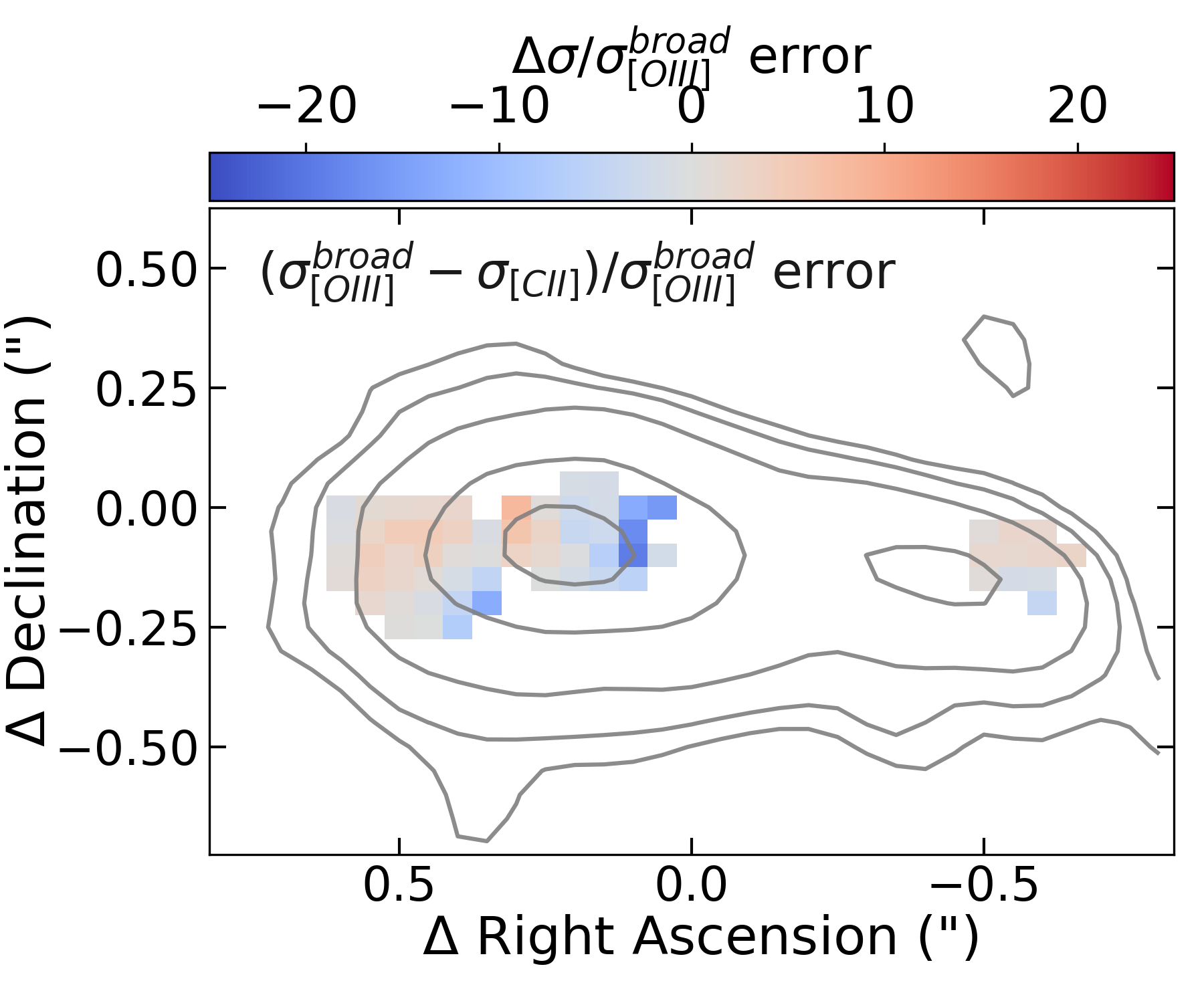}}

\caption{From top to bottom: spaxel-by-spaxel flux attributed to the narrow and broad [\OIII] emission and the corresponding flux ratio, [\OIII] (narrow and broad line emission component) and [\CII] velocity and velocity dispersion maps and their respective comparison maps. Zero velocity corresponds to the rest-frame frequency at $z = 5.6548$ for all maps. The velocity dispersion of the narrow [\OIII] component in most of the spaxels is limited by the JWST/NIRSpec spectral resolution of $R=2700$.
 }\label{fig:velocity-ALMA-JWST}
\end{figure*}

In Fig.~\ref{fig:velocity-ALMA-JWST} we show the corresponding [\OIII] velocity and dispersion maps derived from the best-fit spectral profiles, rebinned and masked [\CII] moment-0 and moment-1 maps and velocity and velocity dispersion comparison maps. Zero velocity is calculated for the rest-frame [\OIII] or [\CII] frequency at $z=5.6548$. For the broad [\OIII] component maps and the velocity dispersion of the narrow [\OIII] component we only show the spaxels where the double Gaussian profile fit is preferred over the single Gaussian profile fit.

In the top panel of Fig.~\ref{fig:velocity-ALMA-JWST} we also compare the [\OIII] flux attributed to the broad and narrow emission line components. 
The flux ratio between the narrow and broad [\OIII] line emission shows that the broad [\OIII] component, which could trace the outflows and/or tidal interactions, contributes significantly to the total [\OIII] flux of the HZ10 system. 
Furthermore, as seen from the spaxel-by-spaxel narrow to broad [\OIII] flux ratio, the narrow [\OIII] emission is surrounded by the broad [\OIII] emission. This effect is most noticeable in the case of HZ10-E.
As expected, the velocity difference between the [\OIII] narrow and broad components reproduces the [\OIII] spectral line asymmetry map in \cite{Jones2024}.

From the comparison of the [\OIII] and [\CII] velocity maps, one can see that the velocity of the [\OIII] narrow emission line is significantly blue-shifted from the [\CII] emission in the central spaxels of HZ10-C and for the entire HZ10-E component. The blue velocity shift for the HZ10-C component reflects the brightness asymmetry in the [\CII] double-gaussian profile in the corresponding spaxels.
On the other hand, the blue shift observed in the HZ10-E component has a more complicated nature and is caused by several factors. Firstly, the ALMA angular resolution leads to a blending of the different [\CII] spectral components from HZ10-C and HZ10-E within the individual spaxels.
Secondly, the weaker [\CII] line emission from HZ10-E compared to HZ10-C elevates this effect resulting in a red shift of the intensity-weighted [\CII] velocity of HZ10-E compared with that of the bright and spatially isolated from HZ10-C [\OIII] narrow emission.

\section{Discussion}\label{sec:discussion}
In the previous sections, we have presented the analysis of the HZ10 system including kinematic modeling of the [\CII]~158$\mu$m emission and comparison of the high spatial and spectral resolution ALMA [\CII] observations with the rest-frame optical emission lines from JWST/NIRSpec. Our main findings are as follows:

(a) We found evidence for at least three [\CII] emission components of the HZ10 system, which are consistent with those recently identified with JWST/NIRSpec rest-frame optical emission line observations;

(b) The HZ10 is a complex system likely showing the ongoing merger between the [\CII] bright HZ10-C$+$HZ10-E component
and HZ10-W;

(c) The HZ10-C$+$HZ10-E system suggests three possible scenarios that are consistent with the currently available observations and modeling: (i) HZ10-C$+$HZ10-E is a disturbed galaxy disk, where HZ10-E is a fainter clump within the extended disk; (ii) HZ10-E is a satellite galaxy merging with the disk-like galaxy HZ10-C; (iii) HZ10-C$+$HZ10-E is triple merger where HZ10-C is itself a close double merger showing ongoing merging with the faint companion galaxy HZ10-E. 

(d) The joint analysis of the [\CII] and [\OIII] spectral profiles and velocity maps indicates that [\CII] and broad [\OIII]  emission kinematically trace each other.

Although [\CII] emission can originate from various phases of the interstellar medium, including the ionized phase, both low-redshift observations and numerical simulations indicate that the majority of the total [\CII] luminosity in galaxies is primarily produced in photodissociation regions \citep[e.g.,][]{Pineda2013,Pallottini2017,Olsen2017}. 
Based on modeling of the [\CII] ~158$\mu$m/[\NII]~205$\mu$m ratio using the photoionization code \textsf{Cloudy}, \cite{Pavesi2016} estimated that only $10-25$\% of the observed [\CII] emission in HZ10 originates from ionized gas. Although this estimate was made for the entire HZ10 system and could only be used for a resolved estimate of the individual components under strict assumptions, the similarity in kinematic properties of the integrated [\CII] and [\OIII] line emission profiles across all three HZ10 components (Fig.~\ref{fig:HZ10-aperture-spectra}) suggests that the neutral and ionized gas in this system trace similar kinematics and appear to be well-mixed.
Moreover, a spaxel-by-spaxel analysis shows that [\CII] velocity and velocity dispersion are particularly similar to those of broad [\OIII] emission (see Fig.~\ref{fig:velocity-ALMA-JWST}), which, as shown by \cite{Jones2024}, trace ionized outflows and tidal interactions in HZ10. Therefore, the kinematic similarity of the [\CII] and [\OIII] line emission points to an interacting nature of the HZ10 system, rather than a multiphase origin of the [\CII] emission.

A recent joint analysis of JWST/NIRSpec and ALMA observations of another high-redshift clumpy merging system, HZ4, revealed a multiphase outflow detectable in [\OIII]~5007$\AA$, H$\alpha$, and [\CII] emission \citep{Parlanti2024}. 
While we do not detect outflow features in [\CII] emission for HZ10, this is likely due to the limited angular resolution of the ALMA data and the close separation between system components.

While the kinematic properties of the [\CII] emission for all three HZ10 components are similar to those of [\OIII], the flux line ratios, calculated in the apertures shown in Fig.~\ref{fig:HZ10-alma-jwst}, differ significantly: for HZ10-E [\OIII]~$5007\AA$/[\CII] is $\gtrsim1.2$, while for HZ10-C and HZ10-E this ratio is $\sim0.5$ and $\sim0.6$, respectively. Despite being the brightest component in the HZ10 system in [\OIII], HZ10-E is very dim in [\CII].

Since the carbon abundance linearly depends on the metallicity, the decreased [\CII] flux for the HZ10-E component can be explained by its relatively low metallicity among the entire HZ10 complex ($Z/Z_\odot =0.53\pm0.09$ in comparison with $Z/Z_\odot =0.64\pm0.11$ and $0.72\pm0.12$ for HZ10-C and HZ10-W \citep{Jones2024}). However, metallicity measurements based on JWST observations for all three components agree with each other within the provided uncertainties. 

Studies in local galaxy clusters have shown that photoevaporation can complement the viscous stripping in removing the cold gas from galaxy disks \citep[e.g.,][and references therein]{Bureau2002,Cortese2021,Villanueva2022}. This may drive lower [\CII] fluxes and consequently higher [\OIII]~5007$\AA$/[\CII]~158$\mu$m flux ratio.
Indeed, recent numerical simulations have shown that the photoevaporation of molecular clouds caused by intense UV radiation from nearby young massive stars can result in a high flux ratio between [\OIII]~88$\mu$m and [\CII]~158$\mu$m \citep{Vallini2017}. However, according to simulations an increase in the [\OIII]/[\CII] flux ratio in the far-infrared is particularly significant in the context of lower metallicity, specifically below a metallicity of about $Z/Z_\odot \sim0.2$, and not necessarily attributed to the increased [\OIII]~5007$\AA$/[\CII]~158$\mu$m flux ratio, which means that this scenario may only be tentatively applicable to HZ10-E.

While negative stellar feedback could in principle lead to an increased [\OIII]/[\CII] flux ratio by clearing out gas around star-forming region~\citep[e.g.][]{Ferrara2019}, the HZ10-E, on the contrary, exhibits a star-formation rate based on H$\alpha$ similar to HZ10-C and HZ10-W~\citep{Jones2024}.

Based on the JWST/NIRSpec data of HZ10, \citet{Jones2024} propose that HZ10-E is a separate galaxy, merging with HZ10-C and HZ10-W.
Indeed, HZ10-E and HZ10-C show spatially distinct optical line emission peaks, slightly different metallicities, line flux ratios, color excess, and UV slopes, supporting scenarios (ii) and (iii). 
However, there are examples where galaxies exhibit metallicity gradient as well as inhomogeneous dust distribution \citep[e.g.][]{Miller2022,RodriguezDelPino2024}.
Therefore, even with the advent of JWST/NIRSpec data, we cannot completely rule out a scenario (i) in which HZ10-E and HZ10-C are clumps of an extended galaxy disk. 
Since we cannot currently discard any of the three possible scenarios, ALMA observations with a higher angular resolution that matches that of JWST data but with a superior spectral resolution are necessary to distinguish between 
them.

\section{Conclusions}\label{sec:conclusions}

We presented a detailed kinematic analysis of the far-infrared bright massive galaxy complex HZ10 at the end of Epoch of Reionization at $z\approx5.65$. The study was based on the high spatial ($\approx 0.3$\arcsec\ or 1.8~kpc at $z=5.65$) and spectral resolution observations of [\CII]~158$\mu$m line emission with ALMA with support of the high spatial resolution ($\approx 0.15$\arcsec\ or 0.9~kpc at $z=5.65$) observations of the [\OIII]~5007\AA\ line emission with JWST/NIRSpec. 
The new observations of the HZ10 system, previously classified as a massive main-sequence galaxy, revealed its complex morphological and kinematical nature, showing that the HZ10 
consists of at least three components in the close projected distance: HZ10-E, HZ10-C, and HZ10-W. The brightest component in [\CII] line emission, HZ10-C (with the dim HZ10-E component in close separation), is characterized by complex kinematics with the possibility of being either a close merger or a disturbed disk.

The high angular resolution JWST/NIRSpec data supports the complex morphology of HZ10 and shows the presence of the broad component emission, which indicates the presence of outflows and/or tidal interactions in the system \citep{Jones2024}. 
Moreover, the broad [\OIII] emission line component contributes significantly to the total [\OIII] flux of the HZ10 complex.

While ALMA and JWST provide insight into the cold molecular and ionized interstellar gas in early galaxies and trace the star formation on different time scales, future deep integral-field Ly$\alpha$ observations would be the keystone in reconstructing the interconnection between interstellar and circum-galactic gas. The latter is required to characterize and disentangle the physical processes -- outflows, inflows, merger-induced shocks -- occurring in overdensity environments. 
Thanks to available ALMA and JWST data, HZ10 was also selected as a part of the ORigin of the [\CII] Halos In Distant Systems (ORCHIDS) JWST Cycle 3 program 5974. This program aims to complement the rest-frame optical JWST/NIRSpec observations with those of dust and [\CII] ALMA and Ly$\alpha$ Keck/KCRM to unveil the nature of [\CII] halos at high redshifts. 
Furthermore, observations of diffuse stellar emission with JWST MIRI, along with higher spatial resolution [\CII] follow-up observations with ALMA, are needed to gain a more detailed view of the dynamical processes occurring in the close HZ10-E$+$HZ10-C system.

\begin{acknowledgements}
K.T. was supported by ALMA ANID grant number 31220026 and by the ANID BASAL project FB210003. K.Tad. acknowledges support from JSPS KAKENHI grant No. 23K03466. A.F. acknowledges support from the ERC Advanced Grant INTERSTELLAR H2020/740120. H.Ü. gratefully acknowledges support by the Isaac Newton Trust and by the Kavli Foundation through a Newton-Kavli Junior Fellowship. M.R. acknowledges support from projects PID2020-114414GB-100 and PID2023-150178NB-I00 financed by MCIN/AEI/10.13039/501100011033. I.D.L. acknowledges funding from the Belgian Science Policy Office (BELSPO) through the PRODEX project “JWST/MIRI Science exploitation” (C4000142239), from the European Research Council (ERC) under the European Union’s Horizon 2020 research and innovation programme DustOrigin (ERC-2019-StG-851622) and from the Flemish Fund for Scientific Research (FWO-Vlaanderen) through the research project G0A1523N. V.V. acknowledges support from the ALMA-ANID Postdoctoral Fellowship under the award ASTRO21-0062. R.J.A. was supported by FONDECYT grant number 1231718 and by the ANID BASAL project FB210003. R.I. is supported by Grants-in-Aid for Japan Society for the Promotion of Science (JSPS) Fellows (KAKENHI Grant Number 23KJ1006). I.L. acknowledges support from PRIN-MUR project "PROMETEUS" (202223XPZM). M.P. acknowledges support from the research project PID2021-127718NB-I00 of the Spanish Ministry of Science and Innovation/State Agency of Research (MCIN/AEI/10.13039/501100011033), and of the INAF Large Grant 2022 "The metal circle: a new sharp view of the baryon cycle up to Cosmic Dawn with the latest generation IFU facilities". T.D.S. acknowledges the research project was supported by the Hellenic Foundation for Research and Innovation (HFRI) under the "2nd Call for HFRI Research Projects to support Faculty Members \& Researchers" (Project Number: 3382).

\end{acknowledgements}

\bibliographystyle{aa}
\bibliography{ref.bib}

\begin{appendix}

\section{Morphological 2D parametric modeling}\label{a:2D_morphological_modeling}
\begin{figure}[h]
\centering
{\includegraphics[trim={2.0cm 0.0cm 0 0},clip, width = 0.9\columnwidth]{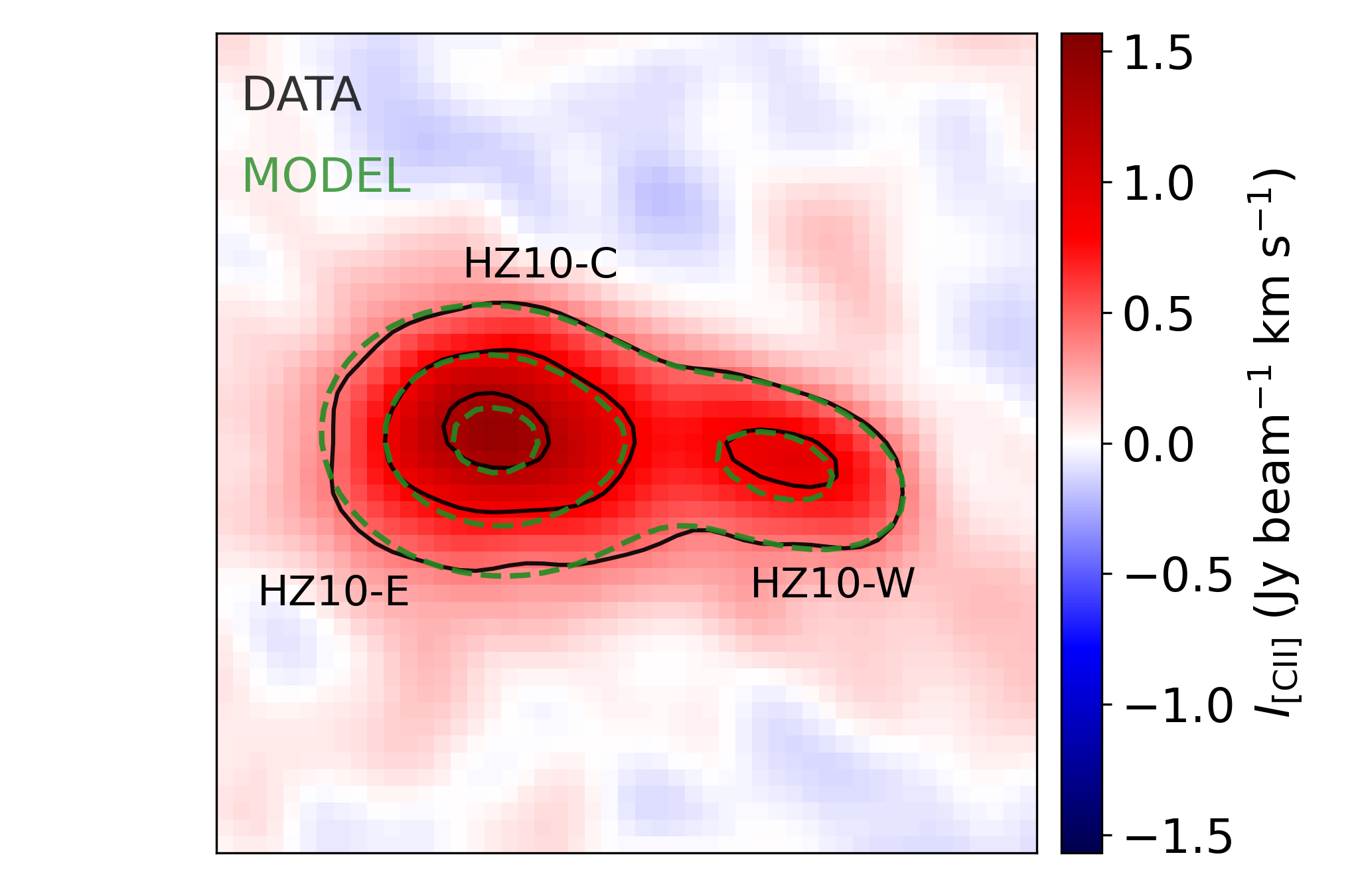}}
{\includegraphics[trim={2.0cm 0.0cm 0 0}, clip, width = 0.9\columnwidth]{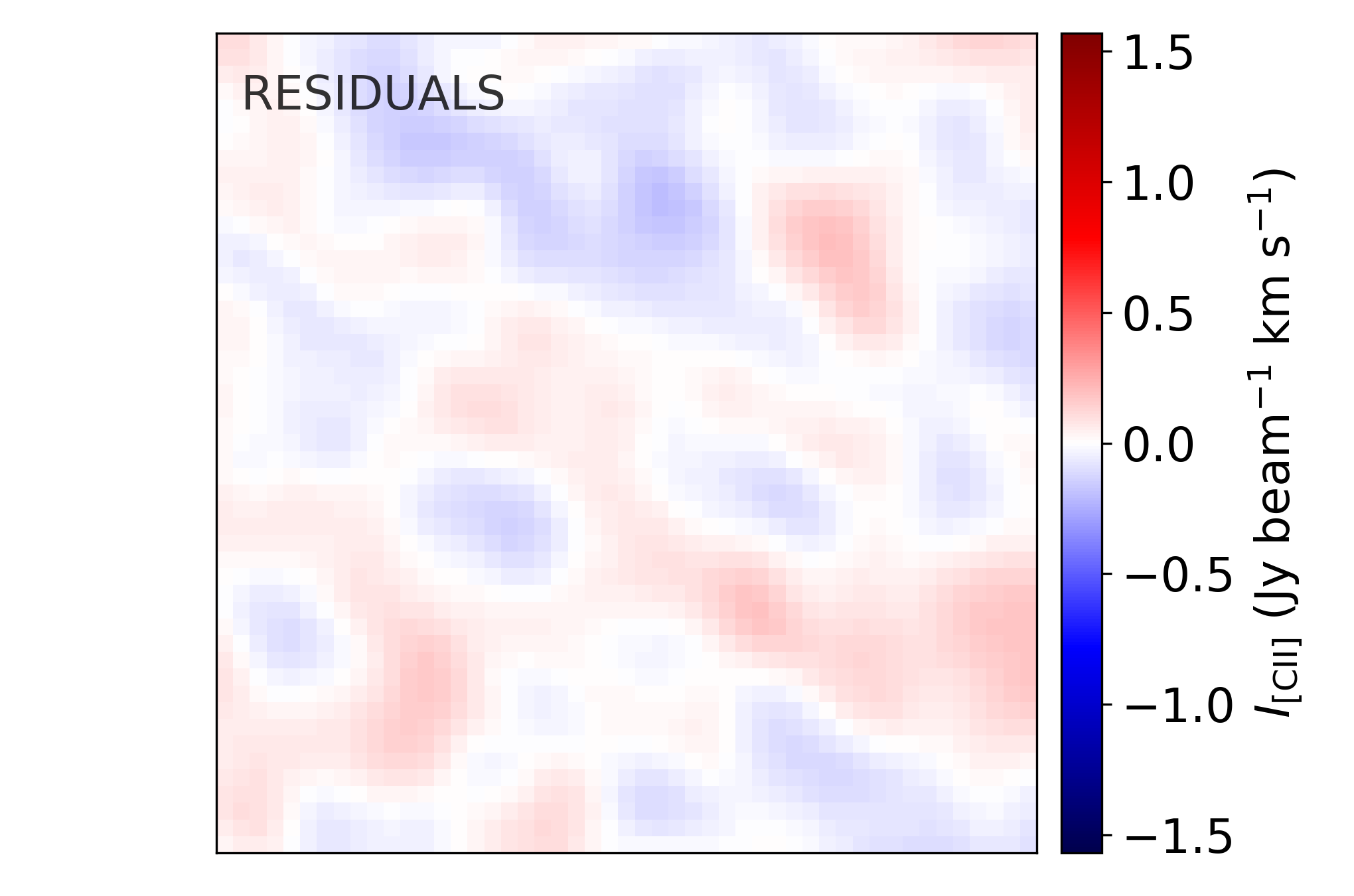}}
\caption{Top: [\CII] integrated intensity (colormap) and 5-, 10-, and 15-$\sigma$ noise levels (black contours). The best-fit two-component 2D S\'ersic profile model is shown by dashed green 5-, 10-, and 15-$\sigma$ contours (see Table~\ref{tab:hz10_properties} for the parameter details). Bottom: Residuals between observed intensity map and best-fit 2D S\'ersic profile.}\label{fig:pyautofit}
\end{figure}

We perform parametric 2D modeling of the [\CII] moment-0 map with a two-component S\'ersic profile using \textsf{PyAutoGalaxy} software. The procedure follows the steps as the 2D modeling of the HZ10 system presented in \citep{Villanueva2024arXiv}. The result of the modeling, along with the residuals between the observations and the model, are shown in Fig.~\ref{fig:pyautofit}.

\section{Aperture spectra fit}\label{a:aperture_spectra_fit}

\begin{figure}[t]
\centering
{\includegraphics[trim={1.0cm 2.5cm 2.5cm 2.5cm},clip,width = 0.99\columnwidth]{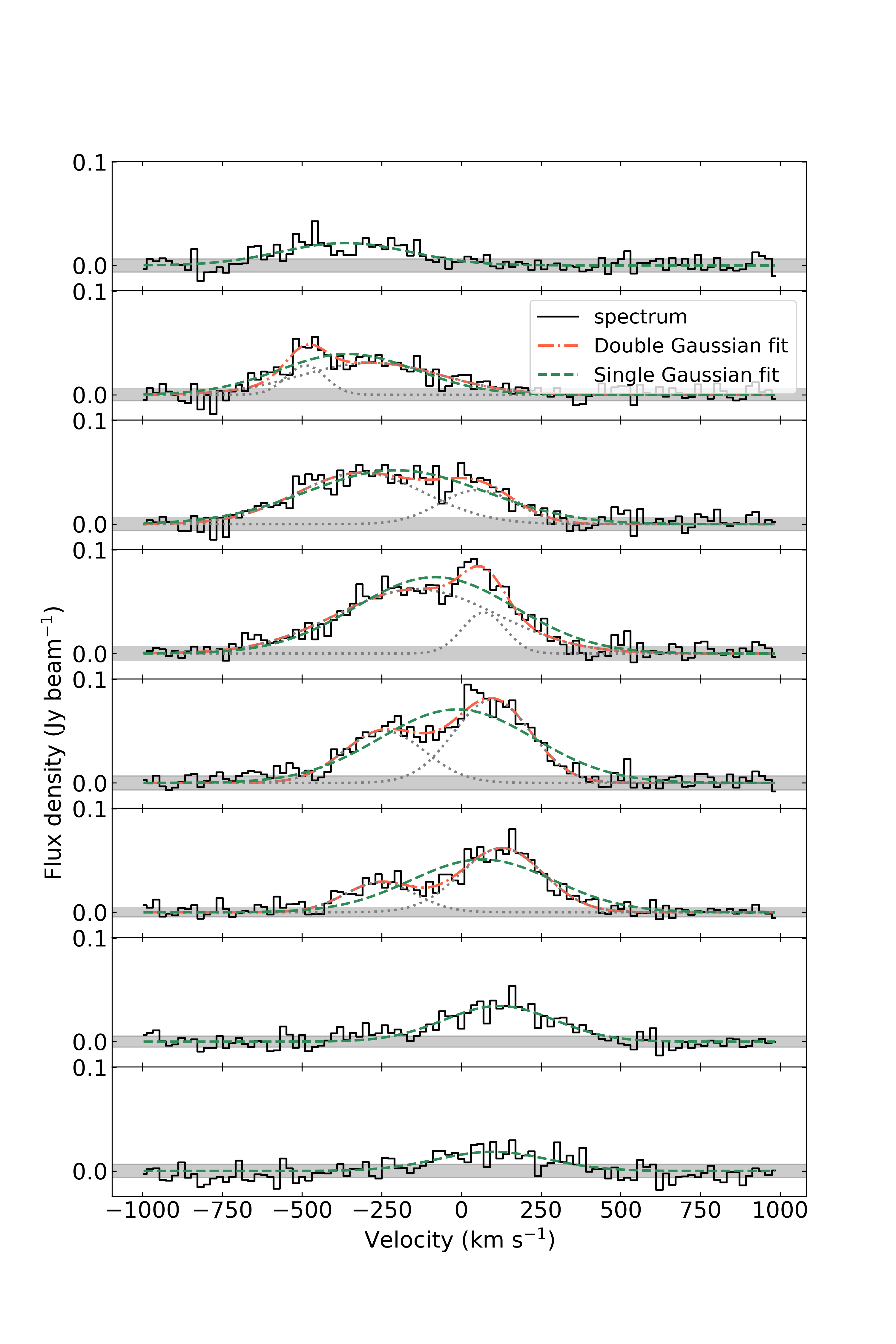}} 
\caption{The HZ10-E$+$HZ10-C aperture [\CII] emission line spectra (black solid) and single (green dashed) or double (red dash-dotted) Gaussian fit profiles. When the double Gaussian fit is performed, the dotted grey curves show the individual components. The spectra (arranged from top to bottom) are extracted from the solid green regions on the moment-0 map in Fig.~\ref{fig:moments-HZ10}, displayed from left to right.}\label{fig:hz10-aperture-spectra-rot}
\end{figure}

\begin{figure}[t]
\centering
{\includegraphics[trim={1.0cm 2.5cm 2.5cm 2.5cm},clip,width = 0.99\columnwidth]{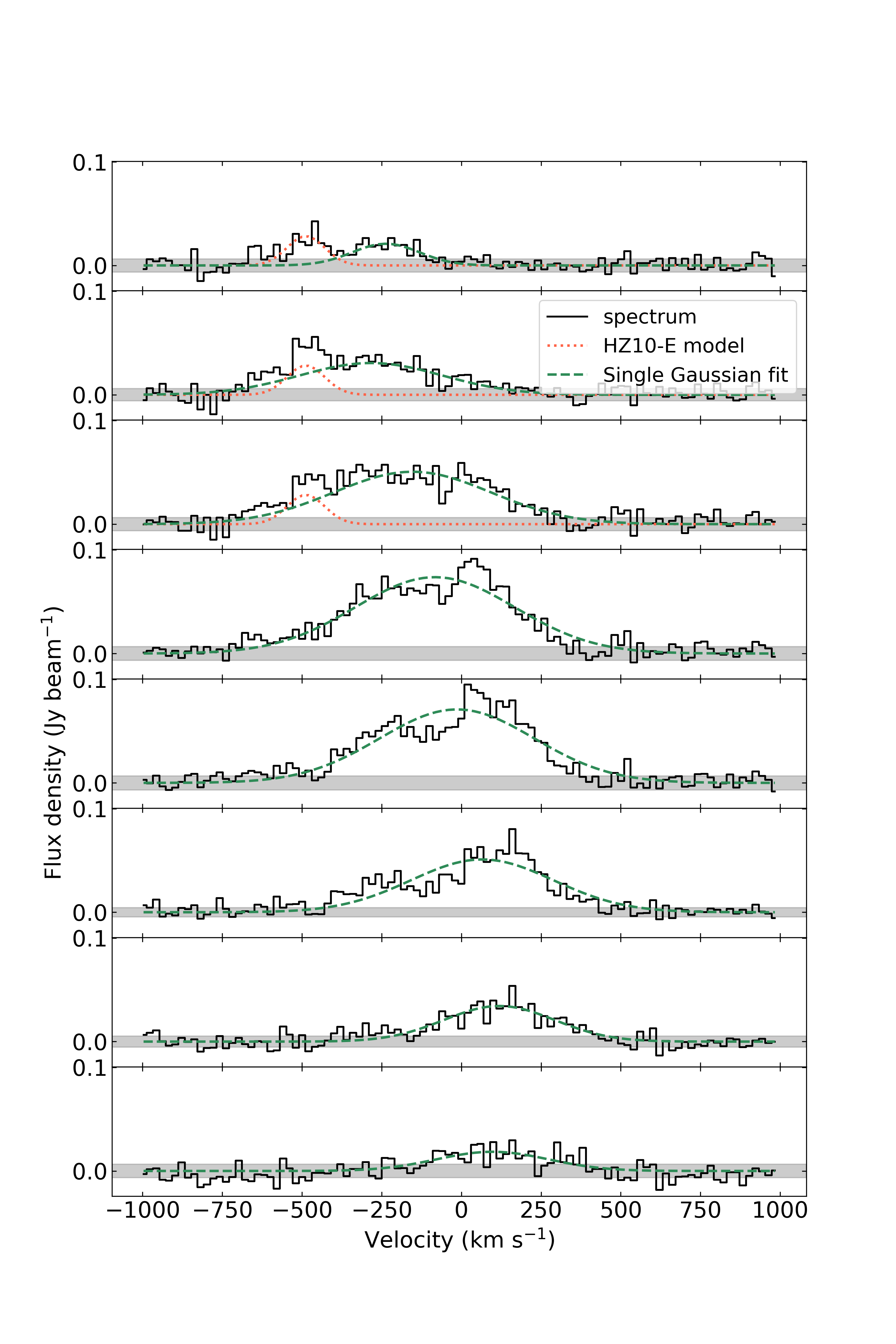}} 
\caption{The HZ10-E$+$HZ10-C aperture [\CII] emission line spectra (black solid) and single Gaussian fit profiles (green dashed) attributed to HZ10-C. The spectra (arranged from top to bottom) are extracted from the solid green regions on the moment-0 map in Fig.~\ref{fig:moments-HZ10}, displayed from left to right. The contribution from HZ10-E in the first three apertures is shown by the red dotted Gaussian profile.}\label{fig:hz10-aperture-spectra-rot-HZ10-C}
\end{figure}

We calculate the rotation curve of the HZ10-C$+$HZ10-E complex by extracting [\CII] spectra along its major axis using circular apertures with a diameter of 0.26\arcsec (solid green regions on the moment-0 map in Fig.~\ref{fig:moments-HZ10}). The corresponding aperture spectra are shown in Fig.~\ref{fig:hz10-aperture-spectra-rot}.
We fit the extracted spectra with single and double Gaussian profiles to calculate velocity centroids and velocity dispersion in each aperture. In cases where the double-component profile fitting effectively results in a single-component solution (i.e. one of the components is poorly constrained), we use only a single-Gaussian model fit.
The resulting fit profiles are shown in the same Fig.~\ref{fig:hz10-aperture-spectra-rot} by green dashed (single Gaussian model) and red dash-dotted (double Gaussian model). Although we performed both single and double Gaussian profile fits, our focus is on the single Gaussian fit, as we aim to model the rotation curves using \textsf{DysmalPy}. \textsf{DysmalPy} is specifically designed to account for observational effects such as beam smearing. As a result, the aperture spectra extracted from the mock cube also exhibit asymmetrical profiles, which are further analyzed using single Gaussian fits. However, the authors conclude that the 1D approach is just as effective as the 2D approach (which uses moment maps for comparison between model and observations) and the 3D approach (where observed and mock cubes are directly compared) (Lee et al., in prep.).
For an additional test, we also separate the HZ10-C spectral component for its independent kinematic analysis with \textsf{DysmalPy}. For that, we model the contribution from the HZ10-E to the total spectra in the first three apertures using a Gaussian profile.
We show the corresponding spectral fits in Fig.~\ref{fig:hz10-aperture-spectra-rot-HZ10-C}.

\section{DysmalPy kinematic modeling}\label{a:dysmal}
\begin{figure*}[h]
\centering
{\includegraphics[trim={0.0cm 0.0cm 0.0cm 0.0cm},clip,width = 0.75\textwidth]{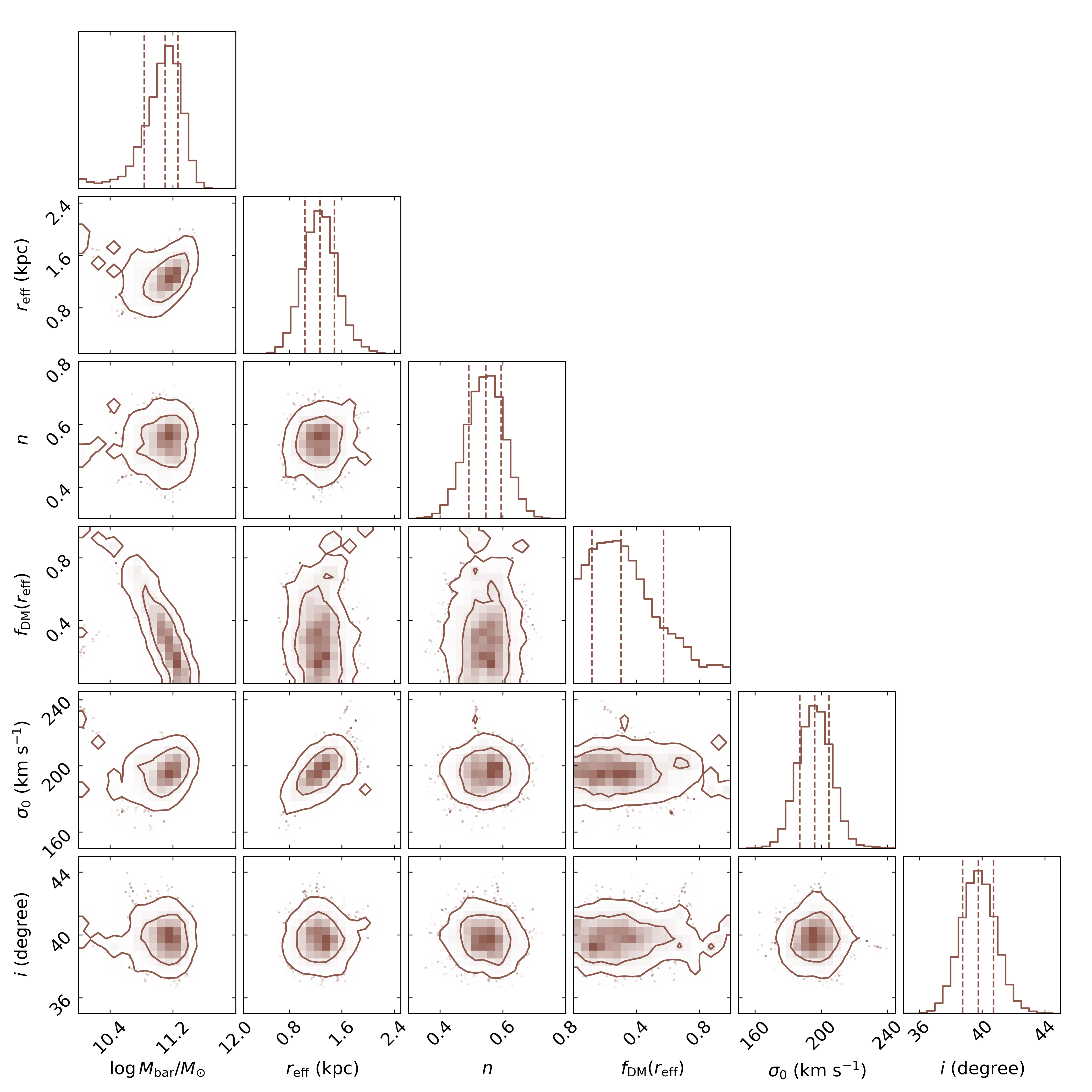}}
\caption{Posterior distributions of the estimated parameters for HZ10-E$+$HZ10-C system. Contours correspond to the 68 and 75 percent highest posterior density credible regions, respectively. Vertical dotted lines show 0.16, 0.5, and 0.84 quantiles of the posterior distributions.}\label{fig:dysmal_corner}
\end{figure*}

\begin{figure*}[h]
\centering
{\includegraphics[width = 0.45\textwidth]
{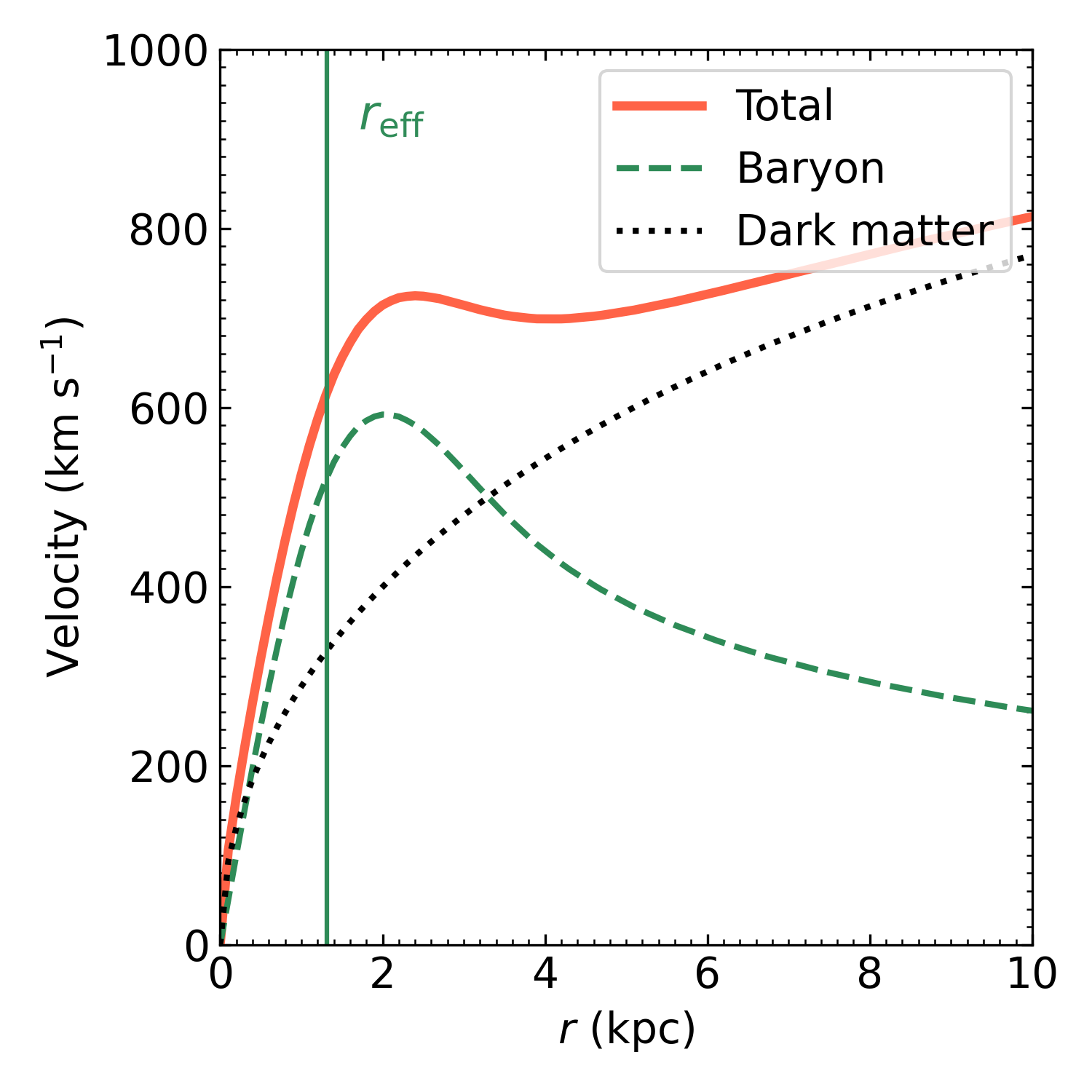}}
\caption{Best-fit results of \textsf{DysmalPy} kinematic modeling of HZ10-C$+$HZ10-E rotation curves. The solid red curve shows the total intrinsic circular velocity corrected by inclination and beam smearing. The baryon and dark matter components are shown by the dashed green and dotted black curves, respectively. The vertical green line corresponds to the disk's effective radius.
}\label{fig:hz10-dysmal}
\end{figure*}

The 1D and 2D marginalized posterior
distributions of the model parameters for HZ10-E$+$HZ10-C system are shown in Fig.~\ref{fig:dysmal_corner}. Circular velocity as a function of radius based on the best-fit results is shown in Fig.~\ref{fig:hz10-dysmal}. 

\begin{figure*}[h]
\centering
{\includegraphics[width = 0.45\textwidth]{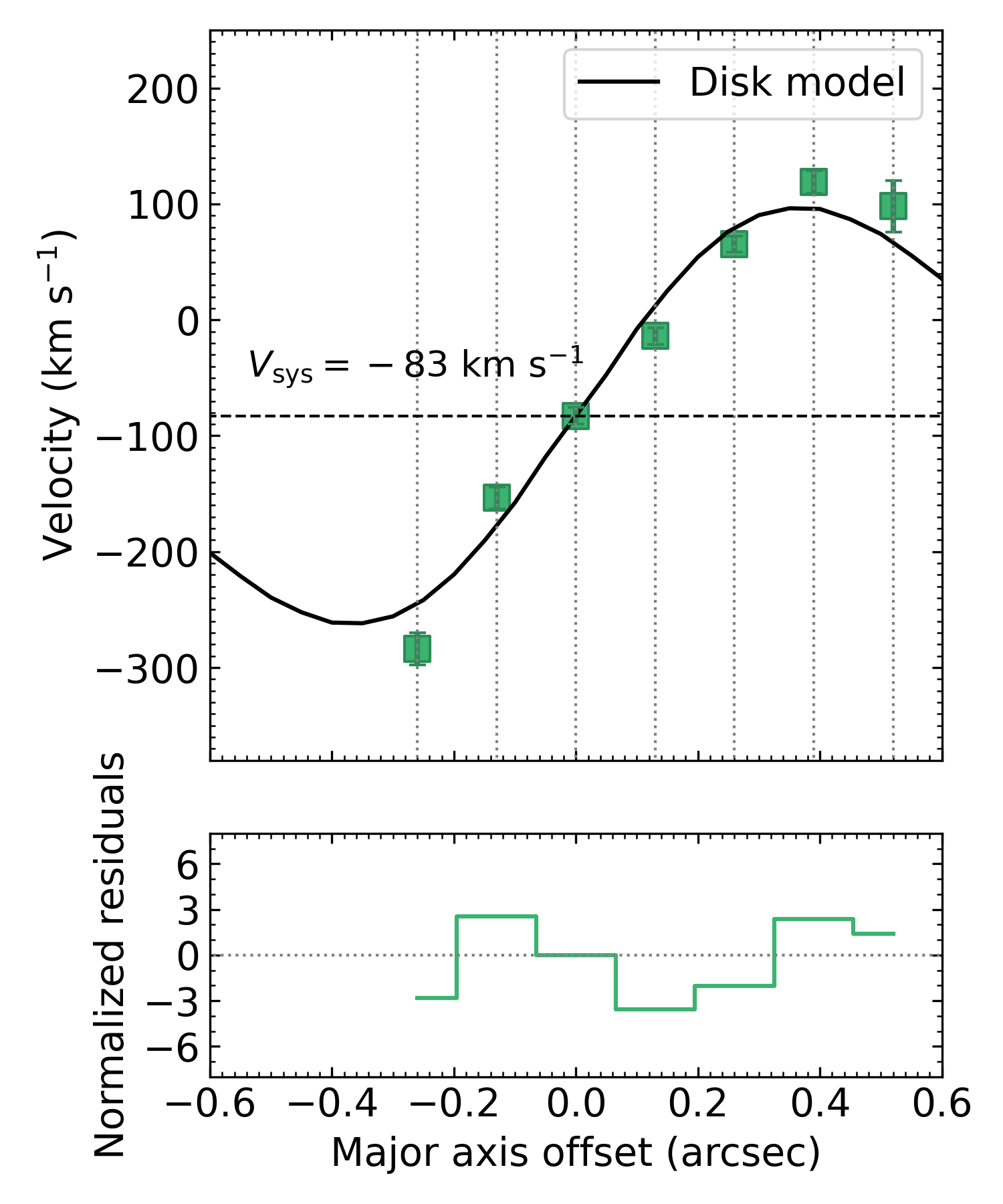}} 
{\includegraphics[width = 0.45\textwidth]{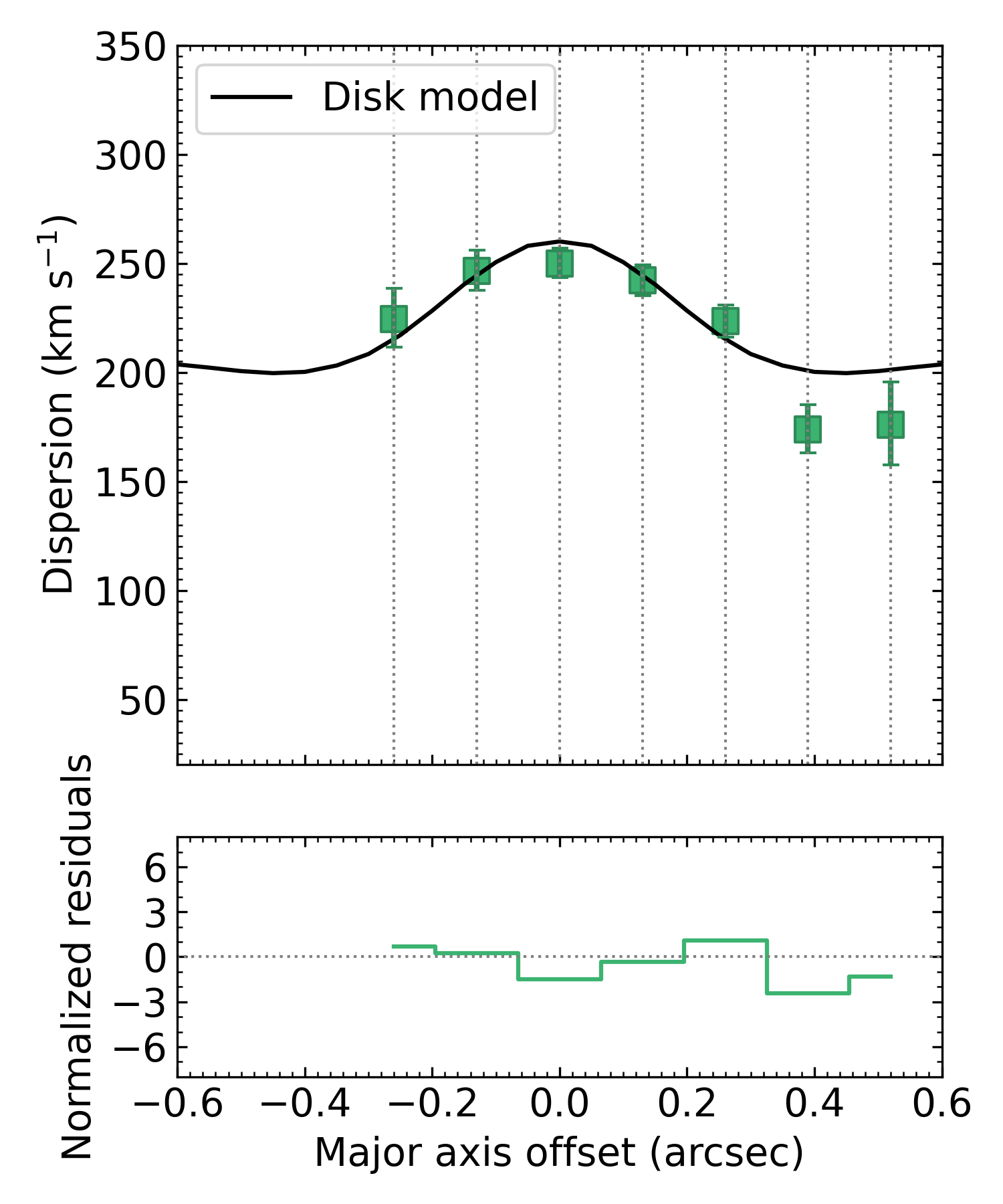}}
\caption{Rotation curves calculated for HZ10-C excluding the impact of HZ10-E component. Data points correspond to the velocity centroids (left panel) and velocity dispersion (right panel) of a single Gaussian fit of the spectra within the circular apertures of a radius of 0.13\arcsec\ as a function of the relative aperture position. Apertures were placed along the major axis with the step of 0.13\arcsec. Green squares result from the single Gaussian spectral fit within the apertures. Zero velocity is calculated for the rest-frame [\CII] frequency at $z=5.6548$. Zero offset corresponds to the aperture position at the same zero point as for the respective PV diagram. Black curves represent the best-fit disk model done with \textsf{DysmalPy}. Normalized residuals represent the difference between data points and the model, divided by the data uncertainties. Note, that we exclude the fist aperture (top panel in Fig.~\ref{fig:hz10-aperture-spectra-rot-HZ10-C}) as the emission in that region is dominated by the HZ10-E component.}\label{fig:rotation-curves-hz10-c}
\end{figure*}

We also show observed and best-fit rotation curves for the HZ10-C component alone after separating in from the HZ10-E in Fig.~\ref{fig:rotation-curves-hz10-c}.

\end{appendix}

\end{document}